\documentclass[11pt,epsfig]{amsart}
\usepackage{amsbsy,amssymb,amscd,amsfonts,latexsym,amstext,delarray,
amsmath,epsfig,graphicx}  \headsep=15pt
\usepackage{hyperref}

\newtheorem{thm}{Theorem}[section]
\newtheorem{prop}[thm]{Proposition}
\newtheorem{cor}[thm]{Corollary}
\newtheorem{lem}[thm]{Lemma}

\newtheorem{defn}[thm]{Definition}
\newtheorem{rem}[thm]{Remark}

\numberwithin{equation}{section}

\def\bA{{\mathbb A}}

\def\bF{{\mathbb F}}

\def\bM{{\mathbb M}}

\def\bW{{\mathbb W}}

\def\C{{\mathbb C}}

\renewcommand{\H}{{\mathbb H}}

\def\Z{{\mathbb Z}}
\def\R{{\mathbb R}}

\def\cA{{\mathcal A}}
\def\cB{{\mathcal B}}
\def\cC{{\mathcal C}}

\def\cE{{\mathcal E}}

\def\cH{{\mathcal H}}

\def\cL{{\mathcal L}}
\def\cM{{\mathcal M}}
\def\cN{{\mathcal N}}

\def\cR{{\mathcal R}}
\def\cS{{\mathcal S}}

\def\cU{{\mathcal U}}

\newcommand{\ie}{{\it i.e.\/}\ }
\newcommand{\eg}{{\it e.g.\/}\ }
\newcommand{\cf}{{\it cf.\/}\ }

\def\text{\hbox}

\def\Diff{{\rm Diff}}

\def\GL{{\rm GL}}

\def\SU{{\rm SU}}

\def\Tr{{\rm Tr}}

\def\qqq{\,,\quad \forall\,}

\def\dirac{{\partial \hspace{-6pt} \slash}}
\def\pslash{{p \!\!\! /}}
\def\cx{{U^{lep}}}
\def\higgs{{\bf H}}

\def\elel{(  \downarrow 1)}
\def\dodo{(   \downarrow 3)}
\def\nunu{(   \uparrow 1)}
\def\upup{(   \uparrow 3)}
\def\mass{Y}

\title
{Gravity and the standard model with   neutrino mixing}

\author[Chamseddine]{Ali H. Chamseddine}
\author[Connes]{Alain Connes}
\author[Marcolli]{Matilde Marcolli}

\address{ A.~Chamseddine: Physics
Department, American University of Beirut, Lebanon}
\email{chams@aub.edu.lb}
\address{A.~Connes: Coll\`ege de France \\
3, rue d'Ulm \\
Paris, F-75005 France\\
I.H.E.S. and Vanderbilt University}
 \email{alain@connes.org}
\address{ M.~Marcolli: Max--Planck Institut f\"ur Mathematik  \\
Vivatsgasse 7, D-53111 Bonn \\ Germany}
\email{marcolli\@@mpim-bonn.mpg.de}

\begin{document}

\maketitle

\begin{abstract}
We present an effective unified theory based on noncommutative
geometry for the standard model with neutrino mixing, minimally
coupled to gravity. The unification is based on the symplectic
unitary group in Hilbert space and on the spectral action. It yields
all the detailed structure of the standard model with several
predictions at unification scale. Besides the familiar predictions
for the gauge couplings  as for GUT theories, it predicts the Higgs
scattering parameter and the sum of the squares of Yukawa couplings.
From these relations one can extract predictions at low energy,
giving in particular a Higgs mass around 170 GeV and a top mass
compatible with present experimental value. The geometric picture
that emerges is that space-time is the product of an ordinary spin
manifold (for which the theory would deliver Einstein gravity) by a
finite noncommutative geometry F. The discrete space F is of
KO-dimension 6 modulo 8 and of metric dimension 0, and accounts for
all the intricacies of the standard model with its spontaneous
symmetry breaking Higgs sector.
\end{abstract}

\tableofcontents

\section{Introduction}

In this paper we present a model based on noncommutative geometry
for the standard model  with massive neutrinos, minimally coupled to
gravity. The model can be thought of as a form of unification, based
on the symplectic unitary group in Hilbert space, rather than on
finite dimensional Lie groups. In particular, the parameters of the
model are set at unification scale and one obtains physical
predictions by running them down through the renormalization group
using the Wilsonian approach. For the renormalizability of the
gravity part of our model one can follow the renormalization
analysis of higher derivatives gravity as in \cite{CoPer} and
\cite{Dono}. Later, we explain in detail how the gravitational
parameters behave.

The input of the model is extremely simple. It consists of the
choice of a finite dimensional algebra, which is natural in the
context of the left--right symmetric models. It is a direct sum
\begin{equation}\label{ALRalg}
\C \oplus \H \oplus \H \oplus M_3(\C),
\end{equation}
where $\H$ is the involutive algebra of quaternions. There is a
natural representation $\cM$ for this algebra, which is the sum of
the irreducible bimodules of odd spin. We show that the fermions of
the standard model can be identified with a basis for a sum of $N$
copies of $\cM$, with $N$ being the number of generations. (We will
restrict ourselves to $N=3$ generations.)

An advantage of working with associative algebras as opposed to Lie
algebras is that the representation theory is more constrained. In
particular a finite dimensional algebra has only a finite number of
irreducible representations, hence a canonical representation in
their sum. The bimodule $\cM$ described above is obtained in this
way by imposing the odd spin condition.

The model we introduce, however, is {\em not} a left--right
symmetric model. In fact, geometric considerations on the form of a
Dirac operator for the algebra \eqref{ALRalg} with the
representation $\cH=\cM^{\oplus 3}$ lead to the identification of a
subalgebra of \eqref{ALRalg} of the form
\begin{equation}\label{AFalg}
\C \oplus \H \oplus M_3(\C) \subset \C \oplus \H \oplus \H\oplus
M_3(\C).
\end{equation}
This will give a model for neutrino mixing which has Majorana mass
terms and a see-saw mechanism.

For this algebra we give a classification of all possible Dirac
operators that give a real spectral triple $(\cA,\cH,D)$, with $\cH$
being the representation described above. The resulting Dirac
operators depend on 31 real parameters, which physically correspond
to the masses for leptons and quarks (including neutrino Yukawa
masses), the angles of the CKM and PMNS matrices, and the Majorana
mass matrix.

This gives a family of geometries $F=(\cA,\cH,D)$ that are
metrically zero-dimensional, but that are of dimension $6 \mod 8$
from the point of view of real $K$-theory.

We consider the product geometry of such a finite dimensional
spectral triple with the spectral triple associated to a
4-dimensional compact Riemannian spin manifold. The bosons of the
standard model, including the Higgs, are obtained as the inner
fluctuations of the Dirac operator of this product geometry. In
particular this gives a geometric interpretation of the Higgs fields
which generate the masses of elementary particles through
spontaneous symmetry breaking. The corresponding mass scale
specifies the inverse size of the discrete geometry $F$. This is in
marked contrast with the grand unified theories where the Higgs
fields are then added by hand to break the GUT symmetry. In our case
the symmetry is broken by a specific choice of the finite geometry,
in the same way as the choice of a specific space-time geometry
breaks the general relativistic invariance group to the much smaller
group of isometries of a given background.

Then we apply to this product geometry a general formalism for
spectral triples, namely the spectral action principle. This is a
universal action functional on spectral triples, which is
``spectral'', in the sense that it depends only on the spectrum of
the Dirac operator and is of the form
\begin{equation}\label{SpAct}
\Tr(f(D/\Lambda)),
\end{equation}
where $\Lambda$ fixes the energy scale and $f$ is a test function.
The function $f$ only plays a role through its momenta $f_0$, $f_2$,
and $f_4$ where $f_k=\int_0^\infty f(v)v^{k-1}dv$ for $k>0$ and
$f_0=f(0)$. (\cf Remark \ref{f0f2f4} below for the relation with the
notations of \cite{cc2}). These give $3$ additional real parameters
in the model. Physically, these are related to the coupling
constants at unification, the gravitational constant, and the
cosmological constant.

The action functional \eqref{SpAct}, applied to inner fluctuations,
only accounts for the bosonic part of the model. In particular, in
the case of classical Riemannian manifolds, where no inner
fluctuations are present, one obtains from \eqref{SpAct} the
Einstein--Hilbert action of pure gravity. This is why gravity is
naturally present in the model, while the other gauge bosons arise
as a consequence of the noncommutativity of the algebra of the
spectral triple.

The coupling with fermions is obtained by including an additional
term
\begin{equation}\label{fermSpAct}
\Tr(f(D/\Lambda))+ \frac{1}{2} \langle J \psi, D\psi\rangle,
\end{equation}
where $J$ is the real structure on the spectral triple, and $\psi$
is an element in the space $\cH$, viewed as a classical fermion, \ie
as a Grassman variable. The fermionic part of the Euclidean
functional integral is given by the Pfaffian of the antisymmetric
bilinear form $\langle J \psi', D\psi\rangle$. This, in particular,
gives a substitute for Majorana fermions in Euclidean signature (\cf
\eg \cite{InSu}, \cite{NiWa}).

We show that the gauge symmetries of the standard model, with the
correct hypercharge assignment, are obtained as a subgroup of the
symplectic unitary group of Hilbert space given by the adjoint
representation of the unimodular unitary group of the algebra.

We prove that the full Lagrangian (in Euclidean signature) of the
standard model minimally coupled to gravity, with neutrino mixing
and Majorana mass terms, is the result of the computation of the
asymptotic formula for the spectral action functional
\eqref{fermSpAct}.

The positivity of the test function $f$ in \eqref{SpAct} ensures the
positivity of the action functional before taking the asymptotic
expansion. In general, this does not suffice to control the sign of
the terms in the asymptotic expansion. In our case, however, this
determines the positivity of the momenta $f_0$, $f_2$, and $f_4$.
The explicit calculation then shows that this implies that the signs
of all the terms are the expected physical ones.

We obtain  the usual Einstein--Hilbert action with a cosmological
term, and in  addition the square of the Weyl curvature and a
pairing of the scalar curvature with the square of the Higgs field.
The Weyl curvature term does not affect gravity at low energies, due
to the smallness of the Planck length.
The coupling of the Higgs to the scalar curvature was discussed by
Feynman in \cite{feynmgrav}.

We show that the general form of the Dirac operator for the finite
geometry gives a see-saw mechanism for the neutrinos (\cf
\cite{MoPa}). The large masses in the Majorana mass matrix are
obtained in our model as a consequence of the equations of motion.

Our model makes three predictions, under the assumption of the ``big
desert'', in running down the energy scale from unification.

The first prediction is the relation $g_2=g_3=\sqrt{5/3}\,g_1$
between the coupling constants at unification scale, exactly as in
the GUT models (\cf \eg \cite{MoPa} \S 9.1 for $SU(5)$ and
\cite{CMGMP} for $SO(10)$). In our model this comes directly from
the computation of the terms in the asymptotic formula for the
spectral action. In fact, this result is a feature of any model that
unifies the gauge interactions, without altering the fermionic
content of the model.

The second prediction is the Higgs scattering parameter $\alpha_h$
at unification scale. From this condition, One obtains a prediction
for the Higgs mass as a function of the $W$ mass, after running it
down through the renormalization group equations. This gives a Higgs
mass of the order of 170 GeV  and agrees with the ``big desert''
prediction of the minimal standard model (\cf \cite{Sher}).

The third prediction is a mass relation between the Yukawa masses of
fermions and the $W$ boson mass, again valid at unification scale.
This is of the form
\begin{equation}\label{MassRel}
\sum_{{\rm generations}}  m_e^2 + m_\nu^2 + 3 m_d^2 + 3 m_u^2 = 8
M_W^2.
\end{equation}
After applying the renormalization group to the Yukawa couplings,
assuming that the Yukawa coupling for the $\nu_\tau$ is comparable
to the one for the top quark, one obtains good agreement with the
measured value.

Moreover, we can extract from the model predictions for the
gravitational constant involving the parameter $f_2/f_0$. The
reasonable assumption that the parameters $f_0$ and $f_2$ are of the
same order of magnitude yields a realistic value for the Newton
constant.

In addition to these predictions, a main advantage of the model is
that it gives a geometric interpretation for all the parameters in
the standard model. In particular, this leaves room for predictions
about the Yukawa couplings, through the geometry of the Dirac
operator.

The properties of the finite geometries described in this paper
suggest possible approaches.

For instance, there are examples of spectral triples of metric
dimension zero with a different $KO$-homology dimension, realized by
homogeneous spaces over quantum groups \cite{dab}.

Moreover, the data parameterizing the Dirac operators of our finite
geometries can be described in terms of some classical moduli spaces
related to double coset spaces of the form $K\backslash (G\times
G)/(K\times K)$ for $G$ a reductive group and $K$ the maximal
compact acting diagonally on the left. The renormalization group
defines a flow on the moduli space.

Finally, the product geometry is $10$-dimensional from the
$KO$-homology point of view and may perhaps be realized as a low
energy truncation, using the type of compact fibers that are
considered in string theory models (\cf \eg \cite{FroGaw}).

Naturally, one does not really expect the ``big desert'' hypothesis
to be satisfied. The fact that the experimental values show that the
coupling constants do not exactly meet at unification scale is an
indication of the presence of new physics. A good test for the
validity of the above approach will be whether a fine tuning of the
finite geometry can incorporate additional experimental data at
higher energies. The present paper shows that the modification of
the standard model required by the phenomenon of neutrino mixing in
fact resulted in several improvements on the previous descriptions
of the standard model via noncommutative geometry.

In summary we have shown that the intricate Lagrangian of the
standard model coupled with gravity can be obtained from a very
simple modification of space-time geometry provided one uses the
formalism of noncommutative geometry. The model contains several
predictions and the corresponding section 5 of the paper can be read
directly, skipping the previous sections. The detailed comparison in
section 4 of the spectral action with the standard model contains
several steps that are familiar to high energy particle physicists
but less to mathematicians. Sections 2 and 3 are more mathematical
but for instance the relation between classical moduli spaces and
the CKM matrices can be of interest to both physicists and
mathematicians.

The results of this paper are a development of the preliminary
announcement of \cite{CoSMneu}.

\bigskip

\noindent {\bf Acknowledgements.} It is a pleasure to acknowledge
the independent preprint by John Barrett \cite{Barrett} with a
solution of the fermion doubling problem. The first author is
supported by NSF Grant Phys-0601213. The second author thanks G.
Landi and T. Schucker, the third author thanks Laura Reina and Don
Zagier for useful conversations. We thank the Newton Institute where
part of this work was done.

\bigskip

\section{The finite geometry}

\subsection{The left-right symmetric algebra}\label{LeftRightSect}\hfill\medskip

The main input for the model we are going to describe is the choice
of a finite dimensional involutive algebra of the form
\begin{equation}\label{algebralr}
\cA_{LR}=\,\C\oplus \H_L\oplus \H_R\oplus M_3(\C).
\end{equation}
This is the direct sum of the matrix algebras $M_N(\C)$ for $N=1, 3$
with two copies of the algebra $\H$ of quaternions, where the
indices L, R are just for book-keeping. We refer to
\eqref{algebralr} as the ``left-right symmetric algebra" \cite{CFF}.

By construction $\cA_{LR}$ is an involutive algebra, with involution
\begin{equation}\label{involution}
(\lambda,q_L,q_R,m)^* = (\bar\lambda,\bar q_L,\bar q_R,m^*),
\end{equation}
where $q\mapsto \bar q$ denotes the involution of the algebra of
quaternions. The algebra $\cA_{LR}$ admits a natural subalgebra
$\C\oplus M_3(\C)$, corresponding to integer spin, which is an
algebra over $\C$. The subalgebra $\H_L\oplus \H_R$, corresponding
to half-integer spin, is an algebra over $\R$.

\subsection{The bimodule $\cM_F$}\label{bimodsect}\hfill\medskip

Let $\cM$ be a bimodule over an involutive algebra $\cA$. For $u\in
\cA$ unitary, \ie such that $uu^*=u^*u=1$, one defines ${\rm Ad}(u)$
 by  ${\rm Ad}(u)\xi=u\xi u^* \,,\forall \xi\in \cM$.

\begin{defn}\label{oddbimod} Let $\cM$ be an $\cA_{LR}$-bimodule. Then $\cM$ is
 {\em odd} iff   the adjoint action of
$s=(1,-1,-1,1)$ fulfills ${\rm Ad}(s)=-1$.
\end{defn}

Let $\cA_{LR}^0$ denote the opposite algebra of $\cA_{LR}$.

\begin{lem}\label{oddred}
An odd bimodule $\cM$ is a representation of the reduction
$\cB=(\cA_{LR}\otimes_\R \cA_{LR}^0)_p$ of $\cA_{LR}\otimes_\R
\cA_{LR}^0$ by the projection $p=\frac 12\,(1-s\otimes s^0)$. This
subalgebra is an algebra over $\C$.
\end{lem}

\proof The result follows directly from the action of
$s=(1,-1,-1,1)$ in Definition \ref{oddbimod}.
\endproof

Since $\cB=(\cA_{LR}\otimes_\R \cA_{LR}^0)_p$ is an algebra over
$\C$, we restrict to consider complex representations.

\smallskip

\begin{defn}\label{contragrad}
One defines the contragredient bimodule of a bimodule $\cM$ as the
complex conjugate space
\begin{equation}\label{oppbim}
\cM^0=\{\bar \xi\;;\;\xi\in \cM\}\,,\quad
a\,\bar\xi\,b=\,\overline{b^*\xi\,a^*}\qqq \,a\,,\;b\in \cA_{LR} .
\end{equation}
\end{defn}

\smallskip

The algebras $M_N(\C)$ and $\H$ are isomorphic to their opposite
algebras (by $m\mapsto m^t$ for matrices and $q\mapsto \bar q $ for
quaternions. We use this antiisomorphism to obtain a representation
$\pi^0$ of the opposite algebra from a representation $\pi$ .

We follow the physicists convention to denote an irreducible
representation by its dimension in boldface. So, for instance, ${\bf
3}^0$ denotes the 3-dimensional irreducible representation of the
opposite algebra $M_3(\C)$.

\begin{prop} \label{propbimdec} Let $\cM_F$  be the direct sum
of all inequivalent irreducible odd $\cA_{LR}$-bimodules.
\begin{itemize}
  \item The dimension of the complex vector space $\cM_F$ is $32$.
  \item  The $\cA_{LR}$-bimodule $\cM_F=\cE\oplus \cE^0$ is the
  direct sum of the bimodule
\begin{equation}\label{bimdec0}
\cE={\bf 2}_L\otimes {\bf 1}^0\oplus {\bf 2}_R\otimes {\bf
1}^0\oplus {\bf 2}_L\otimes {\bf 3}^0\oplus {\bf 2}_R\otimes {\bf
3}^0
\end{equation}
with its contragredient $\cE^0$.
  \item The $\cA_{LR}$-bimodule $\cM_F$ is  isomorphic to the
contragredient bimodule $\cM_F^0$ by the antilinear isometry $J_F$
given by
\begin{equation}\label{mapj}
J_F(\xi,\bar\eta)=(\eta,\bar \xi) \qqq \xi\,,\;\eta \in \cE
\end{equation}
\item One has
\begin{equation}\label{mapj1}
J^2=1\,,\quad \xi\,b=Jb^*J\,\xi \qqq \xi \in \cM_F\,,\;b\in \cA_{LR}
\end{equation}
\end{itemize}
\end{prop}

\proof The first two statements follow from the structure of the
algebra $\cB$ described in the following lemma.

\begin{lem} \label{lembimdec} The algebra $\cB=(\cA_{LR}\otimes_\R \cA_{LR}^0)_p$ is
the direct sum of $4$ copies of the algebra $M_2(\C)\oplus M_6(\C)$.

The sum of irreducible representations of $\cB$ has dimension $32$
and is given by
\begin{equation}\label{bimdec}
{\bf 2}_L\otimes {\bf 1}^0\oplus {\bf 2}_R\otimes {\bf 1}^0\oplus
{\bf 2}_L\otimes {\bf 3}^0\oplus {\bf 2}_R\otimes {\bf 3}^0\oplus
{\bf 1}\otimes {\bf 2}_L^0 \oplus {\bf 1}\otimes {\bf 2}_R^0 \oplus
{\bf 3}\otimes {\bf 2}_L^0 \oplus {\bf 3}\otimes {\bf 2}_R^0
  \end{equation}
\end{lem}

\proof By construction one has
$$
\cB=(\H_L\oplus \H_R)\otimes_\R (\C\oplus
 M_3(\C))^0\oplus (\C\oplus
 M_3(\C))\otimes_\R (\H_L\oplus \H_R)^0
$$
 Thus the first result follows from the isomorphism:
$$
\H \otimes_\R \C=M_2(\C)\,,\quad \H \otimes_\R M_3(\C)=M_6(\C)
$$
The complex algebra $M_N(\C)$ admits only one irreducible
representation and the latter has dimension $N$. Thus the sum of the
irreducible representations of $\cB$ is given by \eqref{bimdec}. The
dimension of the sum of irreducible representations is $ 4\times
2+4\times 6=32 $.
\endproof

\smallskip
To end the proof of Proposition \ref{propbimdec} notice that by
construction $\cM_F$ is the direct sum $\cE\oplus \cE^0$ of the
bimodule \eqref{bimdec0} with its contragredient, and that the map
\eqref{mapj} gives the required antilinear isometry. Note moreover
that one has \eqref{mapj1} using \eqref{oppbim}.
\endproof

\subsection{Real spectral triples}\label{spectripsect}\hfill\medskip

A noncommutative geometry is given by a representation theoretic
datum of spectral nature. More precisely, we have the following
notion.

\begin{defn} \label{spectripdef} A spectral triple $({\mathcal
A},{\mathcal H},D)$ is given by an involutive unital algebra $\cA$
represented as operators in a Hilbert space $\cH$ and a self-adjoint
operator $D$ with compact resolvent such that all commutators
$[D,a]$ are bounded for $a\in \cA$.

A spectral triple is {\em even} if the Hilbert space $\cH$ is
endowed with a $\Z/2$- grading $\gamma$ which commutes with any
$a\in \cA$ and anticommutes with $D$.
\end{defn}

The notion of real structure (\cf \cite{Coreal}) on a spectral
triple $(\cA,\cH,D)$, is intimately related to real $K$-homology
(\cf \cite{Atiyah}) and the properties of the charge conjugation
operator.

\smallskip
\begin{defn}\label{realstr}
A real structure of $KO$-dimension  $n\in \Z/8$ on a spectral triple
$(\cA,\cH,D)$ is an antilinear isometry $J: \cH \to \cH$, with the
property that
\begin{equation}\label{per8}
J^2 = \varepsilon, \ \ \ \ JD = \varepsilon' DJ, \ \ \text{and} \ \
J\gamma = \varepsilon'' \gamma J \, \text{(even case)}.
\end{equation}
The numbers $\varepsilon ,\varepsilon' ,\varepsilon'' \in \{ -1,1\}$
are a function of $n \mod 8$ given by

\begin{center}
\begin{tabular}
{|c| r r r r r r r r|} \hline {\bf n }&0 &1 &2 &3 &4 &5 &6 &7 \\
\hline \hline
$\varepsilon$  &1 & 1&-1&-1&-1&-1& 1&1 \\
$\varepsilon'$ &1 &-1&1 &1 &1 &-1& 1&1 \\
$\varepsilon''$&1 &{}&-1&{}&1 &{}&-1&{} \\  \hline
\end{tabular}
\end{center}

Moreover, the action of $\cA$ satisfies the commutation rule
\begin{equation}\label{comm-rule}
[a,b^0] = 0 \quad \forall \, a,b \in \cA,
\end{equation}
where
\begin{equation}\label{b0}
b^0 = J b^* J^{-1} \qquad \forall b \in \cA,
\end{equation}
and the operator $D$ satisfies the order one condition:
\begin{equation}\label{order1}
[[D,a],b^0] = 0 \qquad \forall \, a,b \in \cA \, .
\end{equation}

A spectral triple endowed with a real structure is called a real
spectral triple.
\end{defn}

\medskip
A key role of the real structure $J$ is in defining the {\em adjoint
action} of the unitary group $\cU$ of the algebra $\cA$ on the
Hilbert space $\cH$. In fact, one defines a right $\cA$-module
structure on $\cH$ by
\begin{equation}\label{bimodule}
\xi\,b=\,b^0\,\xi\qqq \;\xi \in \cH\,,\quad b\in \cA .
\end{equation}
The unitary group of the algebra $\cA$ then acts by the ``adjoint
representation" on $\cH$ in the form
\begin{equation}\label{adjact}
\cH \ni \xi \mapsto {\rm Ad}(u)\,\xi=u\,\xi\,u^*\qqq \;\xi \in
\cH\,,\quad u\in \cA \,,\quad u\,u^*=u^*\,u=1\, .
\end{equation}

\begin{defn}\label{Omega1def}
Let $\Omega^1_D$ denote the $\cA$-bimodule
\begin{equation}\label{Omega1D}
\Omega_D^1=\{ \sum_j a_j [D,b_j]\, | \,  a_j,b_j\in \cA\}.
\end{equation}
\end{defn}

\begin{defn}\label{innfluc}
The inner fluctuations of the metric are given by
\begin{equation}\label{innerfluc1} D\to D_A=D+A+\varepsilon'
\,J\,A\,J^{-1}
\end{equation}
where $A\in\Omega^1_D$, $A=A^*$ is a self-adjoint operator of the
form
\begin{equation}\label{innerfluc2}
A=\sum_j a_j[D,b_j]\,,\quad a_j,b_j\in \cA.
\end{equation}
\end{defn}

\smallskip

For any gauge potential $A\in \Omega^1_D$, $A=A^*$ and any unitary
$u\in \cA$, one has
$$
{\rm Ad}(u)(D+A+\varepsilon' \,J\,A\,J^{-1}){\rm
Ad}(u^*)=\,D+\gamma_u(A)+ \varepsilon' \,J\,\gamma_u(A)\,J^{-1}
$$
where $\gamma_u(A)=\,u\,[D,u^*]+\,u\,A\,u^*$ (\cf \cite{CoSM}).

\medskip
\subsection{The subalgebra and the order one condition}\hfill
\medskip

We let $\cH_F$ be the sum of $N=3$ copies of the $\cA_{LR}$-bimodule
$\cM_F$ of Proposition \ref{propbimdec}, that is,
\begin{equation}\label{HFMF}
\cH_F = \cM_F^{\oplus 3}.
\end{equation}

\begin{rem}\label{Ngener}
The multiplicity $N=3$ here is an input, and it corresponds to the
number of particle generations in the standard model. The number of
generations is not predicted by our model in its present form and
has to be taken as an input datum.
\end{rem}

We define the $\Z/2$-grading $\gamma_F$   by
\begin{equation}\label{gammafc}
 \gamma_F =\,c\,-\,J_F\,c\,J_F \,,\quad
c=(0,1,-1,0)\in \cA_{LR} .
\end{equation}
One then checks that
\begin{equation}\label{sixmodeight}
J_F^2=1\,,\quad J_F\,\gamma_F=-\,\gamma_F\,J_F .
\end{equation}
The relation \eqref{sixmodeight}, together with the commutation of
$J_F$ with the Dirac operators, is characteristic of $KO$-dimension
equal to $6$ modulo $8$ (\cf Definition \ref{realstr}).

\smallskip

By Proposition \ref{propbimdec} one can write $\cH_F$ as the direct
sum
\begin{equation}\label{Hfbarf}
\cH_F=\cH_f\oplus \cH_{\bar f}
\end{equation}
of $N=3$ copies of $\cE$ of \eqref{bimdec0} with the contragredient
bimodule, namely
\begin{equation}\label{HfEE0}
\cH_f = \cE\oplus\cE\oplus \cE, \ \ \ \ \cH_{\bar f} = \cE^0\oplus
\cE^0\oplus \cE^0.
\end{equation}

The left action of $\cA_{LR}$ splits as the sum of a representation
$\pi$ on $\cH_f$ and a representation $\pi'$ on $\cH_{\bar f}$.

These representations of $\cA_{LR}$ are disjoint (\ie they have no
equivalent subrepresentations). As shown in Lemma \ref{lemoffdiag}
below, this precludes the existence of operators $D$ in $\cH_F$ that
fulfill the order one condition \eqref{order1} and intertwine the
subspaces $\cH_f$ and $\cH_{\bar f}$.

We now show that the existence of such intertwining of $\cH_f$ and
$\cH_{\bar f}$ is restored by passing to a unique subalgebra of
maximal dimension in $\cA_{LR}$.

\begin{prop}\label{nameless} Up to an automorphism of $\cA_{LR}$,
there exists a unique subalgebra $\cA_F \subset \cA_{LR}$ of maximal
dimension admitting an off diagonal Dirac operators, namely
operators that intertwine the subspaces $\cH_f$ and $\cH_{\bar f}$
of $\cH_F$. The subalgebra is given by
\begin{equation}\label{subalgF}
\cA_F =\{ (\lambda,q_L,\lambda,m)\;|\;\lambda\in \C\,,\; q_L\in \H
\,,\; m\in M_3(\C)\}\sim \C\oplus \H\oplus M_3(\C).
\end{equation}
\end{prop}

\proof For any operator $T\;:\;\cH_f\to \cH_{\bar f}$ we let
\begin{equation}\label{comrel}
\cA(T)=\{ b\in \cA_{LR}\;|\;
\pi'(b)T=T\pi(b)\,,\;\pi'(b^*)T=T\pi(b^*)\} .
\end{equation}
It is by construction an involutive unital subalgebra of $\cA_{LR}$.

We prove the following preliminary result.

\begin{lem} \label{lemoffdiag}
Let $\cA\subset \cA_{LR}$ be an involutive unital subalgebra of
$\cA_{LR}$. Then the following properties hold.
\begin{enumerate}

\item If the restriction of $\pi$ and $\pi'$ to $\cA$ are disjoint,
then there is no off diagonal Dirac operator for $\cA$.

\item If there exists an off diagonal Dirac for $\cA$, then
there exists a pair $e$, $e'$ of minimal projections in the
commutants of $\pi(\cA_{LR})$ and $\pi'(\cA_{LR})$ and an operator
$T$ such that $e'Te=T\neq 0$ and $\cA\subset \cA(T)$.
\end{enumerate}
\end{lem}

\proof 1) First the order one condition shows that $[D, a^0]$ cannot
have an off diagonal part since it is in the commutant of $\cA$.
Conjugating by $J$ shows that $[D, a]$ cannot have an off diagonal
part. Thus the off diagonal part $D_{off}$ of $D$ commutes with
$\cA$ \ie $[D_{off}, a]=0$, and $D_{off}=0$ since there are no
intertwining operators.

2) By 1) the restrictions of  $\pi$ and $\pi'$ to $\cA$ are not
disjoint and there exists a non-zero operator $T$ such that
$\cA\subset \cA(T)$. For any elements $x$, $x'$ of the commutants of
$\pi$ and $\pi'$, one has
$$
\cA(T)\subset \cA(x'Tx)
$$
since $\pi'(b)T=T\pi(b)$ implies $\pi'(b)x'Tx=x'Tx\pi(b)$. Taking a
partition of unity by minimal projections there exists a pair $e$,
$e'$ of minimal projections in the commutants of $\pi$ and $\pi'$
such that $e'Te\neq 0$ so that one can assume $e'Te=T\neq 0$.
\endproof

We now return to the proof of Proposition \ref{nameless}.

Let $\cA\subset \cA_{LR}$ be an involutive unital subalgebra. If it
admits an off diagonal Dirac, then by Lemma \ref{lemoffdiag} it is
contained in a subalgebra $\cA(T)$ with the support of $T$ contained
in a minimal projection of the commutant of $\pi(\cA_{LR})$ and the
range of $T$ contained in the range of a minimal projection of the
commutant of $\pi'(\cA_{LR})$.

This reduces the argument to two cases, where the representation
$\pi$ is the irreducible representation of $\H$ on $\C^2$ and $\pi'$
is either the representation of $\C$ in $\C$ or the irreducible
representation of $M_3(\C)$ on $\C^3$.

In the first case the support $E$ of $T$ is one dimensional. The
commutation relation \eqref{comrel} defines the subalgebra $\cA(T)$
from the condition $\lambda T \xi = T q \xi$, for all $\xi \in E$,
which implies $\lambda \xi- q \xi =0$. Thus, in this case the
algebra $\cA(T)$ is the pullback of
\begin{equation}\label{graphCH}
\{(\lambda,q)\in \C\oplus \H\;|\;q\,\xi=\lambda\, \xi \qqq \xi \in
E\}
\end{equation}
under the projection on $\C\oplus \H$ from $\cA_{LR}$. The algebra
\eqref{graphCH} is the graph of an embedding of $\C$ in $\H$. Such
an embedding is unique up to inner automorphisms of $\H$. In fact,
the embedding is determined by the image of $i\in \C$ and all
elements in $\H$ satisfying $x^2=-1$ are conjugate.

The corresponding subalgebra $\cA_F\subset \cA_{LR}$ is of real
codimension $4$. Up to the exchange of the two copies of $\H$ it is
given by \eqref{subalgF}.

In the second case the operator $T$ has at most two dimensional
range $\cR(T)$. This range is invariant under the action $\pi'$ of
the subalgebra $\cA$ and so is its orthogonal since $\cA$ is
involutive.

Thus, in all cases the $M_3(\C)$-part of the subalgebra is contained
in the algebra of $2\oplus 1$ block diagonal $3\times 3$ matrices
which is of real codimension $8$ in $M_3(\C)$. Hence $\cA$ is of
codimension at least $8>4$ in $\cA_{LR}$.

It remains to show that the subalgebra \eqref{subalgF} admits off
diagonal Dirac operators. This  follows from Theorem
\ref{diracclassF} below.
\endproof

\subsection{Unimodularity and hypercharges}\hfill\medskip

The unitary group of an involutive algebra $\cA$ is given by
$$
{\rm U}(\cA)=\{u\in \cA\;|\;uu^*=u^*u=1\}
$$
In our context we define the special unitary group $\SU(\cA)\subset
{\rm U}(\cA)$ as follows.

\begin{defn} \label{unimoddef1} We let $\SU(\cA_F)$ be the subgroup of ${\rm U}(\cA_F)$
defined by
$$
\SU(\cA_F)=\{u\in {\rm U}(\cA_F)\;:\;\det(u)=1\}
$$
where $\det(u)$ is the determinant of the action of $u$ in $\cH_F$.
\end{defn}

\medskip

We now describe the group $\SU(\cA_F)$ and its adjoint action.

As before, we denote by ${\bf 2}$ the 2-dimensional irreducible
representation of $\H$ of the form
\begin{equation}\label{qalphabeta}
 \left(\begin{array}{cc} \alpha & \beta \\
-\bar\beta & \bar \alpha \end{array}\right),
\end{equation}
with $\alpha,\beta\in \C$.

\begin{defn}\label{updownarrows}
We let $|\uparrow \rangle$ and $|\downarrow \rangle$ be the basis of
the irreducible representation ${\bf 2}$ of $\H$ of
\eqref{qalphabeta} for which the action of $\lambda \in \C\subset
\H$ is diagonal with eigenvalues $\lambda$ on $|\uparrow\rangle$ and
$\bar \lambda$ on $|\downarrow\rangle$.
\end{defn}

In the following, to simplify notation, we write $\uparrow$ and
$\downarrow$ for the vectors $|\uparrow \rangle$ and $|\downarrow
\rangle$.

\smallskip

\begin{rem}\label{remupdown}{\rm
The notation $\uparrow$ and $\downarrow$ is meant to be suggestive
of ``up" and ``down" as in the first generation of quarks, rather
than refer to spin states. In fact, we will see in Remark
\ref{fermionbasis} below that the basis of $\cH_F$ can be naturally
identified with the fermions of the standard model, with the result
of the following proposition giving the corresponding hypercharges.}
\end{rem}

\smallskip

\begin{prop}\label{smhypercharges}
\begin{enumerate}
  \item Up to a finite abelian group, the group $\SU(\cA_F)$ is of
  the form
\begin{equation}\label{SUAF}
\SU(\cA_F)\sim {\rm U}(1)\times \SU(2)\times\SU(3).
\end{equation}
  \item The  adjoint action  of the ${\rm U}(1)$
  factor is given by multiplication of the basis vectors in $\cH_f$
  by the following powers of $\lambda\in {\rm U}(1)$:
\begin{equation}\label{hyperchargeslist}
\begin{matrix}\hbox{~~~~~}&\uparrow \otimes {\bf 1}^0&\downarrow \otimes {\bf 1}^0&\uparrow \otimes {\bf 3}^0&
\downarrow \otimes {\bf 3}^0\\
&&&&\cr
{\bf 2}_L&-1&-1&\frac{1}{3}&\frac{1}{3}\\
&&&&\\
{\bf 2}_R&0&-2&\frac{4}{3}&-\frac{2}{3}\end{matrix}
\end{equation}
 \end{enumerate}
 \end{prop}

\proof 1) Let $u=(\lambda ,q,m)\in {\rm U}(\cA_F)$. The determinant
of the action of $u$ on the subspace $\cH_f$ is equal to $1$ by
construction since a unitary quaternion has determinant $1$. Thus
$\det(u)$ is the determinant of the action $\pi'(u)$ on $\cH_{\bar
f}$. This representation is given by $4\times 3=12$ copies of the
irreducible representations ${\bf 1}$ of $\C$ and ${\bf 3}$ of
$M_3(\C)$. (The $4$ is from ${\bf 2}_L^0\oplus {\bf 2}_R^0$ and the
$3$ is the additional overall multiplicity of the representation
given by the number $N=3$ of generations.)

Thus, we have
$$
\det(u)=\,\lambda^{12}\,\det(m)^{12} .
$$
Thus, $\SU(\cA_F)$ is the product of the group $\SU(2)$, which is
the unitary group of $\H$, by the fibered product  $G={\rm
U}(1)\times_{12}{\rm U}(3)$ of pairs $(\lambda,m)\in {\rm
U}(1)\times{\rm U}(3)$ such that $\lambda^{12}\,\det(m)^{12}=1$.

One has an exact sequence
\begin{equation}\label{groupG}
1\to \mu_{3}\to {\rm U}(1)\times \SU(3)\to G\stackrel{\mu}{\to}
\mu_{12}\to 1 ,
\end{equation}
where $\mu_{N}$ is the group of roots of unity of order $N$ and the
maps are as follows. The last map $\mu$ is given by
$\mu(\lambda,m)=\,\lambda \,\det(m)$. By definition of $G$, the
image of the map $\mu$ is the group $\mu_{12}$ of 12th roots of
unity. The kernel of $\mu$ is the subgroup $G_0\subset G$ of pairs
$(\lambda,m)\in {\rm U}(1)\times{\rm U}(3)$ such that
$\lambda\,\det(m)=1$.

The map ${\rm U}(1)\times \SU(3)\to G$ is given  by
$(\lambda,m)\mapsto (\lambda^3,\lambda^{-1}\,m)$. Its image is
$G_0$. Its kernel is the subgroup of ${\rm U}(1)\times \SU(3)$ of
pairs $(\lambda,\lambda\,1_3)$ where $\lambda\in \mu_3$ is a cubic
root of $1$ and $1_3$ is the unit $3\times 3$ matrix.

Thus we obtain an exact sequence of the form
\begin{equation}\label{groupG1}
1\to \mu_{3}\to {\rm U}(1)\times \SU(2)\times \SU(3)\to
\SU(\cA_F)\to \mu_{12}\to 1 .
\end{equation}

\smallskip
2) Up to a finite abelian group, the ${\rm U}(1)$ factor of
$\SU(\cA_F)$  is  the subgroup of elements of $\SU(\cA_F)$ of the
form $u(\lambda)=(\lambda,1,\lambda^{-1/3}1_3)$, where $\lambda\in
\C$, with $|\lambda|=1$. We ignore the ambiguity in the cubic root.

Let us compute the action of ${\rm Ad}(u(\lambda))$. One has $ {\rm
Ad}(u)=\,u\,(u^*)^0=\,u\,b^0$ with $b=(\bar\lambda
,1,\lambda^{1/3}1_3)$.

This gives the required table as in \eqref{hyperchargeslist} for the
restriction to the multiples of the left action ${\bf 2}_L$. In
fact, the left action of $u$ is trivial there.

The right action of $b=(\bar\lambda ,1,\lambda^{1/3}1_3)$ is by
$\bar\lambda$ on the multiples of ${\bf 1}^0$ and by
$\lambda^{1/3}1_3^t$ on multiples of ${\bf 3}^0$.

For the restriction to the multiples of the left action ${\bf 2}_R$
one needs to take into account the left action of $u$. This acts by
$\lambda$ on $\uparrow $ and $\bar \lambda$ on $\downarrow$. This
adds a $\pm 1$ according to whether the arrow points up or down.
\endproof

\smallskip

\begin{rem}\label{roots1rem}{\rm
Notice how the finite groups $\mu_3$ and $\mu_{12}$ in the exact
sequence \eqref{groupG1} are of different nature from the physical
viewpoint, the first arising from the center of the color $U(3)$,
while the latter depends upon the presence of three generations. }
\end{rem}

\medskip

We consider the linear basis for the finite dimensional Hilbert
space $\cH_F$ obtained as follows. We denote by
$f^\kappa_{\uparrow,3,L}$ the basis of $\uparrow_L \otimes {\bf
3}^0$, by $f^\kappa_{\uparrow,3,R}$ the basis of $\uparrow_R \otimes
{\bf 3}^0$, by $f^\kappa_{\downarrow,3,L}$ the basis of
$\downarrow_L \otimes {\bf 3}^0$, and by $f^\kappa_{\downarrow,3,R}$
the basis of $\downarrow_R \otimes {\bf 3}^0$. Similarly, we denote
by $f^\kappa_{\uparrow,1,L}$ the basis of $\uparrow_L \otimes {\bf
1}^0$, by $f^\kappa_{\uparrow,1,R}$ the basis of $\uparrow_R \otimes
{\bf 1}^0$, by $f^\kappa_{\downarrow,1,L}$ the basis of
$\downarrow_L \otimes {\bf 1}^0$, and by $f^\kappa_{\downarrow,1,R}$
the basis of $\downarrow_R \otimes {\bf 1}^0$.

Here each $\uparrow_L$, $\uparrow_R$, $\downarrow_L$, $\downarrow_R$
refers to an $N=3$-dimensional space corresponding to the number of
generations. Thus, the elements listed above form a basis of
$\cH_f$, with $\kappa=1,2,3$ the flavor index. We denote by $\bar
f^\kappa_{\uparrow,3,L}$, etc. the corresponding basis of $\cH_{\bar
f}$.

\begin{rem}\label{fermionbasis}{\rm
The result of Proposition \ref{smhypercharges} shows that we can
identify the basis elements $f^\kappa_{\uparrow,3,L}$,
$f^\kappa_{\uparrow,3,R}$ and $f^\kappa_{\downarrow, 3,L}$ and
$f^\kappa_{\downarrow, 3,R}$ of the linear basis of $\cH_F$ with the
quarks, where $\kappa$ is the flavor index. Thus, after suppressing
the chirality index $L,R$ for simplicity, we identify
$f^1_{\uparrow,3}, f^2_{\uparrow, 3}, f^3_{\uparrow, 3}$ with the
up, charm, and top quarks and $f^1_{\downarrow,3},
f^2_{\downarrow,3}, f^3_{\downarrow,3}$ are the down, strange, and
bottom quarks. Similarly, the basis elements $f^\kappa_{\uparrow,1}$
and $f^\kappa_{\downarrow,1}$ are identified with the leptons. Thus,
$f^1_{\uparrow,1}, f^2_{\uparrow,1}, f^3_{\uparrow,1}$ are
identified with the neutrinos $\nu_e$, $\nu_\mu$, and $\nu_\tau$ and
the $f^1_{\downarrow,1}, f^2_{\downarrow,1}, f^3_{\downarrow,1}$ are
identified with the charged leptons $e$, $\mu$, $\tau$. The
identification is dictated by the values of \eqref{smhypercharges},
which agree with the hypercharges of the basic fermions of the
standard model. Notice that, in choosing the basis of fermions there
is an ambiguity on whether one multiplies by the mixing matrix for
the down particles. This point will be discussed more explicitly in
\S \ref{SectLag} below, see \eqref{dlambdaj}.}
\end{rem}

\subsection{The classification of Dirac operators}\hfill \medskip
\label{diracclassif}

\medskip
We now characterize all operators $D_F$ which qualify as Dirac
operators and moreover commute with the subalgebra
\begin{equation}\label{csubf}
\C_F\subset \cA_F\,,\quad \C_F=\{(\lambda,\lambda,0)\,,\lambda\in
\C\}
\end{equation}

\smallskip

\begin{rem}\label{photnomass}{\rm
The physical meaning of the commutation relation of the Dirac
operator with the subalgebra of \eqref{csubf} is to ensure that the
photon will remain massless.}
\end{rem}

\medskip

We have the following general notion of Dirac operator for the
finite noncommutative geometry with algebra $\cA_F$ and Hilbert
space $\cH_F$.

\smallskip
\begin{defn}\label{definediracops}
A Dirac operator is a self-adjoint operator $D$ in $\cH_F$ commuting
with $J_F$, $\C_F$, anticommuting with $\gamma_F$ and fulfilling the
order one condition $[[D,a],b^0] = 0$ for any $a,b \in \cA_F$.
\end{defn}

\medskip
In order to state the classification of such Dirac operators we
introduce the following notation. Let $\mass_{\elel}$,
$\mass_{\nunu}$, $\mass_{\dodo}$, $\mass_{\upup}$ and $\mass_R$ be
$3\times 3$ matrices. We then let $D(\mass)$ be the operator in
$\cH_F$ given by
\begin{equation}\label{dofm}
D(\mass)=\left[\begin{matrix}S &T^*\\
T &\bar S\end{matrix} \right]
\end{equation}
where
\begin{equation}\label{sop}
S = S_1 \, \oplus (S_3 \otimes 1_3) .
\end{equation}
In the decomposition
$(\uparrow_R,\downarrow_R,\uparrow_L,\downarrow_L)$ we have
\medskip
\begin{equation}\label{sops}
S_1 = \left[\begin{matrix}0&0&\mass^{*}_{\nunu}&0\\
0&0&0&\mass^{*}_{\elel}\\
\mass_{\nunu}&0&0&0\\
0&\mass_{\elel}&0&0&\end{matrix} \right]\qquad
S_3 = \left[\begin{matrix}0&0&\mass^{*}_{\upup}&0\\
0&0&0&\mass^{*}_{\dodo}\\
\mass_{\upup}&0&0&0\\
0&\mass_{\dodo}&0&0&\end{matrix} \right].
\end{equation}

The operator $T$ maps the subspace $E_R=\uparrow_R\otimes {\bf
1}^0\subset \cH_F$ to the conjugate $J_F\,E_R$ by the matrix
$\mass_R$, and is zero elsewhere. Namely,
\begin{equation}\label{TmapsMR}
\begin{array}{ll}
T|_{E_R}: E_R \to J_F E_R, & T|_{E_R} f= \mass_R\, J_F f \\[2mm]
T|_{\cH_F \ominus E_R} =0. & \end{array}
\end{equation}

\medskip

We then obtain the classification of Dirac operators as follows.

\medskip
\begin{thm}\label{diracclassF}
\begin{enumerate}
\item Let $D$ be a Dirac operator. There exist $3\times 3$ matrices
$\mass_{\elel}$, $\mass_{\nunu}$, $\mass_{\dodo}$, $\mass_{\upup}$
and $\mass_R$, with $\mass_R$   symmetric, such that $D=D(\mass)$.
\item  All operators
$D(\mass)$ (with $\mass_R$   symmetric) are Dirac operators.
\item
The operators $D(\mass)$ and $D(\mass')$ are conjugate by a unitary
operator commuting with $\cA_F$, $\gamma_F$ and $J_F$ iff there
exists  unitary matrices $V_j$ and $W_j$ such that
  $$
  \mass'_{\elel}=V_1\,\mass_{\elel}V_3^*\,,\;
  \mass'_{\nunu}=V_2\,\mass_{\nunu}\,V_3^*\,,\;$$ $$
  \mass'_{\dodo}=W_1\,\mass_{\dodo}\,W_3^*\,,\; \mass'_{\upup}=W_2\,\mass_{\upup}\,W_3^*\,,\; \mass'_R=V_2\,\mass_R\,\bar V_2^*
  $$
  \end{enumerate}
\end{thm}

\proof The proof relies on the following lemma, which determines the
commutant $\cA'_F$ of $\cA_F$ in $\cH_F$.

\begin{lem} \label{commutantaf} Let $P=\left[\begin{matrix}P_{11} &P_{12}\\
P_{21} &P_{22}\end{matrix} \right]$ be an operator in
$\cH_F=\cH_f\oplus \cH_{\bar f}$. Then $P\in \cA'_F$ iff the
following holds
\begin{itemize}
  \item $P_{11}$ is
block diagonal with three blocks in $M_{12}(\C)$, $M_{12}(\C)$, and
$1_2\otimes M_{12}(\C)$ corresponding to the subspaces where the
action of $(\lambda,q,m)$ is by $\lambda$, $\bar \lambda$ and $q$.
  \item $P_{12}$ has support in ${\bf 1}\otimes {\bf 2}_L^0 \oplus {\bf 1}\otimes {\bf 2}_R^0$ and range in
  $\uparrow_R\otimes {\bf 1}^0\oplus \uparrow_R\otimes {\bf 3}^0 $.
  \item $P_{21}$ has support in $\uparrow_R\otimes {\bf 1}^0\oplus \uparrow_R\otimes {\bf 3}^0 $ and range in
  ${\bf 1}\otimes {\bf 2}_L^0 \oplus {\bf 1}\otimes {\bf 2}_R^0$.
  \item $P_{22}$
is  of the form
\begin{equation}\label{inthecom}
P_{22} = T_{1} \, \oplus (T_2 \otimes 1_3)
\end{equation}
\end{itemize}
\end{lem}

\begin{proof}  The action of $\cA_F$ on $\cH_F=\cH_f\oplus \cH_{\bar f}$
is of the form
\begin{equation}\label{matrixactionofa}
\left[\begin{matrix}\pi(\lambda,q,m) &0\\
0 &\pi'(\lambda,q,m)\end{matrix} \right]
\end{equation}
On the subspace $\cH_f$   and in the decomposition
$(\uparrow_R,\downarrow_R,\uparrow_L,\downarrow_L)$ one has
\begin{equation}\label{matrixaction}
\pi(\lambda,q,m)=\left[\begin{matrix}\lambda&0&0&0\\
0&\overline \lambda&0&0\\
0&0&\alpha&\beta\\
0&0&-\overline {\beta}&\overline {\alpha}\end{matrix} \right]\otimes
1_{12}
\end{equation}
where the $12$ corresponds to $({\bf 1}^0\oplus {\bf 3}^0)\times 3$.
Since \eqref{matrixactionofa} is diagonal the condition $P\in
\cA'_F$ is expressed independently on the matrix elements $P_{ij}$.

Let us consider first the case of the element $P_{11}$. This must
commute with operators of the form $\pi(\lambda,q,m)\otimes 1_{12}$
with $\pi$ as in \eqref{matrixaction}, and $1_{12}$ the unit matrix
in a twelve dimensional space. This means that the matrix of
$P_{11}$ is block diagonal with three blocks in $M_{12}(\C)$,
$M_{12}(\C)$, and $1_2\otimes M_{12}(\C)$, corresponding to the
subspaces where the action of $(\lambda,q,m)$ is by $\lambda$, $\bar
\lambda$ and $q$.

We consider next the case of $P_{22}$. The action of
$(\lambda,q,m)\in\cA_F$ in the subspace $\cH_{\bar f}$ is given by
multiplication by $\lambda$ or by $m$ thus the only condition on
$P_{22}$ is that it is an operator of the form \eqref{inthecom}.

The off diagonal terms $P_{12}$ and $P_{21}$ must intertwine the
actions of $(\lambda,q,m)\in\cA_F$ in $\cH_{f}$ and $\cH_{\bar f}$.
However, the actions of $q$ or $m$ are disjoint in these two spaces,
while only the action  by $\lambda$ occurs in both.  The subspace of
$\cH_f$ on which $(\lambda,q,m)$ acts by $\lambda$ is
$\uparrow_R\otimes {\bf 1}^0\oplus \uparrow_R\otimes {\bf 3}^0 $.
The subspace of $\cH_{\bar f}$ on which $(\lambda,q,m)$ acts by
$\lambda$ is ${\bf 1}\otimes {\bf 2}_L^0 \oplus {\bf 1}\otimes {\bf
2}_R^0$. Thus the conclusion follows from the intertwining
condition.
\end{proof}

Let us now continue with the proof of Theorem \ref{diracclassF}.

\smallskip
1) Let us first consider the off diagonal part of $D(\mass)$ in
\eqref{dofm}, which is of the form $
\left[\begin{matrix}0 &\mass_R^*\\
\mass_R & 0\end{matrix} \right] $. Anticommutation with $\gamma_F$
holds since the operator $\gamma_F$ restricted to $E_R\oplus J_F
E_R$ is of the form $
\left[\begin{matrix}-1 &0\\
0 & 1\end{matrix} \right] $. Moreover the off diagonal part of
$D(\mass)$ commutes with $J_F$ iff
$\overline{(\mass_R\xi)}=\,\mass_R^*\bar \xi$ for all $\xi$ \ie iff
$\mass_R$ is a symmetric matrix. The order one condition is
automatic since in fact the commutator with elements of $\cA_F$
vanishes exactly.

\smallskip
We can now consider the diagonal part $\left[\begin{matrix}S &0\\
0&\bar S\end{matrix} \right]$ of $D(\mass)$. It commutes with $J$
and anticommutes with $\gamma_F$ by construction. It is enough to
check the commutation with $\C_F\subset\cA_F$ and the order one
condition on the subspace $\cH_f$. Since $S$ exactly commutes with
the action of $\cA_F^0$ the order one condition follows.  In fact
for any $b\in \cA_F$, the action of $b^0$ commutes with any operator
of the form \eqref{inthecom} and this makes it possible to check the
order one condition since $P=[S,\pi(a)]$ is of this form. The action
of $\cA_F$ on the subspace $\cH_F$ is given by \eqref{matrixaction}
and one checks that $\pi(\lambda,\lambda,0)$ commutes with $S$ since
the matrix of $S$ has no non-zero element between the $\uparrow$ and
$\downarrow$  subspaces.

\smallskip
2) Let $D$ be a Dirac operator. Since $D$ is self-adjoint and
commutes with $J_F$ it is of the form
$$
D=\left[\begin{matrix}S &T^*\\
T &\bar S\end{matrix} \right]
$$
where $T=T^t$ is symmetric.

Let $v=(-1,1,1)\in \cA_F$. One has
\begin{equation}\label{gammainner}
\gamma_F\,\xi=\,v\,\xi \qqq \xi\in \cH_f .
\end{equation}
Notice that this equality fails on $\cH_{\bar f}$.

The anticommutation of $D$ with $\gamma_F$ implies that $D=-\frac
12\,\gamma_F\,[D,\gamma_F]$. Notice that $\gamma_F$ is given by a
diagonal matrix of the form $$\gamma_F=\left[\begin{matrix}g &0\\
0 &-\bar g\end{matrix} \right].$$ Thus, we get
$$
S=-\frac 12\,g\,[S,g]=-\frac 12\,v\,[S,v]
$$
using \eqref{gammainner}.

The action of $v$ in $\cH_F$ is given by a diagonal matrix
\eqref{matrixactionofa}, hence $v\,[S,v]$ coincides with the
$A_{11}$ block of the matrix of $A=v[D,v]$.

Thus, the order one condition implies that $S$ commutes with all
operators $b^0$, hence that it is of the form \eqref{sop}.

The anticommutation with $\gamma_F$ and the commutation with $\C_F$
then imply that the self-adjoint matrix $S$ can be written in the
form \eqref{sops}.

\smallskip
It remains to determine the form of the matrix $T$. The conditions
on the off diagonal elements of a matrix
$$
P=\left[\begin{matrix}P_{11} &P_{12}\\
P_{21} &P_{22}\end{matrix} \right],
$$
which ensure that $P$ belongs to the commutant of
$\cA_F^0=J_F\cA_F\,J_F$, are
\begin{itemize}
    \item $P_{12}$ has support in ${\bf 1}\otimes {\uparrow}_R^0\oplus
    {\bf 3}\otimes {\uparrow}_R^0$ and range in
  ${\bf 2}_L\otimes {\bf 1}^0\oplus {\bf 2}_R\otimes {\bf 1}^0$.
  \item $P_{21}$ has support in ${\bf 2}_L\otimes {\bf 1}^0\oplus {\bf 2}_R\otimes {\bf 1}^0$ and range in
  ${\bf 1}\otimes {\uparrow}_R^0\oplus
    {\bf 3}\otimes {\uparrow}_R^0$.
  \end{itemize}
This follows from Lemma \ref{commutantaf}, using $J_F$.

Let then $e=(0,1,0)$. One has $\pi'(e)=0$ and $\pi(e)$ is the
projection on the eigenspace $\gamma_F=1$ in $\cH_F$. Thus, since
$[D,e]$ belongs to the commutant of $\cA_F^0=J_F\cA_F\,J_F$ by the
order one condition, one gets that $\pi'(e)T-T\pi(e)=-T\pi(e)$ has
support in ${\bf 2}_L\otimes {\bf 1}^0\oplus {\bf 2}_R\otimes {\bf
1}^0$ and range in ${\bf 1}\otimes {\uparrow}_R^0\oplus {\bf
3}\otimes {\uparrow}_R^0$. In particular $\gamma_F=1$ on the range.

Thus, the anticommutation with $\gamma_F$ shows that the support of
$T\pi(e)$ is in the eigenspace $\gamma_F=-1$, so that $T\pi(e)=0$.

Let $e_3=(0,0,1)\in \cA_F$. Let us show that $T\,e_3^0=0$. By
Definition \ref{definediracops}, $T$ commutes with the actions of
$v(\lambda)=(\lambda,\lambda,0)\in \cA_F$ and of $J
v(\lambda)J^{-1}=v(\lambda)^0$. Thus, it commutes with $e_3^0$. The
action of $e_3^0$ on $\cH_f$ is the projection on the subspace
$\bullet\otimes {\bf 3}^0$. The action of $e_3^0$ on $\cH_{\bar f}$
is zero. Thus, $[T,e_3^0]=T\,e_3^0$ is the restriction of $T$ to the
subspace $ \bullet \otimes {\bf 3}^0$. Since $[T,e_3^0]=0$  we get
$T\,e_3^0=0$. We have shown that the support of $T$ is contained in
${\bf 2}_R\otimes {\bf 1}^0$. Since $T$ is symmetric \ie $T=\bar
T^*$ the range of $T$ is contained in ${\bf 1}\otimes{\bf 2}_R^0$.

The left and right actions of $(\lambda,q,m)$ on these two subspaces
coincide with the left and right actions of $v(\lambda)$. Thus, we
get that $T$ commutes with $\cA_F$ and $\cA_F^0$. Thus, by Lemma
\ref{commutantaf}, it has support in $\uparrow_R\otimes {\bf 1}^0$
and range in ${\bf 1}\otimes \uparrow_R^0$.

This means that $T$ is given by a symmetric $3\times 3$ matrix
$\mass_R$ and the operator $D$ is of the form $D=D(\mass)$.
\smallskip

3) By Lemma \ref{commutantaf}, the commutant of the algebra
generated by $\cA_F$ and $\cA_F^0$ is the algebra of matrices
 $$P=\left[\begin{matrix}P_{11} &P_{12}\\
P_{21} &P_{22}\end{matrix} \right]$$ such that
\begin{itemize}
    \item $P_{12}$ has support in ${\bf 1}\otimes \uparrow_R^0$ and range in
  $\uparrow_R\otimes {\bf 1}^0$.
  \item $P_{21}$ has support in $\uparrow_R\otimes {\bf 1}^0$ and range in
  ${\bf 1}\otimes \uparrow_R^0$.
  \item $P_{jj}$ is   of the form
$$
P_{jj} = P_{jj}^{1} \, \oplus (P_{jj}^3\otimes 1_3)
$$
where
$$
 P_{jj}^{a}=\left[
  \begin{array}{ccc}
    P_j^a(1) & 0 & 0 \\
    0 & P_j^a(2) & 0 \\
    0 & 0 & 1_2\otimes P_j^a(3) \\
  \end{array}
\right]\qquad a=1,3\,,\quad j=1,2 .
$$
\end{itemize}

A unitary operator $U$ acting in $\cH_F$ commuting with $\cA_F$ and
$J$ is in the commutant of the algebra generated by $\cA_F$ and
$\cA_F^0$. If it commutes with $\gamma_F$, then the off diagonal
elements $U_{ij}$ vanish, since $\gamma_F=-1$ on $\uparrow_R\otimes
{\bf 1}^0$ and  $\gamma_F=1$ on ${\bf 1}\otimes \uparrow_R^0$. Thus
 $U$  is determined by the six $3\times 3$ matrices
$U_1^a(k)$ since  it commutes with $J$ so that $U_2^a(k)=\bar
U_1^a(k)$ . One checks that conjugating by $U$ gives the relation 3)
of Theorem \ref{diracclassF}.
\endproof

\begin{rem}\label{unbrokrem}{\rm
It is a consequence of the classification of Dirac operators
obtained in this section that  color is unbroken in our model, as is
physically expected. In fact, this follows from the fact that Dirac
operators are of the form \eqref{dofm}, with the $S$ term of the
form \eqref{sop}.}
\end{rem}

\subsection{The moduli space of Dirac operators and the Yukawa
parameters}\hfill\medskip \label{sectmoduli}

Let us start by considering the moduli space $\cC_3$ of pairs of
invertible $3\times 3$ matrices $(\mass_{\dodo},\mass_{\upup})$
modulo the equivalence relation
\begin{equation}\label{equivalencerel}
\mass'_{\dodo}=W_1\,\mass_{\dodo}\,W_3^*\,,\;
\mass'_{\upup}=W_2\,\mass_{\upup}\,W_3^* ,
\end{equation}
where the $W_j$ are unitary matrices.

\begin{prop}\label{modspGK}
The moduli space $\cC_3$ is the double coset space
\begin{equation}\label{GKcoset}
\cC_3 \cong (U(3)\times U(3))\backslash (\GL_3(\C)\times
\GL_3(\C))/U(3)
\end{equation}
of real dimension $10$.
\end{prop}

\proof This follows from the explicit form of the equivalence
relation \eqref{equivalencerel}. The group $U(3)$ acts diagonally on
the right. \endproof

\smallskip

\begin{rem}\label{rem33}{\rm
Notice that the $3$ in $\cC_3$ corresponds to the color charge for
quarks (like the $1$ in $\cC_1$ below will correspond to leptons),
while in the right hand side of \eqref{GKcoset} the $3$ of
$\GL_3(\C)$ and $U(3)$ corresponds to the number of generations.}
\end{rem}

\smallskip

Each equivalence class under \eqref{equivalencerel} contains a pair
$(\mass_{\dodo},\mass_{\upup})$ where $\mass_{\upup}$ is diagonal
(in the given basis) and with positive entries, while
$\mass_{\dodo}$ is positive.

Indeed, the freedom to chose $W_2$ and $W_3$ makes it possible to
take $\mass_{\upup}$ positive and diagonal and the freedom in $W_1$
then makes it possible to take $\mass_{\dodo}$ positive.

The eigenvalues are the characteristic values (\ie the eigenvalues
of the absolute value in the polar decomposition) of $\mass_{\upup}$
and $\mass_{\dodo}$ and are invariants of the pair.

Thus, we can find diagonal matrices $\delta_\uparrow$ and
$\delta_\downarrow$ and a unitary matrix $C$ such that
$$
\mass_{\upup}=\delta_\uparrow\,,\quad
\mass_{\dodo}=C\,\delta_\downarrow\,C^* .
$$
Since multiplying $C$ by a scalar does not affect the result, we can
assume that $\det(C)=1$. Thus, $C\in \SU(3)$ depends a priori upon
$8$ real parameters. However, only the double coset of $C$ modulo
the diagonal subgroup $\cN\subset\SU(3)$ matters, by the following
result.

\begin{lem} Suppose given diagonal matrices $\delta_\uparrow$
and $\delta_\downarrow$ with positive and distinct eigenvalues. Two
pairs of the form $(\delta_\uparrow,C\,\delta_\downarrow\,C^*)$ are
equivalent iff there exists diagonal unitary matrices $A, B\in \cN$
such that
$$
A\,C=C'\,B .
$$
\end{lem}

\proof For $A\,C=C'\,B$ one has
$$
A\,\mass_{\upup}\,A^*=\mass'_{\upup}\,,\quad
A\,\mass_{\dodo}\,A^*=\mass'_{\dodo}
$$
and the two pairs are equivalent. Conversely, with $W_j$ as in
\eqref{equivalencerel} one gets $W_1=W_3$ from the uniqueness of the
polar decomposition
$$ \delta_\downarrow = (W_1 W_3^*)\, (W_3 \delta_\downarrow W_3^*). $$
Similarly, one obtains $W_2=W_3$. Thus, $W_3=W$ is diagonal and we
get
$$
W\,C\,\delta_\downarrow\,C^*\,W^*=\,C'\,\delta_\downarrow\,C^{\prime
*} ,
$$
so that $W\,C=\,C'\,B$ for some diagonal matrix $B$. Since $W$ and
$B$ have the same determinant one can assume that they both belong
to $\cN$.
\endproof

\smallskip
The dimension of the moduli space is thus $3 + 3 +4=10$ where the $3
+ 3$ comes from the eigenvalues and the $4=8-4$ from the above
double coset space of $C$'s. One way to parameterize the
representatives of the double cosets of the matrix $C$ is by means
of three angles $\theta_i$ and a phase $\delta$,
\begin{equation}\label{CKMmatrixParam}
C= \left[ \begin{array}{ccc} c_1 & -s_1c_3 & -s_1s_3 \\
s_1c_2 & c_1c_2c_3 -s_2s_3 e_\delta & c_1c_2s_3 + s_2c_3 e_\delta \\
s_1s_2 & c_1s_2c_3+c_2s_3e_\delta & c_1s_2s_3-c_2c_3e_\delta
\end{array} \right],
\end{equation}
for $c_i=\cos\theta_i$, $s_i=\sin\theta_i$, and
$e_\delta=\exp(i\delta)$. One has by construction the factorization
\begin{equation}\label{CKMmatrixParam1}
C=R_{23}(\theta_2)\,d(\delta)\,R_{12}(\theta_1)\,R_{23}(-\theta_3)\,
\end{equation}
where $R_{ij}(\theta)$ is the rotation of angle $\theta$ in the
$ij$-plane and $d(\delta)$ the diagonal matrix
$$
d(\delta)=\left[
  \begin{array}{ccc}
    1 & 0 & 0 \\
    0 & 1 & 0 \\
    0 & 0 & -e^{i\delta} \\
  \end{array}
\right]
$$

\smallskip
Let us now consider the moduli space $\cC_1$ of triplets
$(\mass_{\elel},\mass_{\nunu},\mass_R)$, with $\mass_R$ symmetric,
modulo the equivalence relation
\begin{equation}\label{C1eqrel1}
  \mass'_{\elel}=V_1\,\mass_{\elel}V_3^* \, \ \ \ \
\mass'_{\nunu}=V_2\,\mass_{\nunu}\,V_3^*
\end{equation}
\begin{equation}\label{C1eqrel2}
\mass'_R=V_2\,\mass_R\,\bar V_2^* .
\end{equation}

\begin{lem}\label{modC1}
The moduli space $\cC_1$ is given by the quotient
\begin{equation}\label{C1quot}
\cC_1\cong (U(3)\times U(3))\backslash (\GL_3(\C) \times
GL_3(\C)\times \cS)/U(3),
\end{equation}
where $\cS$ is the space of symmetric complex $3\times 3$ matrices
and
\begin{itemize}
\item The action of $U(3)\times U(3)$ on the left is given by left multiplication
on $\GL_3(\C) \times GL_3(\C)$ and  by \eqref{C1eqrel2} on $\cS$.
\item The action of $U(3)$ on the right is trivial on $\cS$ and by
diagonal right multiplication on $\GL_3(\C) \times GL_3(\C)$.
\end{itemize}

 It is of real dimension $21$ and fibers over
$\cC_3$, with generic fiber the quotient of symmetric complex
$3\times 3$ matrices by $U(1)$.
\end{lem}

\proof By construction one has a natural surjective map
$$
\pi: \cC_1\to \cC_3
$$
just forgetting about $\mass_R$. The generic fiber of $\pi$ is the
space of symmetric complex $3\times 3$ matrices modulo the action of
a complex scalar $\lambda$ of absolute value one by
$$
\mass_R\mapsto \lambda^2\,\mass_R .
$$
The (real) dimension of the fiber is $12-1=11$. The total real
dimension of the moduli space $\cC_1$ is then $21$.
\endproof

\medskip

The total $31$-dimensional moduli space of Dirac operators is given
by the product
\begin{equation}\label{prodC1C3}
\cC_1 \times \cC_3.
\end{equation}

\begin{rem}\label{Yukawas}{\rm
The $31$ real parameters of \eqref{prodC1C3} correspond to the
Yukawa parameters in the standard model with neutrino mixing and
Majorana mass terms. In fact, the parameters in $\cC_3$ correspond
to the masses of the quarks and the quark mixing angles of the CKM
matrix, while the additional parameters of $\cC_1$ give the lepton
masses, the angles of the PMNS mixing matrix and the Majorana mass
terms. }
\end{rem}

\subsection{Dimension, $KO$-theory, and Poincar\'e duality}\hfill
\medskip \label{poincare}

In \cite{Co-book} Chapter 6, \S 4, the notion of manifold in
noncommutative geometry was discussed in terms of Poincar\'e duality
in $KO$-homology. In \cite{CoSM} this Poincar\'e duality was shown
to hold rationally for the finite noncommutative geometry used
there. We now investigate how the new finite noncommutative geometry
$F$ considered here behaves with respect to this duality. We first
notice that now, the dimension being equal to $6$ modulo $8$, the
intersection pairing is {\em skew symmetric}. It is given explicitly
as follows.

\begin{prop}
The expression
\begin{equation}\label{oddpairing}
\langle e,\,f\,\rangle=\,\Tr(\gamma\,e\,JfJ^{-1})
\end{equation}
defines an antisymmetric bilinear pairing on $K_0\times K_0$. The
group $K_0(\cA_F)$ is the free abelian group generated by the
classes of $e_1=(1,0,0)$, $e_2=(0,1,0)$ and $f_3=(0,0,f)$, where
$f\in M_3(\C)$ a minimal idempotent.
\end{prop}

\proof The pairing \eqref{oddpairing} is obtained from the
composition of the natural map $$ K_0(\cA_F)\times K_0(\cA_F)\to
K_0(\cA_F\otimes \cA_F^0)
$$
with the graded trace $\Tr(\gamma \,\cdot)$. Since $J$ anticommutes
with $\gamma$, one checks that $$ \langle
f,\,e\,\rangle=\,\Tr(\gamma\,f\,JeJ^{-1}) = -
\,\Tr(\gamma\,J^{-1}f\,J\,e)= -\,\Tr(\gamma\,e\,JfJ^{-1})=-\langle
e,\,f\,\rangle,$$ so that the pairing is antisymmetric.

\smallskip
By construction, $\cA_F$ is the direct sum of the fields $\C$, $\H$
and of the algebra $M_3(\C)\sim \C$ (up to Morita equivalence). The
projections $e_1=(1,0,0)$, $e_2=(0,1,0)$ and $f_3=(0,0,f)$ are the
three minimal idempotents in $\cA_F$.
\endproof

\smallskip
By construction the $KO$-homology class given by the representation
in $\cH_F$ with the $\Z/2$-grading $\gamma$ and the real structure
$J_F$ splits as a direct sum  of two pieces, one for the leptons and
one for the quarks.

\begin{prop} \label{sebigbg}
\begin{enumerate}
  \item The representation of the algebra generated by $(\cA_F,D_F,J_F,\gamma_F)$ in $\cH_F$
  splits as a direct
  sum of two subrepresentations
  $$
  \cH_F=\cH_F^{(1)}\oplus
\cH_F^{(3)} .$$
\item In the generic case (\ie when the matrices
in $D_F$ have distinct eigenvalues) each of these subrepresentations
is irreducible.
\item In the basis $(e_1,e_2, f_3)$ the pairing \eqref{oddpairing}
is (up to an overall multiplicity three corresponding to the number
of generations) given by
\begin{equation}\label{pair1pair2}
\langle \cdot, \cdot \rangle|_{\cH_F^{(1)}}= \left[
  \begin{array}{ccc}
    0 & 2 & 0 \\
    -2 & 0 & 0 \\
    0 & 0 & 0\\
  \end{array}
\right] \qquad \langle\cdot,\cdot\rangle|_{\cH_F^{(3)}}=\left[
  \begin{array}{ccc}
    0 & 0 & 2 \\
    0 & 0 & -2 \\
    -2 & 2 & 0 \\
  \end{array}
\right]
\end{equation}
\end{enumerate}
\end{prop}

\proof 1) Let $\cH_F^{(1)}$ correspond to
\begin{equation}\label{bimdec2}
{\bf 2}_L\otimes {\bf 1}^0\oplus {\bf 2}_R\otimes {\bf 1}^0 \oplus
{\bf 1}\otimes {\bf 2}_L^0 \oplus {\bf 1}\otimes {\bf 2}_R^0
  \end{equation}
  and $\cH_F^{(3)}$ to
  \begin{equation}\label{bimdec3}
{\bf 2}_L\otimes {\bf 3}^0\oplus {\bf 2}_R\otimes {\bf 3}^0 \oplus
{\bf 3}\otimes {\bf 2}_L^0 \oplus {\bf 3}\otimes {\bf 2}_R^0 .
   \end{equation}
By construction, the action of $\cA_F$ in $\cH_F$ is block diagonal
in the decomposition $\cH_F=\cH_F^{(1)}\oplus \cH_F^{(3)}$. Both the
actions of $J_F$ and of $\gamma_F$ are also block diagonal. Theorem
\ref{diracclassF} shows that $D_F$ is also block diagonal, since it
is of the form $D=D(\mass)$.

\smallskip
2) It is enough to show that a unitary operator that commutes with
$\cA_F$, $\gamma_F$, $J_F$ and $D_F$ is a scalar. Let us start with
 $\cH_F^{(3)}$. By Theorem \ref{diracclassF} (3), such a unitary is
given by three unitary matrices $W_j\in M_3(\C)$ such that
$$
\mass_{\dodo}=W_1\,\mass_{\dodo}\,W_3^*\,,\;
\mass_{\upup}=W_2\,\mass_{\upup}\,W_3^*
$$
We can assume that both $\mass_{\upup}$ and $\mass_{\dodo}$ are
positive. Assume also that $\mass_{\upup}$ is diagonal.  The
uniqueness of the polar decomposition shows that
$$
\mass_{\dodo}=(W_1\,W_3^*)\,(W_3\,\mass_{\dodo}\,W_3^*)\,\Rightarrow
W_1\,W_3^*=1\,,\quad W_3\,\mass_{\dodo}\,W_3^*=\mass_{\dodo}
$$
Thus, we get $W_1=W_2=W_3$. Since generically all the eigenvalues of
$\mass_{\upup}$ or $\mass_{\dodo}$ are distinct, we get that the
matrices $W_j$ are diagonal in the basis of eigenvectors of the
matrices $\mass_{\upup}$ and $\mass_{\dodo}$. However, generically
these bases are distinct, hence we conclude that $W_j=1$ for all
$j$. The same result holds ``a fortiori" for $\cH_F^{(1)}$ where the
conditions imposed by Theorem \ref{diracclassF} (3) are in fact
stronger.

\smallskip
3) One computes the pairing directly using the definition of
$\gamma_F$. On $\cH_F^{(1)}$ the subalgebra $M_3(\C)$ acts by zero
which explains why the last line and columns of the pairing matrix
vanish. By antisymmetry one just needs to evaluate $$\langle
e,e_L\rangle=-\langle e_L,e\rangle
=-\Tr(\gamma\,e_L\,JeJ^{-1})=-\Tr(\gamma\,e_L)=2 \times 3,$$ where
$3$ is the number of generations. On $\cH_F^{(3)}$ the same pair
gives $\langle e,e_L\rangle=0$, since now the right action of $e$ is
zero on $\cH_f$. In the same way one gets $\langle
f_3,e_L\rangle=2\times 3$. Finally one has
 $$\langle e,f_3\rangle=\,\Tr(\gamma\,e\,Jf_3J^{-1})=2\times 3.$$
\endproof

Of course an antisymmetric $3\times 3$ matrix is automatically
degenerate since its determinant vanishes. Thus it is not possible
to obtain a non-degenerate Poincar\'e duality pairing with a single
$KO$-homology class. One checks however that the above pair of
$KO$-homology classes suffices to obtain a non-degenerate pairing in
the following way.

\begin{cor}\label{nondegpair}
The pairing $K_0(\cA_F)\oplus K_0(\cA_F)\to \R\oplus \R$ given by
\begin{equation}\label{pair2}
\langle\cdot,\cdot\rangle_{\cH_F}:=\langle
\cdot,\cdot\rangle|_{\cH_F^{(1)}}\oplus \langle
\cdot,\cdot\rangle|_{\cH_F^{(3)}}
\end{equation}
is non-degenerate.
\end{cor}

\proof We need to check that, for any $e$ in $K_0(\cA_F)$ there
exists an $f\in K_0(\cA_F)$ such that $\langle
e,f\rangle_{\cH_F}\neq (0,0)$. This can be seen by the explicit form
of $\langle \cdot,\cdot\rangle|_{\cH_F^{(1)}}$ and $\langle
\cdot,\cdot\rangle|_{\cH_F^{(3)}}$ in \eqref{pair1pair2}.
\endproof

\begin{rem}\label{2genrem}{\rm
The result of Corollary \ref{nondegpair} can be reinterpreted as the
fact that in our case $KO$-homology is not singly generated as a
module over $K_0$, but it is generated by two elements.}
\end{rem}

\section{The spectral action and the standard model}

In this section and in the one that follows we show that the full
Lagrangian of the standard model with neutrino mixing and Majorana
mass terms, minimally coupled to gravity, is obtained as the
asymptotic expansion of the spectral action for the product of the
finite geometry $(\cA_F,\cH_F,D_F)$ described above and a spectral
triple associated to $4$-dimensional spacetime.

\subsection{Riemannian geometry and spectral triples}\hfill\medskip

A spin Riemannian manifold $M$ gives rise in a canonical manner to a
spectral triple. The Hilbert space $\cH$ is the Hilbert space
$L^2(M,S)$ of square integrable spinors on $M$ and the algebra
$\cA=C^\infty(M)$ of smooth functions on $M$ acts in $\cH$ by
multiplication operators:
\begin{equation}\label{mulop}
(f\,\xi)(x)=\,f(x)\,\xi(x)\qqq x\in M .
\end{equation}
The operator $D$ is the Dirac operator
\begin{equation}\label{diraconM}
{\dirac}_M=\sqrt{-1}\,\gamma^\mu\,\nabla^s_\mu
\end{equation}
where $\nabla^s$ is the spin connection which we express in a
vierbein $e$  so that
\begin{align}
\gamma^{\mu} &  =\gamma^{a}e_{a}^{\mu},\nonumber\\
\nabla^s_{\mu} &
=\partial_{\mu}+\frac{1}{4}\omega_{\mu}^{\,ab}\left( e\right)
\gamma_{ab}.\label{spincon}
\end{align}
The grading $\gamma$ is given by the chirality operator which we
denote by $\gamma_5$ in the $4$-dimensional case. The operator $J$
is the charge conjugation operator and we refer to \cite{FGV} for a
thorough treatment of the above notions.

\subsection{The product geometry}\hfill\medskip

We now consider a 4-dimensional smooth compact Riemannian manifold
$M$ with a fixed spin structure. We consider its product with the
finite geometry $(\cA_F,\cH_F,D_F)$ described above.

With $(\cA_j,\cH_j,\gamma_j)$ of $KO$-dimensions $4$ for $j=1$ and
$6$ for $j=2$, the product geometry is given by the rules
$$
\cA = \cA_1 \otimes \cA_2 \ , \ \cH = \cH_1 \otimes \cH_2 \ , \ D =
D_1 \otimes 1 + \gamma_1 \otimes D_2 \, , \ \gamma=\gamma_1\otimes
\gamma_2\,,  \ J=J_1\otimes J_2 .
$$
Notice that it matters here that $J_1$ commutes with $\gamma_1$, in
order to check that $J$ commutes with $D$. One checks that the order
one condition is fulfilled by $D$ if it is fulfilled by the $D_j$.

\smallskip

For the product of the manifold $M$ by the finite geometry $F$ we
then have $\cA = C^{\infty} (M) \otimes \cA_F = C^{\infty}
(M,\cA_F)$, $\cH = L^2 (M,S) \otimes \cH_F = L^2 (M,S \otimes
\cH_F)$ and $D = \dirac_M \otimes 1 + \gamma_5 \otimes D_F$ where
${\dirac}_M$ is the Dirac operator on $M$. It is given by equations
\eqref{diraconM} and \eqref{spincon}.

\subsection{The real part of the product geometry}\label{sectrealpart}\hfill \medskip

The next proposition shows that a noncommutative geometry
automatically gives rise to a commutative one playing in essence the
role of its center (\cf Remark \ref{centerrem} below).

\begin{prop} \label{realequalcentral}
Let $(\cA,\cH,D)$ be a real spectral triple in the sense of
Definition \ref{realstr}. Then the following holds.
\begin{enumerate}
  \item The equality $\cA_J=\,\{x\in\cA\,;\;x\,J=J\,x\}$ defines an
  involutive commutative real subalgebra of the center of $\cA$.
  \item $(\cA_J,\cH,D)$ is a real spectral triple.
  \item Any $a\in \cA_J$ commutes with the algebra generated by the sums
  $\sum a_i[D,b_i]$ for $a_i$, $b_i$ in $\cA$.
\end{enumerate}
\end{prop}

\proof 1) By construction $\cA_J$ is a real subalgebra of $\cA$.
Since $J$ is isometric one has $(JaJ^{-1})^*=Ja^*J^{-1}$ for all
$a$. Thus if $x\in \cA_J$, one has $JxJ^{-1}=x$ and
$Jx^*J^{-1}=x^*$, so that $x^*\in \cA_J$. Let us show that $\cA_J$
is contained in the center of $\cA$. For $x\in \cA_J$ and $b\in \cA$
one has $[b,x^0]=0$ from \eqref{comm-rule}. But $x^0=Jx^*J^{-1}=x^*$
and thus we get $[b,x^*]=0$.

\smallskip
2) This is automatic since we are just dealing with a subalgebra.
Notice that it continues to hold for the complex algebra
$\cA_J\otimes_\R\C$ generated by $\cA_J$.

\smallskip
3)  The order one condition \eqref{order1} shows that $[D,b]$
commutes with $(a^*)^0$ and hence with $a$ since $(a^*)^0=a$ as we
saw above.\endproof

\smallskip
While the real part $\cA_J$ is contained in the center $Z(\cA)$ of
$\cA$, it can be much smaller as one sees in the example of the
finite geometry $F$.  Indeed, one has the following result.

\smallskip
\begin{lem}\label{lemrealpartF} Let $F$ be the finite noncommutative geometry:
\begin{itemize}
  \item The real part of $\cA_F$ is $\R=\{(\lambda,\lambda,\lambda)\,,\lambda\in
  \R\}\subset \cA_F$.
  \item The real part of $C^\infty(M,\cA_F)$ for the product
  geometry $M\times F$ is $C^\infty(M,\R)$.
\end{itemize}
\end{lem}

\proof Let $x=(\lambda,q,m)\in \cA_F$. Then if $x$ commutes with
$J_F$, its action in $\cH_f\subset \cH_F$ coincides with the right
action of $x^*$. Looking at the action on $\cH_F^{(1)}$ , it follows
that $\lambda=\bar\lambda$ and that the action of the quaternion $q$
coincides with that of $\lambda$. Thus $\lambda\in \R$ and
$q=\lambda$. Then looking at the action on $\cH_F^{(3)}$  gives
$m=\lambda$. The same proof applies to $C^\infty(M,\cA_F)$.
\endproof

\smallskip
\begin{rem}\label{centerrem}{\rm
The notion of real part $\cA_J$ can be thought of as a refinement of
the center of the algebra in this geometric context. For instance,
even though the center of $\cA_F$ is non-trivial, this geometry can
still be regarded as ``central" in this perspecive, since the real
part of $\cA_F$ is reduced to just the scalars $\R$.}
\end{rem}

\subsection{The adjoint representation and the gauge symmetries}\hfill\medskip

In this section we display the role of the gauge group
$C^\infty(M,\SU(\cA_F))$ of smooth maps from the manifold $M$ to the
group $\SU(\cA_F)$.

\begin{prop}\label{propautgroup}
Let $(\cA,\cH,D)$ be the real spectral triple associated to $M\times
F$.
\begin{itemize}
\item Let $U$ be a unitary in $\cH$ commuting with $\gamma$ and $J$
and such that $U\,\cA\,U^*=\cA$. Then there exists a unique
diffeomorphism $\varphi\in \Diff(M)$ such that
\begin{equation}\label{UfUstar}
U\,f\,U^*=f\circ\varphi\,\qqq f\in \cA_J .
\end{equation}
  \item Let $U$ be as above and such that $\varphi={\rm id}$. Then,
possibly after passing to a finite abelian cover of $M$, there
exists a unitary $u\in C^\infty(M,\SU(\cA_F))$ such that $U\,{\rm
Ad}(u)^*\in \cC$, where $\cC$ is the commutant of the algebra of
operators in $\cH$ generated by $\cA$ and $J\cA J^{-1}$.
\end{itemize}
\end{prop}

We refer to \cite{schulaz} for finer points concerning the lifting
of diffeomorphisms preserving the given spin structure.

\proof The first statement follows from the functoriality of the
construction of the subalgebra $\cA_J$ and the classical result that
automorphisms of the algebra $C^\infty(M,\R)$ are given by
composition with a diffeomorphism of $M$.

\smallskip
Let us prove the second statement. One has $\cH = L^2 (M,S) \otimes
\cH_F = L^2 (M,S \otimes \cH_F)$. Since $\varphi={\rm id}$, we know
by \eqref{UfUstar} that $U$ commutes with the algebra
$\cA_J=C^\infty(M,\R)$. This shows that $U$ is given by an
endomorphism $x\mapsto U(x)$ of the vector bundle $S \otimes \cH_F$
on $M$. Since $U$ commutes with $J$, the unitary $U(x)$ commutes
with $J_x\otimes J_F$.

The equality $U\,\cA\,U^*=\cA$ shows that, for all $x\in M$, one has
\begin{equation}\label{autofaf}
U(x)\,({\rm id}\otimes \cA_F)\,U^*(x)={\rm id}\otimes \cA_F\,.
\end{equation}
Here we identify $\cA_F$ with a subalgebra of operators on $S\otimes
\cH_F$, through the algebra homomorphism $T\mapsto {\rm id}\otimes
T$.

Let $\alpha$ be an arbitrary automorphism of $\cA_F$. The center of
$\cA_F$ contains three minimal idempotents and the corresponding
reduced algebras $\C$, $\H$, $M_3(\C)$ are pairwise non-isomorphic.
Thus $\alpha$ preserves these three idempotents and is determined by
its restriction to the corresponding reduced algebras $\C$, $\H$,
$M_3(\C)$. In particular, such an automorphism will act on the
subalgebra $\C$ either as the identity or as complex conjugation.

Now consider the automorphism $\alpha_x$ of $\cA_F$ determined by
\eqref{autofaf}. It is unitarily implemented by \eqref{autofaf}. The
action of $\C\subset \cA_F$ on $S \otimes \cH_F$ is not unitarily
equivalent to its composition with complex conjugation. This can be
seen from the fact that, in this representation, the dimension of
the space on which $\C$ acts by $\lambda$ is larger than the one of
the space on which it acts by $\bar \lambda$. It then follows that
the restriction of $\alpha_x$ to $\C\subset \cA_F$ has to be the
identity automorphism.

Similarly, the restriction of $\alpha_x$ to $M_3(\C)\subset \cA_F$
is given by an inner automorphism of the form $ f\to v_x\,f\,v_x^*
$, where $v_x\in \SU(3)$ is only determined modulo the center
$Z_3\sim\mu_3$ of $\SU(3)$. The restriction of $\alpha_x$ to
$\H\subset \cA_F$ is given by an inner automorphism of the form $
f\to q_x\,f\,q_x^* $ where $q_x\in \SU(2)$ is only determined modulo
the center $Z_2\sim\mu_2$ of $\SU(2)$. Thus passing to the finite
abelian cover $\tilde M$ of $M$ corresponding to the morphism
$\pi_1(M)\to Z_2\times Z_3\sim \mu_6$, one gets a unitary element
$u=(1,q,v)\in C^\infty(M,\SU(\cA_F))$ such that $\alpha(f)={\rm
Ad}(u)f{\rm Ad}(u)^*$ for all $f\in C^\infty(M, \cA_F)$. Replacing
$U$ by $U\,{\rm Ad}(u)^*$ one can thus assume that $U$ commutes with
all $f\in C^\infty(M, \cA_F)$, and the commutation with $J$ still
holds so that $ U\,{\rm Ad}(u)^*\in \cC$,  where $\cC$ is the
commutant of the algebra of operators in $\cH$ generated by $\cA$
and $J\cA J^{-1}$.
\endproof

\subsection{Inner fluctuations and bosons}\hfill\medskip

Let us show that the inner fluctuations of the metric give rise to
the gauge bosons of the standard model with their correct quantum
numbers. We first have to compute $A = \Sigma \, a_i [D,a'_i] \quad
a_i ,a'_i \in \cA$. Since $D = {\partial \!\!\! /}_M \otimes 1 +
\gamma_5 \otimes D_F$ decomposes as a sum of two terms, so does $A$
and we first consider the discrete part $A^{(0,1)}$ coming from
commutators with $\gamma_5 \otimes D_F$.

\medskip
\subsubsection{The discrete part $A^{(0,1)}$ of the inner
fluctuations}\label{NCGHiggsbosonsSect}\hfill\medskip

Let $x\in M$ and let $a_i (x) = (\lambda_i , q_i ,m_i)$, $a'_i (x) =
(\lambda'_i ,q'_i ,m'_i)$, the computation of $\sum \, a_i [\gamma_5
\otimes D_F ,a'_i]$ at $x$ on the subspace corresponding to
$\cH_f\subset \cH_F$ gives $\gamma_5$ tensored by the matrices
$A^{(0,1)}_3$ and $A^{(0,1)}_1$ defined below. We set
\begin{equation}\label{innerfluchiggs}
A^{(0,1)}_3=\,\left[
  \begin{array}{cc}
    0 & X \\
    X' & 0 \\
  \end{array}
\right] \otimes 1_3\,,\quad X=\left[
             \begin{array}{cc}
               \mass_{\upup}^*\,\varphi_1 & \mass_{\upup}^*\,\varphi_2 \\
               -\mass_{\dodo}^*\,\bar\varphi_2 & \mass_{\dodo}^*\,\bar\varphi_1 \\
             \end{array}
           \right]
           \,,\quad X'=\left[
             \begin{array}{cc}
               \mass_{\upup}\,\varphi'_1 & \mass_{\dodo}\,\varphi'_2 \\
               -\mass_{\upup}\,\bar\varphi'_2 & \mass_{\dodo}\,\bar\varphi'_1 \\
             \end{array}
           \right]
\end{equation}
for the $\cH_F^{(3)}$ part, with
\begin{equation}\label{innerfluchiggs1}
\varphi_1=\sum \lambda_i(\alpha'_i-\lambda'_i)\,,\quad
\varphi_2=\sum \lambda_i\beta'_i
\end{equation}
\begin{equation}\label{innerfluchiggs2}
\varphi'_1=\sum \alpha_i(\lambda'_i-\alpha'_i)+
\beta_i\bar\beta'_i\,,\quad \varphi'_2=\sum
(-\alpha_i\beta'_i+\beta_i(\bar\lambda'_i-\bar\alpha'_i)) ,
\end{equation}
where we used the notation $$ q=\left[
  \begin{array}{cc}
    \alpha & \beta \\
    -\bar\beta & \bar \alpha \\
  \end{array}
\right]
$$
for quaternions. For the $\cH_F^{(1)}$ part one obtains in the same
way
\begin{equation}\label{innerfluchiggs3}
A^{(0,1)}_1=\,\left[
  \begin{array}{cc}
    0 & Y \\
    Y' & 0 \\
  \end{array}
\right]\,,\quad Y=\left[
             \begin{array}{cc}
               \mass_{\nunu}^*\,\varphi_1 & \mass_{\nunu}^*\,\varphi_2 \\
               -\mass_{\elel}^*\,\bar\varphi_2 & \mass_{\elel}^*\,\bar\varphi_1 \\
             \end{array}
           \right]
           \,,\quad Y'=\left[
             \begin{array}{cc}
               \mass_{\nunu}\,\varphi'_1 & \mass_{\elel}\,\varphi'_2 \\
               -\mass_{\nunu}\,\bar\varphi'_2 & \mass_{\elel}\,\bar\varphi'_1 \\
             \end{array}
           \right] .
\end{equation}
Here the $\varphi$ are given as above by \eqref{innerfluchiggs1} and
\eqref{innerfluchiggs2}.

The off diagonal part of $D_F$, which involves $\mass_R$, does not
contribute to the inner fluctuations, since it exactly commutes with
the algebra $\cA_F$. Since the action of $\cA_F$ on $\cH_{\bar f}$
exactly commutes with $D_F$, it does not contribute to $A^{(0,1)}$.
One lets
\begin{equation}\label{innerfluchiggs4}
q=\varphi_1+\varphi_2\,j\,,\quad q'=\varphi'_1+\varphi'_2\,j
\end{equation}
where $j$ is the quaternion $\left[ \begin{array}{cc}
  0 & 1 \\
   -1 & 0 \\
  \end{array}
   \right]$.

\begin{prop}\label{prophiggsfieldssm} 1) The discrete part $A^{(0,1)}$ of the
inner fluctuations of the metric is parameterized by an arbitrary
quaternion valued function
   $$
   H\in C^\infty(M,\H)\,,\quad H=\varphi_1+\varphi_2\,j\,,\quad
   \varphi_j\in C^\infty(M,\C)
   $$

   2) The role of $H$ in the coupling of the $\uparrow$-part is related to
   its role in the coupling of the $\downarrow$-part by the replacement
$$
H\mapsto \tilde H=j\,H .
$$
\end{prop}

\proof 1) First one checks that there are no linear relations
   between the four terms \eqref{innerfluchiggs1} and
   \eqref{innerfluchiggs2}. We consider a single term $a[D_F,a']$. Taking
   $a=(\lambda,0,0)$ and $a'=(\lambda',0,0)$ gives
   $$
   \varphi_1=-\lambda\lambda'\,,\quad
   \varphi_2=\varphi'_1=\varphi'_2=0
   $$
   Taking $a=(\lambda,0,0)$ and $a'=(0,j\,\bar\beta',0)$ gives
   $$
   \varphi'_1=\lambda\beta'\,,\quad
   \varphi_1=\varphi'_1=\varphi'_2=0
   $$
   Similarly, taking $a=(0,\alpha,0)$ and $a'=(\lambda',0,0)$ gives
   $$
   \varphi'_1=\alpha\lambda'\,,\quad
   \varphi_1=\varphi_2=\varphi'_2=0
   $$
   while taking $a=(0,j\,\bar\beta,0)$ and $a'=(\lambda',0,0)$ gives
   $$
   \varphi'_2=\beta \bar\lambda'\,,\quad
   \varphi_1=\varphi_2=\varphi'_1=0 .
   $$
   This shows that the vector space
   $\Omega^{(0,1)}_D$ of linear combinations $\sum_i a_i[D_F,a'_i]$
   is the space of pairs of quaternion valued functions
   $q( x)$ and $q'(x)$.

The selfadjointness condition $A=A^*$ is equivalent to $q' = q^*$
and we see that the discrete part $A^{(0,1)}$ is exactly given by a
quaternion valued function, $H(x) \in \H$ on $M$.

2) The transition is given by $(\varphi_1,\varphi_2)\mapsto
(-\bar\varphi_2,\bar\varphi_1)$, which corresponds to the
multiplication of $H=\varphi_1+\varphi_2\,j$ by $j$ on the left.
\endproof

\smallskip
For later purposes let us compute the trace of powers of $(D
+A^{(0,1)} +J A^{(0,1)}J)$. Let us define
\begin{equation}\label{diraczeroone}
D^{(0,1)}=D +A^{(0,1)} +J A^{(0,1)}J .
\end{equation}

\begin{lem} 1) On $\cH_F^{(3)}\subset \cH_F$ one has
\begin{equation}\label{traceh24}
\Tr((D^{(0,1)}_3)^2)=\,12\,|1+H|^2\,\Tr(\mass_{\upup}^*\mass_{\upup}+\mass_{\dodo}^*\mass_{\dodo})
\end{equation}
$$
\Tr((D^{(0,1)}_3)^4)=\,12\,|1+H|^4\,\Tr((\mass_{\upup}^*\mass_{\upup})^2+(\mass_{\dodo}^*\mass_{\dodo})^2)
$$

2) On $\cH_F^{(1)}\subset \cH_F$  one has
\begin{equation}\label{traceh24l}
\Tr((D^{(0,1)}_1)^2)=\,4\,|1+H|^2\,\Tr(\mass_{\nunu}^*\mass_{\nunu}+\mass_{\elel}^*\mass_{\elel})+2\,\Tr(\mass_R^*\mass_R)
\end{equation}
$$
\Tr((D^{(0,1)}_1)^4)=\,4\,|1+H|^4\,\Tr((\mass_{\nunu}^*\mass_{\nunu})^2+(\mass_{\elel}^*\mass_{\elel})^2)+2\,\Tr((\mass_R^*\mass_R)^2)
$$ $$+\,8\,|1+H|^2\,\Tr(\mass_R^*\mass_R \mass_{\nunu}^*\mass_{\nunu})
$$
\end{lem}

\proof 1) The left hand side of \eqref{traceh24l} is given by
$2\,\Tr((A^{(0,1)}_3)^2)$, after replacing $H$ by $1+H$ to take into
account the operator $D_3$.  The product $X\,X^*$ is given by the
diagonal matrix
$$
X\,X^*= \left[
  \begin{array}{cc}
    \mass_{\upup}^*\mass_{\upup}\,(\varphi_1\bar\varphi_1 +\varphi_2\bar\varphi_2) & 0 \\
    0 & \mass_{\dodo}^*\mass_{\dodo}\,(\varphi_1\bar\varphi_1 +\varphi_2\bar\varphi_2) \\
  \end{array}
\right] =\,|H|^2\, \left[
  \begin{array}{cc}
    \mass_{\upup}^*\mass_{\upup}\, & 0 \\
    0 & \mass_{\dodo}^*\mass_{\dodo}\,\\
  \end{array}
\right]
$$
One has $\Tr((A^{(0,1)}_3)^2)=3\,\Tr(X\,X^*+X^*X)=6\,\Tr(X\,X^*)$.
This gives the first equality. Similarly, one has
$\Tr((A^{(0,1)}_3)^4)=3\,\Tr((X\,X^*)^2+(X^*X)^2)=6\,\Tr((X\,X^*)^2)$,
which gives the second identity.

\smallskip

2) Let us write the matrix of $(D +A^{(0,1)} +J A^{(0,1)}J)_1$ in
the decomposition
$(\uparrow_R,\downarrow_R,\uparrow_L,\downarrow_L,\bar\uparrow_R,\bar
\downarrow_R,\bar\uparrow_L,\bar \downarrow_L)$. We have
$$
 \left[
   \begin{array}{cccccccc}
     0 & 0 & \mass_{\nunu}^*\,\varphi_1 & \mass_{\nunu}^*\,\varphi_2 & \mass_R^* & 0 & 0 & 0 \\
     0 & 0 &  -\mass_{\elel}^*\,\bar\varphi_2 & \mass_{\elel}^*\,\bar\varphi_1 & 0 & 0& 0 & 0 \\
     \mass_{\nunu} \bar\varphi_1& -\mass_{\elel} \,\varphi_2& 0 & 0 & 0 & 0 & 0 & 0 \\
     \mass_{\nunu} \bar\varphi_2 & \mass_{\elel} \,\varphi_1& 0 & 0 & 0 & 0 & 0 & 0 \\
     \mass_R& 0 & 0 & 0 & 0 & 0 & \bar \mass_{\nunu}^*\bar\varphi_1 & \bar \mass_{\nunu}^*\bar\varphi_2 \\
     0 & 0 & 0 & 0 & 0 & 0 & -\bar \mass_{\elel}^* \varphi_2 & \bar \mass_{\elel}^* \varphi_1 \\
     0 & 0 & 0 & 0 & \bar \mass_{\nunu} \varphi_1& -\bar \mass_{\elel}\bar\varphi_2 & 0 & 0 \\
     0 & 0 & 0 & 0 & \bar \mass_{\nunu} \varphi_2 & \bar \mass_{\elel}\bar\varphi_1 & 0 & 0 \\
   \end{array}
 \right]
$$
The only matrix elements of the square of $(D +A^{(0,1)} +J
A^{(0,1)}J)_1$ involving $\mass_R$ or $\mass_R^*$ are
\begin{small}
$$ \left[
  \begin{array}{cccccccc}
    \mass_R^*\mass_R+\mass_{\nunu}^*\mass_{\nunu}|H|^2 & 0 & 0 & 0 & 0 & 0 & \mass_R^* \bar \mass_{\nunu}^*\bar\varphi_1& \mass_R^*\bar \mass_{\nunu}^*\bar\varphi_2 \\
    0 & 0 & 0 & 0 & 0 & 0 & 0 & 0 \\
    0 & 0& 0& 0& \mass_{\nunu}  \mass_R^* \bar\varphi_1 & 0 & 0 & 0 \\
    0 & 0 & 0 & 0 & \mass_{\nunu}  \mass_R^*\bar\varphi_2 & 0 & 0 & 0 \\
    0 & 0 & \mass_R \mass_{\nunu}^*\,\varphi_1& \mass_R \mass_{\nunu}^*\,\varphi_2& \bar \mass_{\nunu}^*\bar \mass_{\nunu} |H|^2+\mass_R \mass_R^* & 0 & 0 & 0 \\
    0 & 0 & 0 & 0 & 0 & 0 & 0 & 0 \\
    \bar \mass_{\nunu}  \mass_R \varphi_1& 0 & 0 & 0 & 0 & 0 & 0 & 0 \\
    \bar \mass_{\nunu}  \mass_R \varphi_2 & 0 & 0 & 0 & 0 & 0 & 0 & 0 \\
  \end{array}
\right]
$$
\end{small}
This shows that one only gets two additional terms involving
$\mass_R$ for $\Tr((D +A^{(0,1)} +J A^{(0,1)}J)_1^2)$ and each gives
$\Tr(\mass_R \mass_R^*)$. The trace $\Tr((D +A^{(0,1)} +J
A^{(0,1)}J)_1^4)$ is the Hilbert-Schmidt norm square of $(D
+A^{(0,1)} +J A^{(0,1)}J)_1^2$ and we just need to add to the terms
coming from the same computation as  \eqref{traceh24} the
contribution of the terms involving $\mass_R$. The term
$\mass_R^*\mass_R+\mass_{\nunu}^*\mass_{\nunu}|H|^2$ contributes
(after replacing $H\mapsto 1+H$) by $2|1+H|^2\,\Tr(\mass_R^*\mass_R
\mass_{\nunu}^*\mass_{\nunu})$ and $\Tr((\mass_R^*\mass_R)^2)$. The
term $\bar \mass_{\nunu}^*\bar \mass_{\nunu} |H|^2+\mass_R
\mass_R^*$ gives a similar contribution. All the other terms give
simple additive contributions. One gets the result using
$$
\Tr(\bar \mass_{\nunu}  \mass_R \mass_R^* \bar
\mass_{\nunu}^*)=\Tr(\mass_R^*\mass_R \mass_{\nunu}^*\mass_{\nunu})
$$
which follows using complex conjugation from the symmetry of
$\mass_R$ \ie $\bar \mass_R=\mass_R^*$.
\endproof

\medskip

Thus, we obtain for the trace of powers of $D^{(0,1)}$ the formulae
\begin{equation}\label{secondpower}
\Tr((D^{(0,1)})^2)=\,4\,a\,|1+H|^2+2\,c
\end{equation}
and
\begin{equation}\label{fourthpower}
\Tr((D^{(0,1)})^4)= \,4\,b\,|1+H|^4\,+2\,d +\,8\,e\,|1+H|^2\,,
\end{equation}
where
\begin{eqnarray}
  a &=& \,\Tr(\mass_{\nunu}^*\mass_{\nunu}+\mass_{\elel}^*\mass_{\elel}
  +3(\mass_{\upup}^*\mass_{\upup}+\mass_{\dodo}^*\mass_{\dodo}))
  \label{momentslabels}\\
  b &=& \,\Tr((\mass_{\nunu}^*\mass_{\nunu})^2+(\mass_{\elel}^*\mass_{\elel})^2+3(\mass_{\upup}^*\mass_{\upup})^2+3(\mass_{\dodo}^*\mass_{\dodo})^2) \nonumber \\
  c &=& \Tr(\mass_R^*\mass_R)\nonumber  \\
  d &=& \Tr((\mass_R^*\mass_R)^2) \nonumber \\
  e &=& \Tr(\mass_R^*\mass_R \mass_{\nunu}^*\mass_{\nunu}) .\nonumber
\end{eqnarray}
\bigskip

\begin{rem}\label{RGEcoeffs}{\rm
The coefficients in \eqref{momentslabels} appear in the physics
literature in the renormalization group equation for the Yukawa
parameters. For instance, one can recognize the coefficients $a$ and
$b$ respectively as the $Y_2(S)$ and $H(S)$ of \cite{ACKMPRW}.}
\end{rem}

\subsubsection{The vector part $A^{(1,0)}$ of inner fluctuations}\label{NCGhyperSect}\hfill\medskip

Let us now determine the other part $A^{(1,0)}$ of $A$, \ie
\begin{equation}\label{onezeropart}
A^{(1,0)} = \sum \, a_i [(\dirac_M \otimes 1), a'_i] \, .
\end{equation}
We let $a_i = (\lambda_i ,q_i ,m_i)$, $a'_i = (\lambda'_i ,q'_i
,m'_i)$ be elements of $\cA=C^\infty(M,\cA_F)$. We obtain the
following.
\begin{enumerate}
  \item A $U(1)$ gauge field
\begin{equation}\label{U1gf}
  \Lambda = \sum \, \lambda_i \, d \lambda'_i
\end{equation}
  \item An $SU(2)$ gauge field
\begin{equation}\label{SU2gf}
  Q=\sum \, q_i \, d  q'_i
\end{equation}
  \item A $U(3)$ gauge field
\begin{equation}\label{SU3gf}
  V' = \sum \, m_i \, d  m'_i.
\end{equation}
\end{enumerate}

\smallskip
For 1) notice that we have two expressions to compute since there
are two different actions of $\lambda(x)$ in $L^2(M,S)$ given
respectively by
$$
\xi(x)\mapsto \lambda(x)\,\xi(x)\,,\quad \xi(x)\mapsto
\bar\lambda(x)\,\xi(x) .
$$
For the first one, using \eqref{diraconM}, the expression
$\Lambda=\sum \, \lambda_j [(\dirac_M \otimes 1), \lambda'_j] $ is
of the form
$$
\Lambda=\sqrt{-1}\,\sum \, \lambda_j
\partial_\mu\lambda'_j\gamma^\mu=\,\Lambda_\mu\,\gamma^\mu
$$
and it is self-adjoint when the scalar functions
$$
\Lambda_\mu=\sqrt{-1}\,\sum \, \lambda_j
\partial_\mu\lambda'_j
$$
are real valued. It follows then that the second one is given by
$$
\sum \, \bar\lambda_j [(\dirac_M \otimes 1), \bar\lambda'_j]=
\sqrt{-1}\,\sum \, \bar\lambda_j
\partial_\mu\bar\lambda'_j\gamma^\mu=\,-\,\Lambda_\mu\,\gamma^\mu .
$$
Thus, we see that, even though we have two representations of the
$\lambda(x)$, these generate only one $U(1)$ gauge potential. We use
the notation
\begin{equation}\label{uonegaugepot}
\Lambda_\mu=\,\frac{g_1}{2}\,B_\mu
\end{equation}
for this $U(1)$ gauge potential, which will play the role of the
generator of hypercharge (not to be confused with the
electromagnetic vector potential).
\smallskip

For 2) notice that the action of quaternions $\H$ can be represented
in the form
$$
q=\,f_0+\sum\,i\,f_\alpha\,\sigma^{ \alpha}\,,\quad
f_0\,,\;f_\alpha\in C^\infty(M,\R)
$$
where $\sigma^{ \alpha}$ are the Pauli matrices
\begin{equation}\label{paulimatrices}
\sigma_1= \left[
  \begin{array}{cc}
    0 & 1 \\
    1 & 0 \\
  \end{array}
\right] \,,\quad \sigma_2= \left[
  \begin{array}{cc}
    0 & -i \\
    i & 0 \\
  \end{array}
\right] \,,\quad \sigma_3= \left[
  \begin{array}{cc}
    1 & 0 \\
    0 & -1 \\
  \end{array}
\right] \,.\quad
\end{equation}
The Pauli matrices are self-adjoint. Thus the terms of the form
$$
f_0\,[(\dirac_M \otimes 1),\,i\,f'_\alpha\,\sigma^{ \alpha}]\,
$$
are self-adjoint. The algebra of quaternions admits the basis
$(1,i\sigma^{ \alpha})$. Thus, since the elements of this basis
commute with $\dirac_M$, one can rewrite
$$
\sum \, q_i [(\dirac_M \otimes 1), q'_i]=\sum\,f_0\,[(\dirac_M
\otimes 1),\,f'_0]+\sum\,f_\alpha\,[(\dirac_M \otimes
1),\,i\,f'_\alpha\,\sigma^{ \alpha}] ,
$$
where all $f$ and $f'$ are real valued functions. Thus, the
self-adjoint part of this expression is given by
$$
Q=\sum\,f_\alpha\,[(\dirac_M \otimes 1),\,i\,f'_\alpha\,\sigma^{
\alpha}] ,
$$
which is an $SU(2)$ gauge field. We write it in the form
\begin{equation}\label{Wder}
Q=\,Q_\mu\,\gamma^\mu\,,\quad
Q_\mu=\,\frac{g_2}{2}\,W_\mu^\alpha\,\sigma^\alpha .
\end{equation}
Using \eqref{diraconM}, we see that its effect is to generate the
covariant derivatives
\begin{equation}\label{Wderiv}
\partial_\mu -\,\frac i2\,g_2\,W_\mu^\alpha\,\sigma^\alpha .
\end{equation}

\smallskip
For 3), this follows as a special case of the computation of the
expressions of the form
$$
 A=\sum \, a_i [(\dirac_M \otimes 1), a'_i]\,,\quad a_i\,,\;a'_i\in
 C^\infty(M,M_N(\C)) .
$$
One obtains Clifford multiplication by all matrix valued 1-forms on
$M$ in this manner. The self-adjointness condition $A=A^*$ then
reduces them to take values in the Lie algebra of ${\rm U}(N)$
through the identifications $A=\,i\,H$ and
$$
{\rm Lie}({\rm U}(N))=\,\{H\in M_N(\C)\,,\;H^*=-H\}\,.
$$

\smallskip
We now explain how to reduce $V'$ to the Lie subalgebra $\SU(3)$ of
${\rm U}(3)$. We consider the following analogue of Definition
\ref{unimoddef1} of the unimodular subgroup $\SU(\cA_F)$.

\begin{defn} \label{unimod2} A gauge potential $A$ is ``unimodular"
iff $ \Tr(A)=0 $.
\end{defn}

\smallskip
We can now parameterize the unimodular gauge potentials and their
adjoint action, \ie the combination $A + J \, A \, J^{-1}$.

\smallskip
\begin{prop}\label{gaugepotentialssm}
\begin{enumerate}
  \item The unimodular gauge potentials are parameterized by a
  $U(1)$ gauge field $B$, an $\SU(2)$ gauge field $W$ and an
  $\SU(3)$  gauge field $V$.
  \item The adjoint action $A + J \, A \, J^{-1}$ on $\cH_f$ is obtained by
  replacing $\partial_\mu$ by $\partial_\mu+\bA_\mu$ where
  $\bA_\mu=\,(\bA_\mu^q \oplus \bA_\mu^\ell)\otimes 1_3
$ (where the $1_{3}$ is for the three generations), and
\begin{align*}
\bA_\mu^q &  =\left[
\begin{array}
[c]{ccc}%
 -\frac{2i}{3}g_{1}B_{\mu}  \otimes 1_{3} & 0 & 0
\\
0 &   \frac{i}{3}g_{1}B_{\mu}   \otimes1_{3} & 0\\
0 & 0 & (-\frac{i}{2}
g_{2}W_{\mu}^{\alpha}\sigma^{\alpha}-\frac{i}{6}g_{1}B_{\mu}\otimes1_{2})\otimes1_{3}%
\end{array}
\right]  \\
&  +1_{4} \otimes\left(  -\frac{i}{2}g_{3}%
V_{\mu}^{i}\lambda^{i}\right)  ,
\end{align*}
\begin{align*}
\bA_\mu^\ell &  =\left[
\begin{array}
[c]{ccc}%
0 & 0 & 0
\\
0 &    i\,g_{1}B_{\mu}     & 0\\
0 & 0 & (-\frac{i}{2}g_{2}W_{\mu}%
^{\alpha}\sigma^{\alpha}+\frac{i}{2}g_{1}B_{\mu}\otimes1_{2}) %
\end{array}\right]\,.
\end{align*}
\end{enumerate}
\end{prop}

\smallskip
Here the  $\sigma^{ \alpha}$ are the Pauli matrices
\eqref{paulimatrices} and $ \lambda^i$ are the Gell-mann matrices
\begin{equation}\label{gellmatrices}
\lambda_1= \left[
  \begin{array}{ccc}
    0 & 1 & 0 \\
    1 & 0 & 0 \\
    0 & 0 & 0 \\
  \end{array}
\right] \,,\; \lambda_2= \left[
  \begin{array}{ccc}
    0 & i & 0 \\
    -i & 0 & 0 \\
    0 & 0 & 0 \\
  \end{array}
\right] \,,\; \lambda_3= \left[
  \begin{array}{ccc}
    1 & 0 & 0 \\
    0 & -1 & 0 \\
    0 & 0 & 0 \\
  \end{array}
\right] \,,\; \lambda_4= \left[
  \begin{array}{ccc}
    0 & 0 & 1 \\
    0 & 0 & 0 \\
    1& 0 & 0 \\
  \end{array}
\right] \end{equation} $$ \lambda_5= \left[
  \begin{array}{ccc}
    0 & 0 & -i \\
    0 & 0 & 0 \\
    i & 0 & 0 \\
  \end{array}
\right] \,,\; \lambda_6= \left[
  \begin{array}{ccc}
    0 & 0 & 0 \\
    0 & 0 & 1 \\
    0 & 1 & 0 \\
  \end{array}
\right] \,,\; \lambda_7= \left[
  \begin{array}{ccc}
    0 & 0 & 0 \\
    0 & 0 & -i \\
    0 & i & 0 \\
  \end{array}
\right] \,,\; \lambda_8=\frac{1}{\sqrt 3} \left[
  \begin{array}{ccc}
    1 & 0 & 0 \\
    0 & 1 & 0 \\
    0 & 0 & -2 \\
  \end{array}
\right]
$$
which are self-adjoint and satisfy the relation
\begin{equation}\label{gelltrace}
\Tr ( \lambda^i  \lambda^j) = 2\delta^{ij} \, .
 \end{equation}

\proof 1) The action of $A$ on the subspace $\cH_f$ is of the form
$$
\left[
  \begin{array}{cccc}
    \Lambda  &0 &0 &0 \\
    0 &-\Lambda  &0 &0 \\
   0 &0 &Q_{11} &Q_{12}\\
    0 &0 &Q_{21} &Q_{22}  \\
  \end{array}
\right]
$$
on leptons and quarks. Thus, it is traceless, since $Q$ is traceless
as a linear combination of the Pauli matrices. The action of $A$ on
the subspace $\cH_{\bar f}$ is given by $\Lambda$ on the subspace of
leptons and by $V'$ on the space of quarks. One has $4$ leptons and
$4$ quarks per generation (because of the two possible chiralities)
and the color index is taken care of by $V'$. Thus, the
unimodularity condition means that we have
$$
3 \cdot 4\cdot(\Lambda+\Tr(V'))=0.
$$
Thus, we can write $V'$ as a sum of the form
\begin{equation}\label{vprimedec}
V'=\,-\,V-\frac 13 \left[
               \begin{array}{ccc}
                 \Lambda & 0 & 0 \\
                 0 & \Lambda & 0 \\
                 0 & 0 & \Lambda \\
               \end{array}
             \right]=\,-\,V-\frac 13\,\Lambda\,1_3 ,
\end{equation}
where $V$ is traceless, \ie it is an $\SU(3)$ gauge potential.

\smallskip

2)  Since the charge conjugation antilinear operator $J_M$ commutes
with $\dirac_M$, it anticommutes with the $\gamma_\mu$ and the
conjugation by $J$ introduces an additional minus sign in the gauge
potentials.  The computation of $A + J \, A \, J^{-1}$ gives, on
quarks and leptons respectively, the matrices
$$
\left[
  \begin{array}{cccc}
    \Lambda - V' &0 &0 &0 \\
    0 &-\Lambda - V' &0 &0 \\
   0 &0 &Q_{11} - V' &Q_{12}\\
    0 &0 &Q_{21} &Q_{22} - V' \\
  \end{array}
\right]
$$
$$
\left[
  \begin{array}{cccc}
  0&0&0&0\\
0&-2\Lambda &0 &0 \\
0 &0&Q_{11} - \Lambda &Q_{12} \\
0&0 &Q_{21} &Q_{22} -\Lambda \\
 \end{array}
\right]
$$
Thus, using \eqref{vprimedec}, we obtain for the $(1,0)$-part of the
inner fluctuation $A + J \, A \, J^{-1}$ of the metric the matrices
$$
 \left[
  \begin{array}{cccc}
\frac{4}{ 3} \, \Lambda + V &0 &0 &0 \\
0 &-\frac{2}{ 3} \, \Lambda + V &0 &0 \\
0 &0 &Q_{11} + \frac{1}{ 3} \, \Lambda + V &Q_{12} \\
0 &0 &Q_{21} &Q_{22} + \frac{1}{ 3} \, \Lambda + V \\
 \end{array}
\right]
$$
$$
\left[
  \begin{array}{cccc}
  0&0&0&0\\
0&-2\Lambda &0 &0 \\
0 &0&Q_{11} - \Lambda &Q_{12} \\
0&0 &Q_{21} &Q_{22} -\Lambda \\
 \end{array}
\right]
$$
This completes the proof.
\endproof

\begin{rem}\label{bosonshyperch}{\rm
Thus, we have obtained exactly the gauge bosons of the standard
model, coupled with the correct hypercharges $Y_L$, $Y_R$. They are
such that the electromagnetic charge $Q_{em}$ is determined by $2 \,
Q_{em} = Y_R$ for right handed particles. One also has $2 \, Q_{em}
= Y_L + 2 \, I_3$, where $I_3$ is the third generator of the weak
isospin group $SU(2)$. For $Q_{em}$ one gets the same answer for the
left and right components of each particle and $\frac{2}{3}$,
$-\frac{1}{3}$ for the $u,d$ quarks, respectively, and $0$ and $-1$
for the $\nu$ and the $e$ leptons, respectively.}
\end{rem}

\subsubsection{Independence}\label{Indep}\hfill\medskip

It remains to explain why the fields $H=\varphi_1+ j\,\varphi_2$ of
Proposition \ref{prophiggsfieldssm} and $B, W, V$ of Proposition
\ref{gaugepotentialssm} are independent of each other.

\begin{prop}\label{indepfields}
The unimodular inner fluctuations of the metric are parameterized by
independent fields $\varphi_1$, $\varphi_2$, $B$, $W$, $V$, as in
Propositions \ref{prophiggsfieldssm} and \ref{gaugepotentialssm}.
\end{prop}

\proof Let $Z$ be the real vector bundle over $M$, with fiber at $x$
$$ \C\oplus \C\oplus T^*_x M \oplus T^*_x M\otimes {\rm Lie}(\SU(2))
\oplus T^*_x M \otimes {\rm Lie}(\SU(3)). $$ By construction the
inner fluctuations are sections of the bundle $Z$.

The space of sections $\cS$ obtained from inner fluctuations is in
fact not just a linear space over $\R$, but also a module over the
algebra $C^\infty(M,\R)$ which is the real part of
$C^\infty(M,\cA_F)$ (Lemma \ref{lemrealpartF}). Indeed, the inner
fluctuations are obtained as expressions of the form $A=\sum
a_j\,[D,a'_j]$. One has to check that left multiplication by $f\in
C^\infty(M,\R)$ does not alter the self-adjointness condition
$A=A^*$. This follows from Proposition \ref{realequalcentral}, since
we are replacing $a_j$ by $f a_j$, where $f$ commutes with $A$ and
is real so that $f=f^*$.

To show that $\cS=C^\infty(M,Z)$ it is enough to know that one can
find sections in $\cS$ that span the full vector space $Z_x$ at any
given point $x\in M$. Then $C^\infty(M,\R)$-linearity shows that the
same sections continue to span the nearby fibers. Using a partition
of unity one can then express any global section of $Z$ as an
element of $\cS$.

Choose first the elements $a_i (y) = (\lambda_i , q_i ,m_i)$, $a'_i
(y) = (\lambda'_i ,q'_i ,m'_i)$ independent of $y\in N(x)$ in a
neighborhood of $x$. Using Proposition \ref{prophiggsfieldssm}, one
knows that $H(x)$ can be an arbitrary element of $\H$, while $B(x),
W(x), V(x)$ all vanish because they are differential expressions of
the $a_i'$.

The independence of $\lambda$, $q$ and $m$ in the formulae
\eqref{U1gf}, \eqref{SU2gf}, \eqref{SU3gf} implies that one can
construct arbitrary $B(x)$, $W(x)$, $V(x)$ in the form $\sum_i a_i
[D,a_i']$. These, however, will not suffice to give an arbitary
value for $\varphi_1$ and $\varphi_2$, but this can be corrected by
adding an element of the form described above, with vanishing $B$,
$W$, and $V$.
\endproof

\bigskip

\subsection{The Dirac operator and its
square}\label{SMdirac}\hfill\medskip

The Dirac operator $D_A$ that takes the inner fluctuations into
account is given by the sum of two terms
\begin{equation}\label{diracdec}
D_A=D^{(1,0)}+\,\gamma_5\otimes D^{(0,1)} ,
\end{equation}
where $D^{(0,1)}$ is given by \eqref{diraczeroone} and $D^{(1,0)}$
is of the form
\begin{equation}\label{diraconezero}
D^{(1,0)}=\sqrt{-1}\,\gamma^\mu(\nabla^s_\mu+\bA_\mu) ,
\end{equation}
where $\nabla^s$ is the spin connection (\cf\eqref{diraconM}).

\smallskip

The gauge potential $\bA_\mu$ splits as a direct sum in the
decomposition associated to $\cH_F=\cH_f\oplus \cH_{\bar f}$ and its
restriction to $\cH_f$ is given by Proposition
\ref{gaugepotentialssm}.

\medskip
In order to state the next step, \ie the computation of the square
of $D_A$, we introduce the notations
\begin{equation}\label{termcom0}
T(M_1,M_2,\varphi)=\left[
  \begin{array}{cccc}
0 & 0 & M_1^*\,\varphi_1 & M_1^*\,\varphi_2 \\
0 & 0 &  -M_2^*\,\bar\varphi_2 & M_2^*\,\bar\varphi_1\\
M_1 \bar\varphi_1& -M_2 \,\varphi_2& 0 & 0 \\
M_1 \bar\varphi_2 & M_2 \,\varphi_1& 0 & 0\\
  \end{array}
\right]
\end{equation}
with $\varphi=(\varphi_1,\varphi_2)$ and $M_j$ a pair of matrices,
and
\begin{equation}\label{termcomdef}
\bM(\varphi)=T(\mass_{\upup},\mass_{\dodo},\varphi)\otimes 1_3\oplus
T(\mass_{\nunu},\mass_{\elel},\varphi)\oplus
\overline{T(\mass_{\upup},\mass_{\dodo},\varphi)}\otimes
1_3\oplus\overline{T(\mass_{\nunu},\mass_{\elel},\varphi)}
\end{equation}
By construction $\bM(\varphi)$ is self-adjoint and one has
\begin{equation}\label{termcomsquare}
\Tr(\bM(\varphi)^2)=4\, a \,|\varphi|^2\,,\quad
a=\,\Tr(\mass_{\nunu}^*\mass_{\nunu}+\mass_{\elel}^*\mass_{\elel}+3(\mass_{\upup}^*\mass_{\upup}+\mass_{\dodo}^*\mass_{\dodo}))
\end{equation}

\medskip

\begin{lem}\label{squareofda} The square of $D_A$ is given by
\begin{equation}\label{diracsquare}
D_A^2=\nabla^*\nabla-\cE ,
\end{equation}
where $\nabla^*\nabla$ is the connection Laplacian for the
connection
\begin{equation}\label{connectionnabla}
\nabla=\nabla^s+\bA
\end{equation}
and the endomorphism $\cE$ is given, with $s=-R$ the scalar
curvature, by
\begin{equation}\label{endomorphisme}
-\cE=\frac 14 \,s   \otimes {\rm id} + \sum_{\mu<\nu}\,
  \gamma  ^{\mu } \gamma^\nu \otimes  \bF_{\mu \nu}-i\,\gamma_5\,\gamma^\mu\otimes\,\bM(D_{\mu} \, \varphi)+1_4\otimes (D^{0,1})^2
\end{equation}
with $H=\varphi_1+\varphi_2 j$ as above, and
$\varphi=(\varphi_1,\varphi_2)$. Here $\bF_{\mu \nu}$ is the
curvature of the connection $\bA$ and
$\varphi=(\varphi_1,\varphi_2)$ is a row vector. The term $D_\mu
\varphi$ in \eqref{endomorphisme} is of the form
\begin{equation}\label{minimalcoupling}
D_{\mu} \, \varphi  =  \,
\partial_{\mu} \, \varphi + \frac{i }{ 2} \, g_{2} \, W_{\mu}^{ \alpha} \,\varphi\, \sigma^{ \alpha}
 - \frac{i }{ 2} \, g_{1} \, B_{\mu} \, \varphi \, .
\end{equation}
\end{lem}

\proof By construction $D^{1,0}$ anticommutes with $\gamma_5$. Thus,
one has
$$
D_A^2=(D^{1,0})^2+1_4\otimes
(D^{0,1})^2-\gamma_5\,[D^{1,0},1_4\otimes D^{0,1}] .
$$
The last term is of the form
$$
[D^{1,0},1_4\otimes
D^{0,1}]=\,\sqrt{-1}\,\gamma^\mu\,[(\nabla^s_\mu+\bA_\mu),1_4\otimes
D^{0,1}] .
$$
Using \eqref{spincon}, one can replace $\nabla^s_\mu$ by
$\partial_{\mu}$ without changing the result. In order to compute
the commutator $[\bA_\mu,D^{0,1}]$, notice first that the off
diagonal term of $D^{0,1}$ does not contribute, since the
corresponding matrix elements of $\bA_\mu^\ell$ are zero. Thus, it
is enough to compute the commutator of the matrix
\begin{equation}\label{termcom}
\bW=\left[
  \begin{array}{cccc}
    -\frac{i}{2}g_{1}B_{\mu}  & 0 & 0 & 0 \\
    0 & \frac{i}{2}g_{1}B_{\mu}  & 0 & 0 \\
    0 & 0 & -\frac{i}{2}g_{2}W_\mu^3 & -\frac{i}{2}g_{2}(W_\mu^1-iW_\mu^2)\\
    0 & 0 & -\frac{i}{2}g_{2}(W_\mu^1+iW_\mu^2) & \frac{i}{2}g_{2}W_\mu^3 \\
  \end{array}
\right]
\end{equation}
with a matrix of the form
\begin{equation}\label{termcom1}
T(M_1,M_2,\varphi)=\left[
  \begin{array}{cccc}
0 & 0 & M_1^*\,\varphi_1 & M_1^*\,\varphi_2 \\
0 & 0 &  -M_2^*\,\bar\varphi_2 & M_2^*\,\bar\varphi_1\\
M_1 \bar\varphi_1& -M_2 \,\varphi_2& 0 & 0 \\
M_1 \bar\varphi_2 & M_2 \,\varphi_1& 0 & 0\\
  \end{array}
\right]
\end{equation}
One gets
\begin{equation}\label{termcom2}
[\bW,T(M_1,M_2,\varphi)]=T(M_1,M_2,\psi) ,
\end{equation}
where
\begin{equation}\label{termcom3}
(\psi_1,\psi_2)=\,-\frac{i}{2}g_{1}B_{\mu}(\varphi_1,\varphi_2)+\frac{i}{2}g_{2}
W_\mu^\alpha\,(\varphi_1,\varphi_2)\,\sigma^\alpha .
\end{equation}
\endproof

\subsection{The spectral action and the asymptotic expansion}\hfill
\medskip

In this section we compute the spectral action for the inner
fluctuations of the product geometry  $M\times F$.

\medskip

\begin{thm}\label{spectralactmf}
 The spectral  action is given by
\begin{eqnarray}
  S &=& \, \frac{1}{\pi^2}(48\,f_4\,\Lambda^4-f_2\,\Lambda^2\,c+\frac{
f_0}{4}\,d)\,\int \,\sqrt g\,d^4 x \label{bossm}\\
    &+& \, \frac{96\,f_2\,\Lambda^2 -f_0\,c}{ 24\pi^2} \, \int\,R
 \, \sqrt g \,d^4 x  \nonumber\\
    &+& \, \frac{f_0 }{ 10\,\pi^2} \int\,(\frac{11}{6}\,R^* R^* -3 \, C_{\mu
\nu \rho \sigma} \, C^{\mu \nu \rho \sigma})\, \sqrt g \,d^4 x \nonumber \\
 &+&  \, \frac{(- 2\,a\,f_2
  \,\Lambda^2\,+ \,e\,f_0 )}{ \pi^2} \int\,  |\varphi|^2\, \sqrt g \,d^4 x \nonumber \\
    &+&  \, \frac{f_0 }{ 2\,\pi^2} \int\, a\, |D_{\mu} \varphi|^2\, \sqrt g \,d^4 x \nonumber \\
   &-&  \frac{f_0}{ 12\,\pi^2} \int\, a \,R \, |\varphi|^2 \, \sqrt g \,d^4 x
 \nonumber\\
    &+& \, \frac{f_0 }{ 2\,\pi^2} \int\,(g_{3}^2 \, G_{\mu \nu}^i \, G^{\mu \nu i} +  g_{2}^2 \, F_{\mu
\nu}^{ \alpha} \, F^{\mu \nu  \alpha}+\, \frac{5}{ 3} \,
g_{1}^2 \,  B_{\mu \nu} \, B^{\mu \nu})\, \sqrt g \,d^4 x\nonumber \\
    &+&\, \frac{f_0 }{ 2\,\pi^2} \int\,b\, |\varphi|^4 \, \sqrt g \,d^4 x \nonumber
\end{eqnarray}
where
$$ R^*R^*= \frac 14 \epsilon^{\mu\nu\rho\sigma}
\epsilon_{\alpha\beta\gamma\delta} R^{\alpha\beta}_{\mu\nu}
R^{\gamma\delta}_{\rho\sigma} $$ is the topological term that
integrates to the Euler characteristic. The coefficients
$(a,b,c,d,e)$ are defined in \eqref{momentslabels} and
$D_\mu\varphi$ is defined in \eqref{minimalcoupling}.
\end{thm}

\medskip

\proof To prove Theorem \ref{spectralactmf} we use
\eqref{diracsquare} and we apply Gilkey's theorem (see Theorem
\ref{gilkeythm} below) to compute the spectral action. By Remark
\ref{eminusr} below, the relevant term is $-\frac 16 R +\cE$, which
is the sum
\begin{equation}\label{endomorphismer6}
\cE'=-\frac R6 {\rm id} +\cE=(\frac{R}{12}{\rm id}-1_4\otimes
(D^{0,1})^2)- \sum_{\mu<\nu}\,
  \gamma  ^{\mu } \gamma^\nu \otimes  \bF_{\mu \nu}
  +i\,\gamma_5\,\gamma^\mu\otimes\,\bM(D_{\mu} \, \varphi) .
\end{equation}
We need to compute the sum
\begin{equation}\label{gilkeyspectral}
\Sigma=\frac{f_2}{8\,\pi^2} \Lambda^2 \Tr(\cE')+
\frac{f_0}{32\,\pi^2} \Tr((\cE')^2) .
\end{equation}

\medskip
\begin{lem} The term $\Sigma$ in \eqref{gilkeyspectral} is given by
\begin{eqnarray}
  \Sigma &=& \,\frac{4\,f_2}{\pi^2}\,\Lambda^2\,R-\frac{f_2}{2\,\pi^2}
  \,\Lambda^2\,\Tr((D^{0,1})^2)+\,\frac{f_0}{8\,\pi^2}\,\Tr(\bM(D_{\mu}
\, \varphi)^2) \label{gilkeyspectral1} \\
    &+& \,\frac{f_0}{8\,\pi^2}\,\Tr((\frac{R}{12}-
(D^{0,1})^2)^2) +\,\frac{f_0}{16\,\pi^2}\,\Tr(\bF_{\mu \nu}\bF^{\mu
\nu}). \nonumber
 \end{eqnarray}
\end{lem}

\proof The contribution of $\Tr(\cE')$ is only coming from the first
term of \eqref{endomorphismer6}, since the trace of the two others
vanishes due to the Clifford algebra terms. The coefficient of
$\frac{f_2\,\Lambda^2}{\pi^2}\,R$ is
$\frac{1}{8}\cdot\frac{1}{12}\cdot 4\cdot 96=4$. To get the
contribution of $\Tr((\cE')^2)$, notice that the three terms of the
sum \eqref{endomorphismer6} are pairwise orthogonal in the Clifford
algebra, so that the trace of the square is just the sum of the
three contributions from each of these terms. Again the factor of
$4$ comes from the dimension of spinors and the summation on all
indices $\mu\nu$ gives a factor of two in the denominator for
$\frac{f_0}{16\,\pi^2}$.
\endproof

Notice also that the curvature $\Omega_{\mu\nu}$ of the connection
$\nabla$ is independent of the additional term $D^{(0,1)}$. We now
explain the detailed computation of the various terms of the
spectral action.
\medskip

\subsubsection{$\Lambda^4$-terms}\hfill \medskip

The presence of the additional off-diagonal term in the Dirac
operator of the finite geometry adds two contributions to the
cosmological term of \cite{cc2}. Thus while the dimension $N=96$
contributes by the term
$$
\frac{48}{\pi^2}\,f_4\,\Lambda^4\,\int \,\sqrt g\,d^4 x \,,
$$
we get the additional coefficients
$$
-\frac{ f_2}{ \pi^2}\,\Lambda^2\,\Tr(\mass_R^*\mass_R)=-\frac{
c\,f_2}{ \pi^2}\,\Lambda^2 ,
$$
which are obtained from the second term of \eqref{gilkeyspectral1},
using \eqref{secondpower}. Finally, we also get
$$
\frac{ f_0}{4\,\pi^2}\,\Tr((\mass_R^*\mass_R)^2)=\frac{d\,
f_0}{4\,\pi^2}\,,
$$
which comes  from the fifth term in \eqref{gilkeyspectral1}. Thus,
the cosmological term gives
\begin{equation}\label{cosmologicalterm}
\frac{1}{\pi^2}(48\,f_4\,\Lambda^4-f_2\,\Lambda^2\,\,\Tr(\mass_R^*\mass_R)+\frac{
f_0}{4}\,\Tr((\mass_R^*\mass_R)^2))\,\int \,\sqrt g\,d^4 x \, .
\end{equation}

\medskip

\subsubsection{Riemannian curvature terms}\hfill \medskip

The computation of the terms that only depend upon the Riemann
curvature tensor is the same as in \cite{cc2}. It gives the additive
contribution
\begin{equation}\label{riemannianterms}
\frac{1}{\pi^2}\,\int
\,(4\,f_2\,\Lambda^2\,R-\frac{3}{10}\,f_0\,C_{\mu \nu \rho \sigma}
\, C^{\mu \nu \rho \sigma})\sqrt g\,d^4 x \,,
\end{equation}
together with topological terms. Ignoring boundary terms, the latter
is of the form
\begin{equation}\label{riemannianterms1}
\frac{11\,f_0}{60\,\pi^2}\int R^* R^* \,\sqrt g\,d^4 x .
\end{equation}
There is, however, an additional contribution from the fourth term
of \eqref{gilkeyspectral1}. Using \eqref{secondpower}, this gives
\begin{equation}\label{riemannianterms2}
-\frac{f_0}{48\,\pi^2}\,R\,\Tr((D^{0,1})^2)=-\frac{f_0}{12\,\pi^2}\,a\,R\,|\varphi|^2
-\frac{f_0}{24\,\pi^2}\,c\,R\, .
\end{equation}
Notice the presence of the terms in $R\,|\varphi|^2$ (\cf
\cite{feynmgrav} equation 10.3.3).

\medskip

\subsubsection{Scalar minimal coupling}\hfill \medskip

These terms are given by
\begin{equation}\label{minimalcouplingterms}
\frac{f_0}{8\,\pi^2}\,\Tr(\bM(D_{\mu} \,
\varphi)^2)=\,\frac{f_0}{2\,\pi^2}\,a\,|D_{\mu} \, \varphi|^2
\end{equation}
using \eqref{termcomsquare} and \eqref{gilkeyspectral1}.

\smallskip
Notice that there is a slight change of notation with respect to
\cite{cc1} since we are using the Higgs doublet $\tilde H$ instead
of $H$ with the notations of \cite{cc1}.

\medskip

\subsubsection{Scalar mass terms}\hfill \medskip

There are two contributions with opposite signs. The second term in
\eqref{gilkeyspectral1} \ie
$$
-\frac{f_2}{2\,\pi^2}
  \,\Lambda^2\,\Tr((D^{0,1})^2)
  $$
  gives, using \eqref{secondpower}, a term in
  $$
-\frac{2\,f_2}{\pi^2}
  \,\Lambda^2\,a\,|\varphi|^2\,.
  $$
The fourth term in \eqref{gilkeyspectral1} gives, using
\eqref{fourthpower},
$$
\frac{f_0}{8\,\pi^2}\,8\,e\,|\varphi|^2=\,\frac{\,e\,f_0}{
\pi^2}\,|\varphi|^2 .
$$
Thus, the mass term gives
\begin{equation}\label{higgsmassterms}
\frac{1}{\pi^2}(- 2\,a\,f_2
  \,\Lambda^2\,+ \,e\,f_0 )\,|\varphi|^2  \, .
\end{equation}
\medskip

\subsubsection{Scalar quartic potential}\hfill \medskip

The only contribution, in this case, comes from the fourth term in
\eqref{gilkeyspectral1} \ie from the term
$$
\frac{f_0}{8\,\pi^2}\,\Tr((D^{0,1})^4) .
$$
Using \eqref{fourthpower}, this gives
\begin{equation}\label{higgsquarticterms}
\frac{f_0}{2\,\pi^2}\,b\,|\varphi|^4 \, .
\end{equation}

\medskip

\subsubsection{Yang-Mills terms}\hfill \medskip

For the Yang-Mills terms the computation is the same as in
\cite{cc2}. Thus, we get the coefficient $\frac{f_0}{24\pi^2}$ in
front of the trace of the square of the curvature. For the gluons,
\ie the term $ G_{\mu \nu}^i \, G^{\mu \nu i}$, we get the
additional coefficient $3\cdot 4\cdot 2=24$, since there are three
generations, $4$ quarks per generation $(u_R,d_R,u_L,d_L)$, and a
factor of two coming from the sectors $\cH_f$ and $\cH_{\bar f}$. In
other words, because of the coefficient $\frac{g_3}{2}$, we get
$$
\frac{f_0\,g_3^2}{4\pi^2}\,\Tr(G_{\mu \nu} \, G^{\mu \nu })=\,2 \,
\frac{f_0\,g_3^2}{4\pi^2}\,G^i_{\mu \nu} \, G_i^{\mu \nu }=\,
\frac{f_0\,g_3^2}{2\pi^2}\,G^i_{\mu \nu} \, G_i^{\mu \nu } ,
$$
where we use \eqref{gelltrace}. For the weak interaction bosons
$W^\alpha$ we get the additional coefficient $3\cdot 4\cdot 2=24$
with the $3$ for $3$ generations, the $4$  for the $3$ colors of
quarks and $1$ lepton per isodoublet and per generation
$(u_{jL},d_{jL},\nu_L,e_L)$, and the factor of $2$  from the sectors
$\cH_f$ and $\cH_{\bar f}$. Thus, using
$\Tr(\sigma_a\sigma_b)=2\delta_{ab}$, we obtain the similar term
$$
\frac{f_0\,g_2^2}{4\pi^2}\,\Tr(F_{\mu \nu} \, F^{\mu \nu
})=\,2\,\frac{f_0\,g_2^2}{4\pi^2}\,F^a_{\mu \nu} \, F_a^{\mu \nu }=
\,\frac{f_0\,g_2^2}{2\pi^2}\,F^a_{\mu \nu} \, F_a^{\mu \nu } .
$$
For the hypercharge generator $B_\mu$ we get the additional
coefficient
$$
2\cdot(((\frac{4}{3})^2+(\frac{2}{3})^2+2\,(\frac{1}{3})^2)\cdot 3
+(2^2+2))\cdot 3=80 ,
$$
which gives an additional coefficient of $\frac {10}{3}$ in the
corresponding term
$$
\frac {10}{3}\frac{f_0\,g_1^2}{4\pi^2}\, B_{\mu \nu} \, B^{\mu \nu}=
\frac 53\,\frac{f_0\,g_1^2}{2\pi^2}\, B_{\mu \nu} \, B^{\mu \nu} .
$$

This completes the proof of Theorem \ref{spectralactmf}.
\endproof

\bigskip

\section{The Lagrangian}\label{SectLag}

The $KO$-dimension of the finite space $F$ is $6\in \Z/8$, hence the
$KO$-dimension of the product geometry $M\times F$ (for $M$ a spin
$4$-manifold) is now $2\in \Z/8$. In other words, according to
Definition \ref{realstr}, the commutation rules are
\begin{equation}\label{per82}
J^2 = -1, \ \ \ \ JD =   DJ, \ \ \text{and} \ \ J\gamma = - \gamma J
\, .
\end{equation}
Let us now explain how these rules define a natural antisymmetric
bilinear form on the even part
\begin{equation}\label{Hplus}
\cH^+=\{\xi\in \cH\;,\;\gamma\,\xi=\xi\}
\end{equation}
of $\cH$.

\smallskip
\begin{prop} \label{functspec0} On a real spectral triple of $KO$-dimension $2\in
\Z/8$, the expression
\begin{equation}\label{antisybil}
A_D(\xi',\xi)= \langle J\,\xi',D\,\xi\rangle \qqq \xi, \xi'\in \cH^+
\end{equation}
defines an antisymmetric bilinear form on $\cH^+=\{\xi\in
\cH\;,\;\gamma\,\xi=\xi\}$. The trilinear pairing \eqref{antisybil}
between $D$, $\xi$ and $\xi'$ is gauge invariant under the adjoint
action of the unitary group of $\cA$, namely
\begin{equation}\label{antisybil1}
A_D(\xi',\xi)=A_{D_u}({\rm Ad}(u)\xi',{\rm Ad}(u)\xi)\,,\quad
D_u={\rm Ad}(u)\,D\,{\rm Ad}(u^*) .
\end{equation}
\end{prop}

\proof 1) We use an inner product which is antilinear in the first
variable. Thus, since $J$ is antilinear, $A$ is a bilinear form. Let
us check that $A$ is antisymmetric. One has
$$
A_D(\xi,\xi')=\langle\,J\,\xi,D\,\xi'\rangle=-\langle\,J\,\xi,J^2\,D\,\xi'\rangle
=-\langle\,J\,D\,\xi',\,\xi\rangle=-\langle\,D\,J\,\xi',\,\xi\rangle
=-\langle\,J\,\xi',\,D\,\xi\rangle
$$
where we used the unitarity of $J$, \ie the equality
\begin{equation}\label{JJinnprod}
\langle\,J\,\xi,J\,\eta\rangle=\langle\eta,\xi\rangle\qqq
\xi\,,\eta\in \cH .
\end{equation}
Finally, one can restrict the antisymmetric form $A_D$ to $\cH^+$
without automatically getting zero since one has
$$
\gamma\,JD=\,JD\,\gamma\,.
$$

\smallskip

2) Let us check that ${\rm Ad}(u)$ commutes with $J$. By definition
${\rm Ad}(u)=u\,(u^*)^0=u\,JuJ^{-1}$. Thus
$$
J\,{\rm Ad}(u)=J\,u\,JuJ^{-1}=\,u\,J\,u\,JJ^{-1}=\,u\,J\,u={\rm
Ad}(u)\,J\,,
$$
where we used the commutation of $u$ with $J\,u\,J$. Since ${\rm
Ad}(u)$ is unitary, one gets \eqref{antisybil1}.
\endproof

\medskip

Now the Pfaffian of an antisymmetric bilinear form is best expressed
in terms of the functional integral involving anticommuting
``classical fermions" (\cf \cite{Ramond}, \S 5.1) At the formal
level, this means that we write
\begin{equation}\label{PfA}
{\rm Pf}(A)=\int\,e^{-\frac 12 A(\tilde\xi)}\,D[\tilde\xi]
\end{equation}
Notice that $A(\xi,\xi)=0$ when applied to a vector $\xi$, while
$A(\tilde\xi,\tilde\xi)\neq 0$ when applied to anticommuting
variables $\tilde\xi$. We define
\begin{equation}\label{Hpluscl}
\cH^+_{cl}=\{\tilde\xi \, : \,\xi\in \cH^+\}
\end{equation}
to be the space of classical fermions (Grassman variables)
corresponding to $\cH^+$ of \eqref{Hplus}.

\smallskip

As the simplest example let us consider a two dimensional vector
space $E$ with basis $e_j$ and the antisymmetric bilinear form $$
A(\xi',\xi)=a(\xi'_1\xi_2-\xi'_2\xi_1)$$ For $\tilde\xi_1$
anticommuting with $\tilde\xi_2$, using the basic rule (\cf
\cite{Ramond}, \S 5.1)
$$
\int\,\tilde\xi_j\,d\tilde\xi_j=1
$$
one gets
$$
\int\,e^{-\frac 12
A(\tilde\xi)}\,D[\tilde\xi]=\int\,e^{-a\,\tilde\xi_1\tilde\xi_2}
\,d\tilde\xi_1d\tilde\xi_2=\,a .
$$

\medskip

\begin{rem}\label{ferm2}{\rm
It is the use of the Pfaffian as a square root of the determinant
that makes it possible to solve the Fermion doubling puzzle which
was pointed out in \cite{lizzi}. We discuss this in \S
\ref{sectFerm2} below. The solution obtained by a better choice of
the $KO$-dimension of the space $F$ and hence of $M\times F$ is not
unrelated to the point made in \cite{gbis}.}
\end{rem}

\medskip
We now state our main result as follows.

\begin{thm}\label{maintheorem}
Let $M$ be a Riemannian spin $4$-manifold and $F$ the finite
noncommutative geometry of $KO$-dimension $6$ described above. Let
$M\times F$ be endowed with the product metric.
\begin{enumerate}
\item The unimodular subgroup of the unitary group acting by the adjoint
representation ${\rm Ad}(u)$ in $\cH$ is the group of gauge
transformations of SM.
\item The unimodular inner fluctuations of the metric give the gauge
bosons of SM.
\item The full standard model (with neutrino mixing and seesaw
mechanism) minimally coupled to Einstein gravity is given in
Euclidean form by the action functional
\begin{equation}\label{functspec}
S=\,\Tr(f(D_A/\Lambda))+\frac
12\,\langle\,J\,\tilde\xi,D_A\,\tilde\xi\rangle\,,\quad\tilde\xi\in
\cH^+_{cl} ,
\end{equation}
where $D_A$ is the Dirac operator with the unimodular inner
fluctuations.
\end{enumerate}
\end{thm}

\begin{rem}\label{alldata}{\rm
Notice that the action functional \eqref{functspec} involves all the
data of the spectral triple, including the grading $\gamma$ and the
real structure $J$.}
\end{rem}

\medskip
\proof We split the proof of the theorem in several subsections.

To perform the comparison, we look separately at the terms in the SM
Lagrangian. After dropping the ghost terms, one has five different
groups of terms.
\begin{enumerate}
  \item Yukawa coupling $\cL_{Hf}$ \index{coupling!Yukawa}
\index{Yukawa coupling}
  \item Gauge fermion couplings $\cL_{gf}$ \index{coupling!gauge-fermion}
  \item Higgs self-coupling $\cL_{H}$ \index{coupling!Higgs}
  \item Self-coupling of gauge fields $\cL_{g}$ \index{coupling!gauge fields}
  \item Minimal coupling of Higgs fields $\cL_{Hg}$
\end{enumerate}

\subsection{Notation for the standard model}\hfill
\medskip \label{sectstandardmodel}

The spectral action naturally gives a Lagrangian for matter
minimally coupled with gravity, so that we would obtain the standard
model Lagrangian on a curved spacetime. By covariance, it is in fact
sufficient to check that we obtain the standard model Lagrangian in
flat spacetime. This can only be done by a direct calculation, which
occupies the remaining of this section.

In flat space and in Lorentzian signature the Lagrangian of the
standard model with neutrino mixing and Majorana mass terms, written
using the Feynman gauge fixing, is of the form
\begin{small}
\begin{center}
\begin{math}
{\mathcal
L}_{SM}=-\frac{1}{2}\partial_{\nu}g^{a}_{\mu}\partial_{\nu}g^{a}_{\mu}
-g_{s}f^{abc}\partial_{\mu}g^{a}_{\nu}g^{b}_{\mu}g^{c}_{\nu}
-\frac{1}{4}g^{2}_{s}f^{abc}f^{ade}g^{b}_{\mu}g^{c}_{\nu}g^{d}_{\mu}g^{e}_{\nu}
-\partial_{\nu}W^{+}_{\mu}\partial_{\nu}W^{-}_{\mu}-M^{2}W^{+}_{\mu}W^{-}_{\mu}
-\frac{1}{2}\partial_{\nu}Z^{0}_{\mu}\partial_{\nu}Z^{0}_{\mu}-\frac{1}{2c^{2}_{w}}
M^{2}Z^{0}_{\mu}Z^{0}_{\mu}
-\frac{1}{2}\partial_{\mu}A_{\nu}\partial_{\mu}A_{\nu}
-igc_{w}(\partial_{\nu}Z^{0}_{\mu}(W^{+}_{\mu}W^{-}_{\nu}-W^{+}_{\nu}W^{-}_{\mu})
-Z^{0}_{\nu}(W^{+}_{\mu}\partial_{\nu}W^{-}_{\mu}-W^{-}_{\mu}\partial_{\nu}W^{+}_{\mu})
+Z^{0}_{\mu}(W^{+}_{\nu}\partial_{\nu}W^{-}_{\mu}-W^{-}_{\nu}\partial_{\nu}W^{+}_{\mu}))
-igs_{w}(\partial_{\nu}A_{\mu}(W^{+}_{\mu}W^{-}_{\nu}-W^{+}_{\nu}W^{-}_{\mu})
-A_{\nu}(W^{+}_{\mu}\partial_{\nu}W^{-}_{\mu}-W^{-}_{\mu}\partial_{\nu}W^{+}_{\mu})
+A_{\mu}(W^{+}_{\nu}\partial_{\nu}W^{-}_{\mu}-W^{-}_{\nu}\partial_{\nu}W^{+}_{\mu}))
-\frac{1}{2}g^{2}W^{+}_{\mu}W^{-}_{\mu}W^{+}_{\nu}W^{-}_{\nu}+\frac{1}{2}g^{2}
W^{+}_{\mu}W^{-}_{\nu}W^{+}_{\mu}W^{-}_{\nu}
+g^2c^{2}_{w}(Z^{0}_{\mu}W^{+}_{\mu}Z^{0}_{\nu}W^{-}_{\nu}-Z^{0}_{\mu}Z^{0}_{\mu}W^{+}_{\nu}
W^{-}_{\nu})
+g^2s^{2}_{w}(A_{\mu}W^{+}_{\mu}A_{\nu}W^{-}_{\nu}-A_{\mu}A_{\mu}W^{+}_{\nu}
W^{-}_{\nu})
+g^{2}s_{w}c_{w}(A_{\mu}Z^{0}_{\nu}(W^{+}_{\mu}W^{-}_{\nu}-W^{+}_{\nu}W^{-}_{\mu})
-2A_{\mu}Z^{0}_{\mu}W^{+}_{\nu}W^{-}_{\nu})
-\frac{1}{2}\partial_{\mu}H\partial_{\mu}H-2M^2\alpha_{h}H^{2}
-\partial_{\mu}\phi^{+}\partial_{\mu}\phi^{-}
-\frac{1}{2}\partial_{\mu}\phi^{0}\partial_{\mu}\phi^{0}
-\beta_{h}\left(\frac{2M^{2}}{g^{2}}+\frac{2M}{g}H+\frac{1}{2}(H^{2}+\phi^{0}\phi^{0}+2\phi^{+}\phi^{-
})\right)
  +\frac{2M^{4}}{g^{2}}\alpha_{h}
  -g\alpha_h
M\left(H^3+H\phi^{0}\phi^{0}+2H\phi^{+}\phi^{-}\right)
-\frac{1}{8}g^{2}\alpha_{h}
\left(H^4+(\phi^{0})^{4}+4(\phi^{+}\phi^{-})^{2}
+4(\phi^{0})^{2}\phi^{+}\phi^{-}
+4H^{2}\phi^{+}\phi^{-}+2(\phi^{0})^{2}H^{2}\right)
-gMW^{+}_{\mu}W^{-}_{\mu}H-\frac{1}{2}g\frac{M}{c^{2}_{w}}Z^{0}_{\mu}Z^{0}_{\mu}H
-\frac{1}{2}ig\left(W^{+}_{\mu}(\phi^{0}\partial_{\mu}\phi^{-}
-\phi^{-}\partial_{\mu}\phi^{0})
-W^{-}_{\mu}(\phi^{0}\partial_{\mu}\phi^{+}
-\phi^{+}\partial_{\mu}\phi^{0})\right)
+\frac{1}{2}g\left(W^{+}_{\mu}(H\partial_{\mu}\phi^{-}
-\phi^{-}\partial_{\mu}H)
 +W^{-}_{\mu}(H\partial_{\mu}\phi^{+}-\phi^{+}\partial_{\mu}H)\right)
+\frac{1}{2}g\frac{1}{c_{w}}(Z^{0}_{\mu}(H\partial_{\mu}\phi^{0}-\phi^{0}\partial_{\mu}H)
+M\,(\frac{1}{c_{w}}Z^{0}_{\mu}\partial_{\mu}\phi^{0}+W^{+}_{\mu}
\partial_{\mu}\phi^{-}+W^{-}_{\mu}
\partial_{\mu}\phi^{+})
-ig\frac{s^{2}_{w}}{c_{w}}MZ^{0}_{\mu}(W^{+}_{\mu}\phi^{-}-W^{-}_{\mu}\phi^{+})
   +igs_{w}MA_{\mu}(W^{+}_{\mu}\phi^{-}-W^{-}_{\mu}\phi^{+})
-ig\frac{1-2c^{2}_{w}}{2c_{w}}Z^{0}_{\mu}(\phi^{+}\partial_{\mu}\phi^{-}
-\phi^{-}\partial_{\mu}\phi^{+})
+igs_{w}A_{\mu}(\phi^{+}\partial_{\mu}\phi^{-}-\phi^{-}\partial_{\mu}\phi^{+})
-\frac{1}{4}g^{2}W^{+}_{\mu}W^{-}_{\mu}
\left(H^{2}+(\phi^{0})^{2}+2\phi^{+}\phi^{-}\right) -\frac{1}{8}
g^{2}\frac{1}{c^{2}_{w}}Z^{0}_{\mu}Z^{0}_{\mu}
\left(H^{2}+(\phi^{0})^{2}+2(2s^{2}_{w}-1)^{2}\phi^{+}\phi^{-}\right)
-\frac{1}{2}g^{2}\frac{s^{2}_{w}}{c_{w}}Z^{0}_{\mu}\phi^{0}(W^{+}_{\mu}\phi^{-}+W^{-}_{\mu}\phi^{+})
-\frac{1}{2}ig^{2}\frac{s^{2}_{w}}{c_{w}}Z^{0}_{\mu}H(W^{+}_{\mu}\phi^{-}-W^{-}_{\mu}\phi^{+})
+\frac{1}{2}g^{2}s_{w}A_{\mu}\phi^{0}(W^{+}_{\mu}\phi^{-}+W^{-}_{\mu}\phi^{+})
+\frac{1}{2}ig^{2}s_{w}A_{\mu}H(W^{+}_{\mu}\phi^{-}-W^{-}_{\mu}\phi^{+})
-g^{2}\frac{s_{w}}{c_{w}}(2c^{2}_{w}-1)Z^{0}_{\mu}A_{\mu}\phi^{+}\phi^{-}
-g^{2}s^{2}_{w}A_{\mu}A_{\mu}\phi^{+}\phi^{-}
+\frac 12 i
g_s\,\lambda_{ij}^a(\bar{q}^{\sigma}_{i}\gamma^{\mu}q^{\sigma}_{j})g^{a}_{\mu}
-\bar{e}^{\lambda}(\gamma\partial+m^{\lambda}_{e})e^{\lambda}
-\bar{\nu}^{\lambda}(\gamma\partial+m^{\lambda}_{\nu})\nu^{\lambda}
-\bar{u}^{\lambda}_{j}(\gamma\partial+m^{\lambda}_{u})u^{\lambda}_{j}
-\bar{d}^{\lambda}_{j}(\gamma\partial+m^{\lambda}_{d})d^{\lambda}_{j}
+igs_{w}A_{\mu}\left(-(\bar{e}^{\lambda}\gamma^{\mu}
e^{\lambda})+\frac{2}{3}(\bar{u}^{\lambda}_{j}\gamma^{\mu}
u^{\lambda}_{j})-\frac{1}{3}(\bar{d}^{\lambda}_{j}\gamma^{\mu}
d^{\lambda}_{j})\right)
   +\frac{ig}{4c_{w}}Z^{0}_{\mu}
\{(\bar{\nu}^{\lambda}\gamma^{\mu}(1+\gamma^{5})\nu^{\lambda})+
(\bar{e}^{\lambda}\gamma^{\mu}(4s^{2}_{w}-1-\gamma^{5})e^{\lambda})
     +(\bar{d}^{\lambda}_{j}\gamma^{\mu}(\frac{4}{3}s^{2}_{w}-1-\gamma^{5})d^{\lambda}_{j})+
(\bar{u}^{\lambda}_{j}\gamma^{\mu}(1-\frac{8}{3}s^{2}_{w}+\gamma^{5})u^{\lambda}_{j})
\}
+\frac{ig}{2\sqrt{2}}W^{+}_{\mu}\left((\bar{\nu}^{\lambda}\gamma^{\mu}(1+\gamma^{5})\cx_{\lambda\kappa}e^{\kappa})
+(\bar{u}^{\lambda}_{j}\gamma^{\mu}(1+\gamma^{5})C_{\lambda\kappa}d^{\kappa}_{j})\right)
+\frac{ig}{2\sqrt{2}}W^{-}_{\mu}\left((\bar{e}^{\kappa}\cx^{\dagger}_{\kappa\lambda}\gamma^{\mu}(1+\gamma^{5})\nu^{\lambda})
+(\bar{d}^{\kappa}_{j}C^{\dagger}_{\kappa\lambda}\gamma^{\mu}(1+\gamma^{5})u^{\lambda}_{j})\right)
   +\frac{ig}{2M\sqrt{2}}\phi^{+}
\left(-m^{\kappa}_{e}(\bar{\nu}^{\lambda}\cx_{\lambda\kappa}(1-\gamma^{5})e^{\kappa})
+m^{\lambda}_{\nu}(\bar{\nu}^{\lambda}\cx_{\lambda\kappa}(1+\gamma^{5})e^{\kappa}\right)
   +\frac{ig}{2M\sqrt{2}}\phi^{-}
\left(m^{\lambda}_{e}(\bar{e}^{\lambda}\cx^{\dagger}_{\lambda\kappa}(1+\gamma^{5})\nu^{\kappa})
-m^{\kappa}_{\nu}(\bar{e}^{\lambda}\cx^{\dagger}_{\lambda\kappa}(1-\gamma^{5})\nu^{\kappa}\right)
-\frac{g}{2}\frac{m^{\lambda}_{\nu}}{M}H(\bar{\nu}^{\lambda}\nu^{\lambda})
-\frac{g}{2}\frac{m^{\lambda}_{e}}{M}H(\bar{e}^{\lambda}e^{\lambda})
+\frac{ig}{2}\frac{m^{\lambda}_{\nu}}{M}\phi^{0}(\bar{\nu}^{\lambda}\gamma^{5}\nu^{\lambda})
-\frac{ig}{2}\frac{m^{\lambda}_{e}}{M}\phi^{0}(\bar{e}^{\lambda}\gamma^{5}e^{\lambda})
-\frac 14\,\bar
\nu_\lambda\,M^R_{\lambda\kappa}\,(1-\gamma_5)\hat\nu_\kappa -\frac
14\,\overline{\bar
\nu_\lambda\,M^R_{\lambda\kappa}\,(1-\gamma_5)\hat\nu_\kappa}
+\frac{ig}{2M\sqrt{2}}\phi^{+}
\left(-m^{\kappa}_{d}(\bar{u}^{\lambda}_{j}C_{\lambda\kappa}(1-\gamma^{5})d^{\kappa}_{j})
+m^{\lambda}_{u}(\bar{u}^{\lambda}_{j}C_{\lambda\kappa}(1+\gamma^{5})d^{\kappa}_{j}\right)
   +\frac{ig}{2M\sqrt{2}}\phi^{-}
\left(m^{\lambda}_{d}(\bar{d}^{\lambda}_{j}C^{\dagger}_{\lambda\kappa}(1+\gamma^{5})u^{\kappa}_{j})
-m^{\kappa}_{u}(\bar{d}^{\lambda}_{j}C^{\dagger}_{\lambda\kappa}(1-\gamma^{5})u^{\kappa}_{j}\right)
-\frac{g}{2}\frac{m^{\lambda}_{u}}{M}H(\bar{u}^{\lambda}_{j}u^{\lambda}_{j})
-\frac{g}{2}\frac{m^{\lambda}_{d}}{M}H(\bar{d}^{\lambda}_{j}d^{\lambda}_{j})
+\frac{ig}{2}\frac{m^{\lambda}_{u}}{M}\phi^{0}(\bar{u}^{\lambda}_{j}\gamma^{5}u^{\lambda}_{j})
-\frac{ig}{2}\frac{m^{\lambda}_{d}}{M}\phi^{0}(\bar{d}^{\lambda}_{j}\gamma^{5}d^{\lambda}_{j})
\end{math}
\end{center}
\end{small}

Here the notation is as in \cite{VDiag}, as follows.
\begin{itemize}
\item Gauge bosons: $A_\mu, W_\mu^{\pm}, Z_\mu^0, g^{a}_{\mu}$
\item Quarks: $u^{\kappa}_{j}, d^{\kappa}_{j}$, collective : $ q^{\sigma}_{j}$
\item Leptons: $e^{\lambda}, \nu^{\lambda}$
\item Higgs fields: $H, \phi^{0}, \phi^{+}, \phi^{-}$
\item Ghosts: $G^a, X^{0}, X^{+}, X^{-}, Y$,
\item Masses: $m^{\lambda}_{d}, m^{\lambda}_{u}, m^{\lambda}_{e}, m_h,
M$ (the latter is the mass of the $W$)
\item Coupling constants $g=\sqrt{4\pi \alpha}$ (fine structure), $g_s=$
strong, $\alpha_h=\frac{m_h^2}{4M^2}$
\item Tadpole Constant $\beta_h$
\item Cosine and sine of the weak mixing angle $c_w, s_w$
\index{mixing angles}
\item Cabibbo--Kobayashi--Maskawa mixing matrix: $C_{\lambda\kappa}$
\index{Cabibbo-Kobayashi-Maskawa matrix}
\item Structure constants of ${\rm SU}(3)$: $f^{abc}$
\item The Gauge is the Feynman gauge.
\end{itemize}

\begin{rem}\label{massC}
Notice that, for simplicity, we use for leptons the same convention
usually adopted for quarks, namely to have the up particles in
diagonal form (in this case the neutrinos) and the mixing matrix for
the down particles (here the charged leptons). This is different
from the convention usually adopted in neutrino physics (\cf \eg
\cite{MoPa} \S 11.3), but it is convenient here, in order to write
the Majorana mass matrix in a simpler form.
\end{rem}

\smallskip

Our goal is to compare this Lagrangian with the one we get from the
spectral action, when dealing with flat space and Euclidean
signature. All the results immediately extend to curved space since
our formalism is fully covariant.

\subsection{The asymptotic formula for the spectral action}\hfill
\medskip \label{sectchangevar}

The change of variables from the standard model to the spectral
model is summarized in Table \ref{smtospec}.

\medskip
\begin{table}
\begin{center}

  \medskip

  \begin{tabular}{|c||c|c||c|}
  \hline
  Standard Model & notation & notation & Spectral Action\\
  \hline
  & & &\\
 Higgs Boson & $\varphi=(\frac{2M}{g}+ H-i\phi^{0},-i\sqrt{2}\phi^{+})$ & $\higgs=\frac{1}{\sqrt 2}\frac{\sqrt{a}}
 {g} (1+\psi)$& Inner metric$^{(0,1)}$ \\
 &&&\\
  \hline
  & & &\\
  Gauge bosons \index{gauge!bosons} & $A_\mu,Z^0_\mu, W^\pm_\mu,
g_\mu^a$ & $(B,W,V)$ & Inner metric$^{(1,0)}$ \\
  &&&\\
  \hline
  & & &\\
   Fermion masses&   $m_u,m_\nu$ &   $\mass_{\upup}=
   \delta_{\upup},\mass_{\nunu}=\delta_{\nunu}$  &  Dirac$^{(0,1)}$ in $\uparrow$\\
    $u,\nu$&&&\\
  \hline
  & && \\
  CKM matrix & $C_\lambda^\kappa, m_d$ & $\mass_{\dodo}=C\,\delta_{3,\,
  \downarrow}\,C^\dagger$ &  Dirac$^{(0,1)}$ in $\dodo$\\
  Masses down &&&\\
  \hline
   & && \\
  Lepton mixing & $\cx_{\lambda\kappa}, m_e$
   & $\mass_{\elel}=\cx\,\delta_{\elel}\,\cx^\dagger$ &  Dirac$^{(0,1)}$ in $\elel$\\
  Masses leptons $e$ &&&\\
  \hline
  & && \\
  Majorana & $M_R$ & $\mass_R$ & Dirac$^{(0,1)}$ on \\
   mass matrix&&&  
$E_R\oplus J_F E_R$ \\
  \hline
  & && \\
  Gauge couplings \index{coupling!gauge fields} & $g_1=g\tan (\theta_w),g_2=g,g_3=g_s$ & $g_{3}^2= g_{2}^2 = \frac{5}{ 3} \,
g_{1}^2$ & Fixed at \\
   &&&unification\\
  \hline
  & && \\
  Higgs scattering \index{Higgs} \index{scattering} & $\frac
18\,g^2\,\alpha_h, \alpha_h=\frac{m_h^2}{4M^2} $ &
  $\lambda_0  =  g^2\,\frac{b}{ a^2}$ & Fixed at  \\
   parameter&&&unification\\
  \hline
  & && \\
  Tadpole constant \index{tadpole!constant} & $\beta_h,
(-\alpha_{h}\,M^2\,+\frac{\beta_h}{2})\,|\varphi|^2$ &
  $\mu_0^2=2\frac{f_2\Lambda^2}{f_0}  -\frac ea$ & $- \mu_0^2\, |\higgs|^2$\\
   &&&\\
  \hline
  & && \\
  Graviton & $g_{\mu\nu}$ & $\dirac_M$  & Dirac$^{(1,0)}$ \\
   &&&\\
  \hline
\end{tabular}
\bigskip
\bigskip
\caption{Conversion from Spectral Action to Standard
Model\label{smtospec}}
\end{center}
\end{table}

We first perform a trivial rescaling of the Higgs field $\varphi$ so
that kinetic terms are normalized. To normalize the Higgs fields
kinetic energy we have to rescale $\varphi$ to:
\begin{equation}\label{rescalehiggs}
\higgs   =\frac{\sqrt{a\,f_0}}{ \pi}  \, \varphi \, ,
\end{equation}
so that the kinetic term becomes
$$
\int\,\frac 12|D_\mu \higgs|^2\,\sqrt g \,d^4x
$$
The normalization of the kinetic terms, as in Lemma
\ref{namelesslemma} below, imposes a relation between the coupling
constants $g_1$, $g_2$, $g_3$ and the coefficient $f_0$, of the form
\begin{equation}\label{coeffymterm}
  \frac{g_{3}^2 \, f_0 }{ 2\pi^2} =\frac 14 , \ \ \ \ \ \
  g_3^2=g_2^2 = \frac{5}{ 3} \, g_1^2 \, .
\end{equation}

The bosonic action \eqref{bossm} then takes the form
\begin{equation}\label{rescaledbosaction}
\begin{array}{rl}
  S = & \int  \ \biggl( \frac{1}{
2\kappa_0^2} \, R + \alpha_0 \, C_{\mu \nu \rho \sigma} \, C^{\mu
\nu \rho \sigma} + \ \ \gamma_0 + \tau_0 \, R^*
R^*  \\[3mm]
    + &  \frac{1}{ 4} \, G_{\mu
\nu}^i \, G^{\mu \nu i} + \frac{1}{ 4} \, F_{\mu \nu}^{ \alpha} \,
F^{\mu \nu  \alpha}+\ \frac{1}{ 4} \, B_{\mu \nu}\, B^{\mu
\nu} \\[2mm]
    +&   \,\frac 12\vert D_{\mu} \, \higgs\vert^2 - \mu_0^2 |\higgs|^2 - \xi_0
\, R \,\vert \higgs \vert^2 + \lambda_0 |\higgs|^4 \biggl)\sqrt g\,
d^4 x,
\end{array}
\end{equation}
where
\begin{equation}\label{list}
\begin{array}{rl}
\frac{1}{\kappa_0^2} =& \ \frac{96\,f_2\,\Lambda^2
-f_0\,c}{ 12\,\pi^2} \\[2mm]
 \mu_0^2 =& \ 2\,\frac{f_2\,\Lambda^2}{ f_0}-\,\frac{e}{a} \\[2mm]
  \alpha_0 =& -\frac{3\,f_0 }{ 10\,\pi^2} \\[2mm]
   \tau_0  =& \frac{11\,f_0 }{ 60\,\pi^2} \\[2mm]
  \gamma_0 =& \ \frac{1}{\pi^2}(48\,f_4\,\Lambda^4-f_2\,\Lambda^2\,c+\frac{
f_0}{4}\,d) \\[2mm]
  \lambda_0  =& \frac{\pi^2 }{2\,
f_0}\frac{b}{ a^2}\\[2mm]
  \xi_0 =& \frac{1}{ 12}
\end{array}
\end{equation}

Notice that the matrices $\mass_{\upup}$, $\mass_{\dodo}$,
$\mass_{\nunu}$ and $\mass_{\elel}$ are only relevant up to an
overall scale. Indeed they only enter in the coupling of the Higgs
with fermions and because of the rescaling \eqref{rescalehiggs} only
by the terms
\begin{equation}\label{rescaledyukawamasses}
k_x=\,\frac{\pi}{\sqrt{a\,f_0}}\,\mass_x , \ \ \ \
x\in\{(\uparrow\downarrow, j)\}
\end{equation}
which are dimensionless matrices by construction. In fact, by
\eqref{momentslabels}
$$
a=\,\Tr(\mass_{\nunu}^*\mass_{\nunu}+\mass_{\elel}^*\mass_{\elel}+3(\mass_{\upup}^*\mass_{\upup}+\mass_{\dodo}^*\mass_{\dodo}))
$$
has the physical dimension of a $(mass)^2$.

Using \eqref{coeffymterm} to replace $\frac{\sqrt{a\,f_0}}{\pi}$ by
$\frac{1}{\sqrt 2}\frac{\sqrt{a}}{g}$, the change of notations for
the Higgs fields is
\begin{equation}\label{higgstranslate}
\higgs=\frac{1}{\sqrt 2}\frac{\sqrt{a}}
 {g} (1+\psi)=(\frac{2M}{g}+
H-i\phi^{0},-i\sqrt{2}\phi^{+})\,,
\end{equation}

\medskip
\subsection{The mass relation}\label{sectmrel}
\hfill\medskip

The relation between the mass matrices comes from the equality of
the Yukawa coupling terms $\cL_{Hf}$. For the standard model these
terms are given by Lemma \ref{lemmalgfandlhf} below. For the
spectral action they are given by $\gamma_5\,\bM(\varphi)$ with the
notations of \eqref{diracdec} and \eqref{termcomdef}.

After Wick rotation to Euclidean and the chiral transformation $U=
e^{i\frac{\pi}{4}\gamma_5}\otimes 1$ they are the same (\cf Lemma
\ref{unitarylemma} below) provided the following equalities hold
\begin{eqnarray}
  (k_{\upup})_{\sigma\kappa} &=&
\frac{g}{2M}\,m_u^\sigma\,\delta_\sigma^\kappa
\label{massrelations}\\
  (k_{\dodo})_{\sigma\kappa} &=& \frac{g}{2M}\,m_d^\mu\,C_{\sigma\mu}\delta_\mu^\rho C^\dagger_{\rho\kappa}\nonumber\\
   (k_{\nunu})_{\sigma\kappa} &=& \frac{g}{2M}\,m_\nu^\sigma\,\delta_\sigma^\kappa\nonumber\\
  (k_{\elel})_{\sigma\kappa} &=& \frac{g}{2M}\,m_e^\mu\,\cx_{\sigma\mu}\delta_\mu^\rho \cx^\dagger_{\rho\kappa}\nonumber
\end{eqnarray}
Here the symbol $\delta_i^j$ is the Kronecker delta (not to be
confused with the previous notation $\delta_{\uparrow\downarrow}$).

\begin{lem}\label{lemmassrel}
The mass matrices of \eqref{massrelations} satisfy the constraint
\begin{equation}\label{massrelation2}
\sum_\sigma\,(m^\sigma_\nu)^2+(m^\sigma_e)^2+3\,(m^\sigma_u)^2+3\,
(m^\sigma_d)^2=\,8\,M^2 .
\end{equation}
\end{lem}

\proof It might seem at first sight that one can simply use
\eqref{massrelations} to define the matrices $k_x$ but this
overlooks the fact that \eqref{rescaledyukawamasses} implies the
constraint
\begin{equation}\label{massrelation1}
\Tr(k_{\nunu}^*k_{\nunu}+k_{\elel}^*k_{\elel}+3(k_{\upup}^*k_{\upup}+k_{\dodo}^*k_{\dodo}))=\,2\,
g^2\,,
\end{equation}
where we use \eqref{coeffymterm} to replace $\frac{\pi^2}{f_0}$ by $
2\, g^2$. Using \eqref{massrelations}, we then obtain the constraint
\eqref{massrelation2}, where summation is performed with respect to
the flavor index $\sigma$. Notice that $g^2$ appeared in the same
way on both sides and drops out of the equation.
\endproof

We discuss in \S \ref{sectmassrel} below the physical interpretation
of the imposition of this constraint at unification scale.

\bigskip

\subsection{The coupling of fermions}\hfill \medskip
\label{sectfermions}

Let us isolate the Yukawa coupling part of the standard model
Lagrangian, ignoring first the right handed neutrinos (\ie using the
minimal standard model as in \cite{VDiag}). We consider the
additional terms later in Lemma \ref{lemaddedterms}. In the minimal
case, one has
\begin{equation}
\cL_{Hf}=\,
\end{equation}
\begin{center}
\begin{math}
-\bar{e}^{\lambda}\,m^{\lambda}_{e}\,e^{\lambda}
-\bar{u}^{\lambda}_{j}\,m^{\lambda}_{u}\,u^{\lambda}_{j}
-\bar{d}^{\lambda}_{j}\,m^{\lambda}_{d}\,d^{\lambda}_{j}
+\frac{ig}{2\sqrt{2}}\frac{m^{\lambda}_{e}}{M}
\left(-\phi^{+}(\bar{\nu}^{\lambda}(1-\gamma^{5})e^{\lambda})
+\phi^{-}(\bar{e}^{\lambda}(1+\gamma^{5})\nu^{\lambda})\right)
-\frac{g}{2}\frac{m^{\lambda}_{e}}{M}\left(H(\bar{e}^{\lambda}e^{\lambda})
+i\phi^{0}(\bar{e}^{\lambda}\gamma^{5}e^{\lambda})\right)
+\frac{ig}{2M\sqrt{2}}\phi^{+}
\left(-m^{\kappa}_{d}(\bar{u}^{\lambda}_{j}C_{\lambda\kappa}(1-\gamma^{5})d^{\kappa}_{j})
+m^{\lambda}_{u}(\bar{u}^{\lambda}_{j}C_{\lambda\kappa}(1+\gamma^{5})d^{\kappa}_{j}\right)
   +\frac{ig}{2M\sqrt{2}}\phi^{-}
\left(m^{\lambda}_{d}(\bar{d}^{\lambda}_{j}C^{\dagger}_{\lambda\kappa}(1+\gamma^{5})u^{\kappa}_{j})
-m^{\kappa}_{u}(\bar{d}^{\lambda}_{j}C^{\dagger}_{\lambda\kappa}(1-\gamma^{5})u^{\kappa}_{j}\right)
-\frac{g}{2}\frac{m^{\lambda}_{u}}{M}H(\bar{u}^{\lambda}_{j}u^{\lambda}_{j})
-\frac{g}{2}\frac{m^{\lambda}_{d}}{M}H(\bar{d}^{\lambda}_{j}d^{\lambda}_{j})
+\frac{ig}{2}\frac{m^{\lambda}_{u}}{M}\phi^{0}(\bar{u}^{\lambda}_{j}\gamma^{5}u^{\lambda}_{j})
-\frac{ig}{2}\frac{m^{\lambda}_{d}}{M}\phi^{0}(\bar{d}^{\lambda}_{j}\gamma^{5}d^{\lambda}_{j})
\end{math}
\end{center}

The matrix $C_{\lambda\kappa}$ is the mixing matrix. It does enter
in the Lagrangian elsewhere but only in the two gauge coupling terms
  where the down and up fermions are
involved together and which are part of the expression

\begin{equation}\label{termslgf}
\cL_{gf}=\,
\end{equation}
\begin{center}
\begin{math}
\frac 12 i
g_s\,\lambda^{ij}_a(\bar{q}^{\sigma}_{i}\gamma^{\mu}q^{\sigma}_{j})g^{a}_{\mu}
-\bar{e}^{\lambda}(\gamma\partial)e^{\lambda}
-\bar{\nu}^{\lambda}\gamma\partial\nu^{\lambda}
-\bar{u}^{\lambda}_{j}(\gamma\partial)u^{\lambda}_{j}
-\bar{d}^{\lambda}_{j}(\gamma\partial)d^{\lambda}_{j}
+igs_{w}A_{\mu}\left(-(\bar{e}^{\lambda}\gamma^{\mu}
e^{\lambda})+\frac{2}{3}(\bar{u}^{\lambda}_{j}\gamma^{\mu}
u^{\lambda}_{j})-\frac{1}{3}(\bar{d}^{\lambda}_{j}\gamma^{\mu}
d^{\lambda}_{j})\right)
   +\frac{ig}{4c_{w}}Z^{0}_{\mu}
\{(\bar{\nu}^{\lambda}\gamma^{\mu}(1+\gamma^{5})\nu^{\lambda})+
(\bar{e}^{\lambda}\gamma^{\mu}(4s^{2}_{w}-1-\gamma^{5})e^{\lambda})
     +
(\bar{d}^{\lambda}_{j}\gamma^{\mu}(\frac{4}{3}s^{2}_{w}-1-\gamma^{5})d^{\lambda}_{j})+
(\bar{u}^{\lambda}_{j}\gamma^{\mu}(1-\frac{8}{3}s^{2}_{w}+\gamma^{5})u^{\lambda}_{j})\}
+\frac{ig}{2\sqrt{2}}W^{+}_{\mu}\left((\bar{\nu}^{\lambda}\gamma^{\mu}(1+\gamma^{5})e^{\lambda})
+(\bar{u}^{\lambda}_{j}\gamma^{\mu}(1+\gamma^{5})C_{\lambda\kappa}d^{\kappa}_{j})\right)
+\frac{ig}{2\sqrt{2}}W^{-}_{\mu}\left((\bar{e}^{\lambda}\gamma^{\mu}(1+\gamma^{5})\nu^{\lambda})
+(\bar{d}^{\kappa}_{j}C^{\dagger}_{\kappa\lambda}\gamma^{\mu}(1+\gamma^{5})u^{\lambda}_{j})\right)
\end{math}
\end{center}
  Since the matrix $C_{\lambda\kappa}$ is
unitary the quadratic expressions in $d^{\lambda}_{j}$ are unchanged
by the change of variables given by
\begin{equation}\label{dlambdaj}
d_{\lambda j}=\,C_{\lambda\kappa}\,d^{\kappa}_{j}\,,\quad \bar
d_{\lambda j}=\,\bar{C_{\lambda\kappa}}\,\bar
d^{\kappa}_{j}=\,C^\dagger_{\kappa\lambda}\,\bar d^{\kappa}_{j}
\end{equation}
and in this way one can eliminate $C_{\lambda\kappa}$ in $\cL_{gf}$.

Once written in terms of the new variables, the term $\cL_{gf}$
reflects the kinetic terms of the fermions and their couplings to
the various gauge fields. The latter is simple for the color fields,
where it is of the form
$$
\frac 12 i g_s\,\lambda^{ij}_a(\bar{q}^{\sigma}_{i}
\gamma^{\mu}q^{\sigma}_{j})g^{a}_{\mu},
$$
where the $\lambda$ are the Gell'mann matrices
\eqref{gellmatrices}).

It is more complicated for the $(A,W^\pm,Z^0)$. This displays in
particular the complicated hypercharges assigned to the different
fermions, quarks and leptons, which depend upon their chirality. At
the level of electromagnetic charges themselves, the assignment is
visible in the coupling with $A_{\mu}$. There one sees that the
charge of the electron is $-1$, while it is $\frac 23$ for the up
quark and $-\frac 13$ for the down quark.

\begin{lem}\label{lemmalgfandlhf}
Let the fermions $f$ be obtained from the quarks and leptons by
performing the change of basis \eqref{dlambdaj} on the down quarks.
Then the following holds.
\begin{enumerate}
  \item The terms $\cL_{gf}$ are of the
form
 \begin{eqnarray}\label{fermionminimal}
  \cL_{gf} &=& -\sum_f[\overline f_L\gamma^{\mu}(\partial_{\mu} -i g
\frac{{\sigma_a}}{2}W_{\mu
  a}-ig^{\prime} \frac{{Y_L}}{2}B_{\mu}-ig^{\prime\prime}\frac{\lambda_b}{2}V_{\mu
  b})f_L \\
     &+& \overline f_R\gamma^{\mu}(\partial_{\mu}-ig^{\prime}
\frac{{Y_R}}{2}B_{\mu}-ig^{\prime\prime}\lambda_bV_{\mu
b})f_R]\nonumber
 \end{eqnarray}
and similar terms for the leptons, with $W_\mu^+=\frac{W_{\mu
1}-iW_{\mu 2}}{\sqrt 2}$, $W_\mu^-=\frac{W_{\mu 1}+iW_{\mu 2}}{\sqrt
2}$ and
\begin{equation}\label{gaugeform1}
g'=g\, \tan (\theta_w)\,,\quad g''= g_s \,,\quad
B_\mu=\,c_w\,A_\mu-s_w\,Z^0_\mu\,\,,\quad W_{\mu 3}=\,s_w\,A_\mu
+\,c_w\,Z_\mu^0
\end{equation}
  \item The terms $\cL_{Hf}$ are given with the notation
  \eqref{termcom0} by
  \begin{equation}\label{higgsform1}
\cL_{Hf}=-\bar f\;T(0,K_e,\varphi)\;f -\bar f\;T(K_u,K_d,\varphi)\;f
\end{equation}
where
\begin{equation}\label{phi12change}
\varphi_1 =\frac{2M}{g}+ H-i\phi^{0}, \ \ \ \
\varphi_2=-i\sqrt{2}\phi^{+}
\end{equation}
and
\begin{equation}\label{Keud}
(K_e)_{\mu\rho} =\frac{g}{2M} m_e^\mu \delta^\rho_\mu \ \ \ \
(K_u)_{\mu\rho} =\frac{g}{2M} m_u^\mu \delta^\rho_\mu \ \ \ \
(K_d)_{\mu\rho} =\frac{g}{2M} m_d^\lambda C_{\mu\lambda}
\delta^\kappa_\lambda C^\dag_{\kappa\rho} .
\end{equation}
\end{enumerate}
\end{lem}

\proof  1) In Minkowski space a quark $q$ is represented by a column
vector and one has the relation
\begin{equation}\label{relationqqbar}
\bar{q}= q^*\,\gamma_0
\end{equation}
between $q$ and $\bar{q}$. Thus, $q$ and $\bar{q}$ have opposite
chirality.

Since the $\gamma^{\mu}$ switch the chirality to its opposite and
all the terms in \eqref{termslgf} involve the $\gamma^{\mu}$, they
can be separated as a sum of terms only involving $f_L$, $\bar f_{
L}$ and terms only involving $f_R$, $\bar f_{R}$. The neutrinos
$\nu^{\lambda}$ only appear as left handed, \ie as the combination
$(1+\gamma^{5})\nu^{\lambda}$.

The last two lines of \eqref{termslgf} correspond to the terms in
$ig \frac{{\sigma^a}}{2}W_{\mu a}$ for the off diagonal Pauli
matrices $\sigma_1$, $\sigma_2$. The first line of \eqref{termslgf}
corresponds to the gluons and the kinetic terms. The terms involving
the gluons $g^{a}_{\mu}$ in \eqref{termslgf} give the strong
coupling constant $g''= g_s$. The second and third lines of
\eqref{termslgf} use the electromagnetic field $A_{\mu}$ related to
$B_\mu$ by
\begin{equation}\label{weinberg0}
g\,s_w\,(A_\mu-\tan (\theta_w)\,Z^0_\mu)=\,g'\,B_\mu .
\end{equation}
This gives (note the sign $-\times-=+$ in \eqref{fermionminimal})
the terms
\begin{equation}\label{weinberg1}
ig^{\prime}
\frac{{Y_R}}{2}B_{\mu}=ig\,s_w\,A_\mu\,Q_{em}-\,ig\,\frac{s_w^2}{c_w}\,Z^0_\mu\,Q_{em}
\end{equation}
for the right handed part. On the left handed sector one has
$$
Q_{em}=\frac{{Y_L}}{2}+\frac{{\sigma_3}}{2} .
$$
The diagonal terms for the left-handed part
$$
ig \frac{{\sigma_3}}{2}W_{\mu 3}+ig^{\prime} \frac{{Y_L}}{2}B_{\mu}
$$
are then of the form
$$
ig \frac{{\sigma_3}}{2}W_{\mu 3}+i g\,s_w\,(A_\mu-\tan
(\theta_w)\,Z^0_\mu)(Q_{em}-\frac{{\sigma_1}}{2})=
$$
$$
ig\,s_w\,A_\mu\,Q_{em}-\,ig\,\frac{s_w^2}{c_w}\,Z^0_\mu\,Q_{em}+ (ig
W_{\mu 3}-i g\,s_w\,(A_\mu-\tan
(\theta_w)\,Z^0_\mu))\frac{{\sigma_3}}{2} .
$$
The relation
\begin{equation}\label{weinberg1.5}
(ig W_{\mu 3}-i g\,s_w\,(A_\mu-\tan
(\theta_w)\,Z^0_\mu))=\frac{ig}{c_w}\,Z^0_\mu
\end{equation}
then determines $W_{\mu 3}$ as a function of $A_\mu$ and $Z^0_\mu$.
It gives
$$
 W_{\mu 3}=\,s_w\,(A_\mu-\tan (\theta_w)\,Z^0_\mu)+\frac{1}{c_w}\,Z^0_\mu
$$
\ie
\begin{equation}\label{weinberg2}
 W_{\mu 3}=\,s_w\,A_\mu +\,c_w\,Z_\mu^0
\end{equation}
The diagonal terms for the left-handed sector can then be written in
the form
\begin{equation}\label{weinberg3}
ig\,s_w\,A_\mu\,Q_{em}-\,ig\,\frac{s_w^2}{c_w}\,Z^0_\mu\,Q_{em}+
\frac{ig}{c_w}\,Z^0_\mu\,\frac{{\sigma_3}}{2} .
\end{equation}
This matches with the factor $\frac{ig}{4c_{w}}$ in \eqref{termslgf}
multiplying $(1+\gamma^5)$. The latter is twice the projection on
the left handed particles. This takes care of one factor of two,
while the other comes from the denominator in
$\frac{{\sigma_3}}{2}$.

The term
$$
\frac{ig}{4c_{w}}Z^{0}_{\mu}
\{(\bar{\nu}^{\lambda}\gamma^{\mu}(1+\gamma^{5})\nu^{\lambda})+
(\bar{e}^{\lambda}\gamma^{\mu}(4s^{2}_{w}-1-\gamma^{5})e^{\lambda})
$$
is fine, since the neutrino has no electromagnetic charge and one
gets the term $-\,ig\,\frac{s_w^2}{c_w}\,Z^0_\mu\,Q_{em}$ for the
electron, while the left handed neutrino has $\sigma_3=1$ and the
left handed electron has $\sigma_3=-1$. The other two terms
$$
\frac{ig}{4c_{w}}Z^{0}_{\mu} \{
(\bar{d}^{\lambda}_{j}\gamma^{\mu}(\frac{4}{3}s^{2}_{w}-1-\gamma^{5})d^{\lambda}_{j})+
(\bar{u}^{\lambda}_{j}\gamma^{\mu}(1-\frac{8}{3}s^{2}_{w}+\gamma^{5})u^{\lambda}_{j})\}
$$
give the right answer, since the electromagnetic charge of the down
is $-\frac 13$ and it has $\sigma_3=-1$, while for the up the
electromagnetic charge is $\frac 23$ and $\sigma_3=1$.

\smallskip
2) We rely on \cite{MoPa} equation (2.14) at the conceptual level,
while we perform the computation in full details. The first thing to
notice is that, by \eqref{relationqqbar}, the $\bar q$ have opposite
chirality. Thus, when we spell out the various terms in terms of
chiral ones, we always get combinations of the form $\bar q_L\,X\,
q_R$ or $\bar q_R\,X\, q_L$. We first look at the lepton sector.
This gives
$$
-\bar{e}^{\lambda}\,m^{\lambda}_{e}\,e^{\lambda}
+\frac{ig}{2\sqrt{2}}\frac{m^{\lambda}_{e}}{M}
\left(-\phi^{+}(\bar{\nu}^{\lambda}(1-\gamma^{5})e^{\lambda})
+\phi^{-}(\bar{e}^{\lambda}(1+\gamma^{5})\nu^{\lambda})\right)
-\frac{g}{2}\frac{m^{\lambda}_{e}}{M}\left(H(\bar{e}^{\lambda}e^{\lambda})
+i\phi^{0}(\bar{e}^{\lambda}\gamma^{5}e^{\lambda})\right) .
$$
The terms in $\bar e\,X\,e$ are of two types. The first gives
$$
-\bar{e}^{\lambda}\,m^{\lambda}_{e}\,(1+\frac{g\,H}{2\,M})\,e^{\lambda}=
-\bar{e}_L^{\lambda}\,m^{\lambda}_{e}\,(1+\frac{g\,H}{2\,M})\,e^{\lambda}_R
-\bar{e}_R^{\lambda}\,m^{\lambda}_{e}\,(1+\frac{g\,H}{2\,M})\,e^{\lambda}_L
$$
The second type gives
$$
-\frac{g}{2}\frac{m^{\lambda}_{e}}{M}
i\phi^{0}(\bar{e}^{\lambda}\gamma^{5}e^{\lambda}) =
\frac{g}{2}\frac{m^{\lambda}_{e}}{M}
i\phi^{0}(\bar{e}_L^{\lambda}e^{\lambda}_R)
-\frac{g}{2}\frac{m^{\lambda}_{e}}{M}
i\phi^{0}(\bar{e}_R^{\lambda}e_L^{\lambda}) .
$$
Thus, they combine together using the complex field
\begin{equation}\label{higgsrecast1}
\psi_1=H-i\phi^{0}
\end{equation}
and give
$$
-\bar{e}_L^{\lambda}\,m^{\lambda}_{e}\,(1+\frac{g\,\psi_1}{2\,M})\,e^{\lambda}_R
-\bar{e}_R^{\lambda}\,m^{\lambda}_{e}\,(1+\frac{g\,\bar\psi_1}{2\,M})\,e^{\lambda}_L
$$
The terms where both $e$ and $\nu$ appear involve only $\nu_L$,
hence only $e_R$. The fields $\phi^{\pm}$ are complex fields that
are complex conjugates of each other. We let
\begin{equation}\label{higgsrecast2}
\psi_2=\,-\,i\,\sqrt{2}\,\phi^{+} .
\end{equation}
The contribution of the terms involving both $e$ and $\nu$ is then
$$
\bar{\nu}_L^{\lambda}\,m^{\lambda}_{e}\,(\frac{g\,\psi_2}{2\,M})\,e^{\lambda}_R
+\,\bar{e}_R^{\lambda}\,m^{\lambda}_{e}\,(\frac{g\,\bar\psi_2}{2\,M})\,\nu^{\lambda}_L
$$
We use the notation \eqref{termcom0}, that is,
$$
T(K_1,K_2,\varphi)=\left[
  \begin{array}{cccc}
0 & 0 & K_1^*\,\varphi_1 & K_1^*\,\varphi_2 \\
0 & 0 &  -K_2^*\,\bar\varphi_2 & K_2^*\,\bar\varphi_1\\
K_1 \bar\varphi_1& -K_2 \,\varphi_2& 0 & 0 \\
K_1 \bar\varphi_2 & K_2 \,\varphi_1& 0 & 0\\
  \end{array}
\right].
$$
We then get that, for the lepton sector, the terms $\cL_{Hf}$ are of
the form
\begin{equation}\label{higgsrecast3}
-\,\bar f\;T(0,K_e,\varphi)\;f \,,\quad
\varphi_1=\psi_1+\frac{2\,M}{g}\,,\quad \varphi_2=\psi_2 ,
\end{equation}
where $K_e$ is the diagonal matrix with diagonal entries the
$\frac{g}{2M}\,m^{\lambda}_{e}$.

\smallskip Let us now look at the quark sector, \ie at the terms
\begin{center}
\begin{math}
-\bar{u}^{\lambda}_{j}\,m^{\lambda}_{u}\,u^{\lambda}_{j}
-\bar{d}^{\lambda}_{j}\,m^{\lambda}_{d}\,d^{\lambda}_{j}
+\frac{ig}{2M\sqrt{2}}\phi^{+}
\left(-m^{\kappa}_{d}(\bar{u}^{\lambda}_{j}C_{\lambda\kappa}(1-\gamma^{5})d^{\kappa}_{j})
+m^{\lambda}_{u}(\bar{u}^{\lambda}_{j}C_{\lambda\kappa}(1+\gamma^{5})d^{\kappa}_{j}\right)
   +\frac{ig}{2M\sqrt{2}}\phi^{-}
\left(m^{\lambda}_{d}(\bar{d}^{\lambda}_{j}C^{\dagger}_{\lambda\kappa}(1+\gamma^{5})u^{\kappa}_{j})
-m^{\kappa}_{u}(\bar{d}^{\lambda}_{j}C^{\dagger}_{\lambda\kappa}(1-\gamma^{5})u^{\kappa}_{j}\right)
-\frac{g}{2}\frac{m^{\lambda}_{u}}{M}H(\bar{u}^{\lambda}_{j}u^{\lambda}_{j})
-\frac{g}{2}\frac{m^{\lambda}_{d}}{M}H(\bar{d}^{\lambda}_{j}d^{\lambda}_{j})
+\frac{ig}{2}\frac{m^{\lambda}_{u}}{M}\phi^{0}(\bar{u}^{\lambda}_{j}\gamma^{5}u^{\lambda}_{j})
-\frac{ig}{2}\frac{m^{\lambda}_{d}}{M}\phi^{0}(\bar{d}^{\lambda}_{j}\gamma^{5}
d^{\lambda}_{j}).
\end{math}
\end{center}
Notice that we have to write it in terms of the $d_{\lambda j}$
given by \eqref{dlambdaj} instead of the $d^{\lambda}_{j}$. The
terms of the form $\bar u \,X\,u$ are
$$
-\bar{u}^{\lambda}_{j}\,m^{\lambda}_{u}\,u^{\lambda}_{j}
-\frac{g}{2}\frac{m^{\lambda}_{u}}{M}H(\bar{u}^{\lambda}_{j}u^{\lambda}_{j})
+\frac{ig}{2}\frac{m^{\lambda}_{u}}{M}\phi^{0}(\bar{u}^{\lambda}_{j}\gamma^{5}
u^{\lambda}_{j}).
$$
They are similar to the terms in $\bar e \,X\,e$ for the leptons but
with an opposite sign in front of $\phi^{0}$. Thus, if we let $K_u$
be the diagonal matrix with diagonal entries the
$\frac{g}{2M}\,m^{\lambda}_{u}$, we get the terms depending on
$\varphi_1$ and $K_u$ in the expression
\begin{equation}\label{higgsrecast4}
-\,\bar f\;T(K_u,K_d,\varphi)\;f ,
\end{equation}
where $K_d$ remains to be determined. There are two other terms
involving the $m^{\lambda}_{u}$, which are directly written in terms
of the $d_{\lambda j}$. They are of the form
$$
\frac{ig}{2M\sqrt{2}}\,\phi^{+}\,
m^{\lambda}_{u}(\bar{u}^{\lambda}_{j}(1+\gamma^{5})d_{\lambda j})
   -\frac{ig}{2M\sqrt{2}}\,\phi^{-}\,
 m^{\kappa}_{u}(\bar{d_{\kappa j}}(1-\gamma^{5})u^{\kappa}_{j}) .
$$
This is the same as
$$
-\bar{d}_{\lambda j
L}\,m^{\lambda}_{u}\,(\frac{g\,\bar\psi_2}{2\,M})\,u^{\lambda}_{j R}
-\,\bar{u}^{\lambda}_{j
R}\,m^{\lambda}_{u}\,(\frac{g\,\psi_2}{2\,M})\,{d}_{\lambda j L} ,
$$
which corresponds to the other terms involving $K_u$ in
\eqref{higgsrecast4}.

\smallskip
The remaining terms are
\begin{equation}\label{higgsrecast5}
-\bar{d}^{\lambda}_{j}\,m^{\lambda}_{d}\,d^{\lambda}_{j}
+\frac{ig\,m^{\kappa}_{d}}{2M\sqrt{2}}(-\phi^{+}\,
(\bar{u}^{\lambda}_{j}C_{\lambda\kappa}(1-\gamma^{5})d^{\kappa}_{j})
   +\phi^{-}\,
(\bar{d}^{\lambda}_{j}C^{\dagger}_{\lambda\kappa}(1+\gamma^{5})u^{\kappa}_{j}))\end{equation}
$$
-\frac{g}{2}\frac{m^{\lambda}_{d}}{M}(H(\bar{d}^{\lambda}_{j}d^{\lambda}_{j})
+\,i\,\phi^{0}(\bar{d}^{\lambda}_{j}\gamma^{5}d^{\lambda}_{j})) .
$$
Except for the transition to the the $d_{\lambda j}$, these terms
are the same as for the lepton sector. Thus, we define the matrix
$K_d$ in such a way that it satisfies
$$
\bar{d}_{\lambda j L}\,K_d^{\lambda\kappa}\,\,{d}_{\kappa j
R}+\bar{d}_{\lambda j R}\,K_d^{\dagger\lambda\kappa}\,\,{d}_{\kappa
j L}=\,\frac{g}{2M}\,
\bar{d}^{\lambda}_{j}\,m^{\lambda}_{d}\,d^{\lambda}_{j} .
$$

We can just take the positive matrix obtained as the conjugate
\begin{equation}\label{higgsrecast6}
(K_d)_{\mu\rho} =\frac{g}{2M} m_d^\lambda C_{\mu\lambda}
\delta^\kappa_\lambda C^\dag_{\kappa\rho}
\end{equation}
as in \eqref{Keud}.

The only terms that remain to be understood are then the cross terms
(with up and down quarks) in \eqref{higgsrecast5}. It might seem at
first that one recognizes the expression for $d_{\lambda
j}=C_{\lambda\kappa}\,d^{\kappa}_{j}$, but this does not hold, since
the summation index $\kappa$ also appears elsewhere, namely in
$m^{\kappa}_{d}$. One has in fact
$$
\frac{g}{2M} m^{\kappa}_{d}C_{\lambda\kappa}d^{\kappa}_{j}=
\frac{g}{2M} m_d^\mu C_{\lambda\mu} \delta^\kappa_\mu
C^\dag_{\kappa\rho} d_{\rho j} =(K_d\,d)_{\lambda j} .
$$
Thus, the cross terms  in \eqref{higgsrecast5} can be written in the
form
$$
\frac{i}{\sqrt{2}}(-\phi^{+}\,
(\bar{u}^{\lambda}_{j}K_d^{\lambda\kappa}(1-\gamma^{5})d_{\kappa j})
   +\phi^{-}\,
(\bar{d}_{\lambda
j}(K_d^{\dagger})^{\lambda\kappa}(1+\gamma^{5})u^{\kappa}_{j})) .
$$
Thus, we get the complete expression \eqref{higgsrecast4}.
\endproof

\medskip
We still  need to add the new terms that account for neutrino masses
and mixing. We have the following result.

\begin{lem}\label{lemaddedterms}
The neutrino masses and mixing are obtained in two additional steps.
The first is the replacement
$$
T(0,K_e,\varphi)\mapsto T(K_\nu,K_e',\varphi) ,
$$
where the $K_e$ of \eqref{Keud} is replaced by
\begin{equation}\label{Knew}
(K_e)_{\lambda\kappa}= \frac{g}{2M}m_e^\mu
\cx_{\lambda\mu}\delta_\mu^\rho \cx^\dagger_{\rho\kappa} ,
\end{equation}
while $K_\nu$ is the neutrino Dirac mass matrix
\begin{equation}\label{Knu}
(K_\nu)_{\lambda \kappa}= \frac{g}{2M} m_\nu^\lambda
\delta_\lambda^\kappa.
\end{equation}
The second step is the addition of the Majorana mass term
\begin{equation}\label{Majmassterm}
\cL_{mass}=\,-\frac 14\,\bar
\nu_\lambda\,(M_R)_{\lambda\kappa}\,(1-\gamma_5)\hat\nu_\kappa
\,-\frac 14\,\bar{ \hat\nu}_\lambda\,\bar
(M_R)_{\lambda\kappa}\,(1+\gamma_5)\nu_\kappa.
\end{equation}
\end{lem}

\proof After performing the  inverse of the change of variables
\eqref{dlambdaj} for the leptons, using the matrix $\cx$ instead of
the CKM matrix, the new Dirac Yukawa coupling terms for the leptons
imply the replacement of
$$
-\frac{g}{2}\frac{m^{\lambda}_{e}}{M}\left(H(\bar{e}^{\lambda}e^{\lambda})
+i\phi^{0}(\bar{e}^{\lambda}\gamma^{5}e^{\lambda})\right)
$$
by
$$
-\frac{g}{2}\frac{m^{\lambda}_{\nu}}{M}H(\bar{\nu}^{\lambda}\nu^{\lambda})
-\frac{g}{2}\frac{m^{\lambda}_{e}}{M}H(\bar{e}^{\lambda}e^{\lambda})
+\frac{ig}{2}\frac{m^{\lambda}_{\nu}}{M}\phi^{0}(\bar{\nu}^{\lambda}\gamma^{5}\nu^{\lambda})
-\frac{ig}{2}\frac{m^{\lambda}_{e}}{M}\phi^{0}(\bar{e}^{\lambda}\gamma^{5}e^{\lambda})
$$
and of
$$
\frac{ig}{2\sqrt{2}}\frac{m^{\lambda}_{e}}{M}
\left(-\phi^{+}(\bar{\nu}^{\lambda}(1-\gamma^{5})e^{\lambda})
+\phi^{-}(\bar{e}^{\lambda}(1+\gamma^{5})\nu^{\lambda})\right)
$$
by
\begin{center}
\begin{math}
\frac{ig}{2M\sqrt{2}}\phi^{+}
\left(-m^{\kappa}_{e}(\bar{\nu}^{\lambda}\cx_{\lambda\kappa}(1-\gamma^{5})e^{\kappa})
+m^{\lambda}_{\nu}(\bar{\nu}^{\lambda}\cx_{\lambda\kappa}(1+\gamma^{5})e^{\kappa}\right)
   +\frac{ig}{2M\sqrt{2}}\phi^{-}
\left(m^{\lambda}_{e}(\bar{e}^{\lambda}\cx^{\dagger}_{\lambda\kappa}(1+\gamma^{5})\nu^{\kappa})
-m^{\kappa}_{\nu}(\bar{e}^{\lambda}\cx^{\dagger}_{\lambda\kappa}(1-\gamma^{5})\nu^{\kappa}\right),
\end{math}
\end{center}
where the matrix $\cx$ plays the same role as the CKM matrix. Since
the structure we obtained in the lepton sector is now identical to
that of the quark sector, the result follows from Lemma
\ref{lemmalgfandlhf}.

\smallskip
The Majorana mass terms are of the form \eqref{Majmassterm}, where
the coefficient $\frac 14$ instead of $\frac 12$ comes from the
chiral projection $(1-\gamma_5)=2 R$. The mass matrix $M_R$ is a
symmetric matrix in the flavor space.
\endproof

In order to understand the Euclidean version of the action
considered above, we start by treating the simpler case of the free
Dirac field.

It is given in Minkowski space by the action functional associated
to the Lagrangian
\begin{equation}\label{freefermimink}
-\bar{u}\,\gamma\partial\,u\,-\bar{u}\,m\,u .
\end{equation}
In Euclidean space the action functional becomes (\cf
\cite{Coleman}, ``The use of instantons'', \S 5.2)
\begin{equation}\label{freefermieuc}
S=-\int \bar{\psi}\,(i\,\gamma^\mu\,\partial_\mu-im)\,\psi\,d^4x ,
\end{equation}
where the symbols $\psi$ and $\bar \psi$ now stand for {\em
classical} fermions \ie independent anticommuting Grassman
variables.

Notice that, in \eqref{freefermieuc}, the gamma matrices
$\gamma^\mu$ are self-adjoint and the presence of $i=\sqrt {-1}$ in
the mass term is crucial to ensure that the Euclidean propagator is
of the form
$$
\frac{\pslash + im}{p^2+m^2} .
$$

\medskip
In our case, consider the Dirac operator $D_A$ that incorporates the
inner fluctuations. Recall that $D_A$ is given by the sum of two
terms
\begin{equation}\label{diracdec2}
D_A=D^{(1,0)}+\,\gamma_5\otimes D^{(0,1)} ,
\end{equation}
where $D^{(0,1)}$ is given by \eqref{diraczeroone} and $D^{(1,0)}$
is of the form
\begin{equation}\label{diraconezero2}
D^{(1,0)}=\sqrt{-1}\,\gamma^\mu(\nabla^s_\mu+\bA_\mu) ,
\end{equation}
where $\nabla^s$ is the spin connection (\cf\eqref{diraconM}), while
the $\bA_\mu$ are as in Proposition \ref{gaugepotentialssm}.

\begin{lem}\label{unitarylemma}  The  unitary operator
$$
U= e^{i\frac{\pi}{4}\gamma_5}\otimes 1
$$
commutes with $\cA$ and $\gamma$. One has $JU=U^*J$ and
\begin{equation}\label{diracgammafive}
U\,D_A\,U = D^{(1,0)} + i \otimes D^{(0,1)} .
\end{equation}
\end{lem}

\proof Since $\gamma_5$ anticommutes with the  $\gamma^\mu$, one has
$D^{(1,0)}
\,e^{i\frac{\pi}{4}\gamma_5}=\,e^{-i\frac{\pi}{4}\gamma_5}\,D^{(1,0)}$.
Moreover
$$
U\,(\gamma_5\otimes D^{(0,1)})\,U=
(\gamma_5\,e^{i\frac{\pi}{2}\gamma_5})\otimes D^{(0,1)}=i\,\otimes
D^{(0,1)}
$$
\endproof

\smallskip

The result of Lemma \ref{unitarylemma} can be restated as the
equality of antisymmetric bilinear forms
\begin{equation}\label{eqbilin}
\langle J U \xi', D_A U\xi \rangle = \langle J \xi', (D^{(1,0)} + i
\otimes D^{(0,1)}) \xi \rangle .
\end{equation}

\medskip

\subsubsection{The Fermion doubling
problem}\label{sectFerm2}\hfill\medskip

We can now discuss the Fermion doubling issue of \cite{lizzi}. As
explained there the number of fermion degrees of freedom when one
simply writes the Euclidean action $\langle \bar\psi,D\psi\rangle$
in our context is in fact $4$ times what it should be. The point is
that we have included one Dirac fermion for each of the chiral
degrees of freedom such as $e_R$ and that we introduced the mirror
fermions $\bar f$ to obtain the Hilbert space $\cH_F$.

Thus, we now need to explain how the action functional
\eqref{functspec} divides the number of degrees of freedom by $4$ by
taking a $4$'th root of a determinant.

By Proposition \ref{functspec0} we are dealing with an antisymmetric
bilinear form and the functional integral involving anticommuting
Grassman variables delivers the Pfaffian, which takes care of a
square root.

Again by Proposition \ref{functspec0}, we can restrict the
functional integration to the chiral subspace $\cH^+_{cl}$ of
\eqref{Hpluscl}, hence gaining another factor of two.

\smallskip
Let us spell out what happens first with quarks. With the basis
$q_L, q_R, \bar q_L,\bar q_R$ in $\cH_F$, the reduction to $\cH_+$
makes it possible  to write a generic vector as
\begin{equation}\label{hplusgeneric}
\zeta=\xi_L\otimes q_L+\xi_R\otimes q_R+\eta_R\otimes \bar
q_L+\eta_L \otimes \bar q_R ,
\end{equation}
where the subscripts $L$ and $R$ indicate the chirality of the usual
spinors $\xi_L\ldots \in L^2(M,S)$. Similarly, one has
\begin{equation}\label{hplusgeneric1}
J\,\zeta'=J_M\xi'_L\otimes \bar q_L+J_M\xi'_R\otimes \bar
q_R+J_M\eta'_R\otimes   q_L+J_M\eta'_L \otimes  q_R
\end{equation}
and
\begin{equation}\label{hplusgeneric2}
\zeta''=(\dirac_M\otimes 1)\,J\,\zeta'=\dirac_M\,J_M\xi'_L\otimes
\bar q_L+\dirac_M\,J_M\xi'_R\otimes \bar q_R
+\dirac_M\,J_M\eta'_R\otimes q_L+\dirac_M\,J_M\eta'_L
 \otimes q_R .
\end{equation}
Thus, since the operator $\dirac_M\,J_M$ anticommutes with
$\gamma_5$ in $L^2(M,S)$, we see that the vector $\xi''$ still
belongs to $\cH_+$ \ie is of the form \eqref{hplusgeneric}. One gets
\begin{center}
\begin{math}
\langle (\dirac_M\otimes 1)\,J\,\zeta',\zeta \rangle=\langle
\dirac_M\,J_M\xi'_L,\eta_R\rangle+\langle\dirac_M\,J_M\xi'_R,\eta_L\rangle+
\langle\dirac_M\,J_M\eta'_R,\xi_L\rangle+\langle\dirac_M\,J_M\eta'_L,
\xi_R\rangle .
\end{math}
\end{center}
The right hand side can be written, using the spinors
$\xi=\xi_L+\xi_R$ etc, as
\begin{equation}\label{hplusgeneric3}
\langle (\dirac_M\otimes 1)\,J\,\zeta',\zeta \rangle= \langle
\dirac_M\,J_M\xi'  ,\eta \rangle+ \langle\dirac_M\,J_M\eta' ,\xi
\rangle .
\end{equation}
This is an antisymmetric bilinear form in $L^2(M,S)\oplus L^2(M,S)$.
Indeed if $\zeta'=\zeta$ \ie $\xi'=\xi$ and $\eta'=\eta$ one has
\begin{equation}\label{hplusgeneric3.5}
\langle \dirac_M\,J_M\xi  ,\eta \rangle=- \langle\dirac_M\,J_M\eta
,\xi \rangle ,
\end{equation}
since $J_M$ commutes with $\dirac_M$ and has square $-1$.

At the level of the fermionic functional integral the classical
fermions $\tilde\xi$ and $\tilde\eta$ anticommute. Thus, up to the
factor $2$ taken care of by the $\frac 12$ in front of the fermionic
term, one gets
$$
\int\,e^{\langle J_M \tilde\eta ,\;\dirac_M\,\tilde\xi
\rangle}\,D[\tilde\eta]D[\tilde\xi] ,
$$
where $\tilde\xi$ and $\tilde\eta$ are independent anticommuting
variables. (Here we use the same notation as in \eqref{Hpluscl}).

This coincides with the prescription for the Euclidean functional
integral given in \cite{Coleman} (see ``The use of instantons'', \S
5.2) when using $J_M$ to identify $L^2(M,S)$ with its dual.

\smallskip

The Dirac Yukawa terms simply replace $\dirac_M\otimes 1$ in the
expression above by an operator of the form
$$
\dirac_M\otimes 1+\gamma_5\otimes \,T ,
$$
where $T=T(x)$ acts as a matrix valued function on the bundle
$S\otimes \cH_F$.

By construction, $T$ preserves $\cH_f$ and anticommutes with
$\gamma_F$. Thus, one gets an equation of the form
$$
(\gamma_5\otimes \,T)J\,\zeta'= T_1\,J_M\xi'_L\otimes \bar
q_R+T_2\,J_M\xi'_R\otimes \bar q_L+T_3\,J_M\eta'_R\otimes
q_R+T_4\,J_M\eta'_L \otimes  q_L ,
$$
where the $T_j$ are endomorphisms of the spinor bundle commuting
with the $\gamma_5$ matrix. In particular, it is a vector in
$\cH_+$. Thus, one gets
\begin{center}
\begin{math}
\langle (\gamma_5\otimes \,T)\,J\,\zeta',\zeta \rangle=\langle
T_1\,J_M\xi'_L,\eta_L\rangle+\langle T_2\,J_M\xi'_R,\eta_R\rangle+
\langle T_3\,J_M\eta'_R,\xi_R\rangle+\langle T_4\,J_M\eta'_L,
\xi_L\rangle .
\end{math}
\end{center}

The expression \eqref{hplusgeneric3} remains valid for the Dirac
operator with Yukawa couplings, with the $J_M\xi'$, $J_M\eta'$ on
the left, paired with the $\eta$ and $\xi$ respectively. Thus, the
Pfaffian of the corresponding classical fermions as Grassman
variables delivers the determinant of the Dirac operator.

\smallskip

We now come to the contribution of the piece of the operator $D$
which in the subspace $\nu_R,\bar\nu_R$ is of the form
$$
T=\left[
  \begin{array}{cc}
    0 & M^*_R \\
    M_R & 0 \\
  \end{array}
\right]
$$
where $M_R$ is a symmetric matrix in the flavor space. We use
\eqref{hplusgeneric} and \eqref{hplusgeneric1}, replacing quarks by
leptons, and we take a basis in which the matrix $M_R$ is diagonal.
We denote the corresponding eigenvalues still by $M_R$. We get
$$
\zeta=\xi_L\otimes \nu_L+\xi_R\otimes \nu_R+\eta_R\otimes \bar
\nu_L+\eta_L \otimes \bar \nu_R
$$
$$
J\,\zeta'=J_M\xi'_L\otimes \bar \nu_L+J_M\xi'_R\otimes \bar
\nu_R+J_M\eta'_R\otimes   \nu_L+J_M\eta'_L \otimes  \nu_R
$$
so that
$$
(\gamma_5\otimes T)\,J\,\zeta'=\gamma_5 M_R J_M\xi'_R\otimes  \nu_R
+\gamma_5 M_R J_M\eta'_L \otimes  \bar\nu_R
$$
\begin{equation}\label{hplusgeneric4}
\langle(\gamma_5\otimes T)\,J\,\zeta',\zeta\rangle= M_R \langle
\gamma_5
 J_M\xi'_R,\xi_R\rangle +M_R \langle
\gamma_5
 J_M\eta'_R,\eta_R\rangle
\end{equation}

The only effect of the $\gamma_5$ is an overall sign. The charge
conjugation operator $J_M$ is now playing a key role in the terms
\eqref{hplusgeneric4}, where it defines an antisymmetric bilinear
form on spinors of a given chirality (here right handed ones).

Notice also that one needs an overall factor of $\frac 12$ in front
of the fermionic action, since in the Dirac sector the same
expression repeats itself twice, see \eqref{hplusgeneric3.5}.

Thus, in the Majorana sector we get a factor $\frac 12$ in front of
the kinetic term. This corresponds to equation (4.20) of
\cite{MoPa}. For the treatment of Majorana fermions in Euclidean
functional integrals see \eg \cite{InSu}, \cite{NiWa}.
\endproof

\bigskip

\subsection{The self interaction of the gauge bosons}\hfill\medskip
\label{sectselfbosons}

The self-interaction terms for the gauge fields have the form
\begin{equation}\label{smgaugeym}
\cL_{g}=
\end{equation}
\begin{center}
\begin{math}
-\frac{1}{2}\partial_{\nu}g^{a}_{\mu}\partial_{\nu}g^{a}_{\mu}
-g_{s}f^{abc}\partial_{\mu}g^{a}_{\nu}g^{b}_{\mu}g^{c}_{\nu}
-\frac{1}{4}g^{2}_{s}f^{abc}f^{ade}g^{b}_{\mu}g^{c}_{\nu}g^{d}_{\mu}g^{e}_{\nu}
-\partial_{\nu}W^{+}_{\mu}\partial_{\nu}W^{-}_{\mu}-M^{2}W^{+}_{\mu}W^{-}_{\mu}
-\frac{1}{2}\partial_{\nu}Z^{0}_{\mu}\partial_{\nu}Z^{0}_{\mu}-\frac{1}{2c^{2}_{w}}
M^{2}Z^{0}_{\mu}Z^{0}_{\mu}
-\frac{1}{2}\partial_{\mu}A_{\nu}\partial_{\mu}A_{\nu}
-igc_{w}(\partial_{\nu}Z^{0}_{\mu}(W^{+}_{\mu}W^{-}_{\nu}-W^{+}_{\nu}W^{-}_{\mu})
-Z^{0}_{\nu}(W^{+}_{\mu}\partial_{\nu}W^{-}_{\mu}-W^{-}_{\mu}\partial_{\nu}W^{+}_{\mu})
+Z^{0}_{\mu}(W^{+}_{\nu}\partial_{\nu}W^{-}_{\mu}-W^{-}_{\nu}\partial_{\nu}W^{+}_{\mu}))
-igs_{w}(\partial_{\nu}A_{\mu}(W^{+}_{\mu}W^{-}_{\nu}-W^{+}_{\nu}W^{-}_{\mu})
-A_{\nu}(W^{+}_{\mu}\partial_{\nu}W^{-}_{\mu}-W^{-}_{\mu}\partial_{\nu}W^{+}_{\mu})
+A_{\mu}(W^{+}_{\nu}\partial_{\nu}W^{-}_{\mu}-W^{-}_{\nu}\partial_{\nu}W^{+}_{\mu}))
-\frac{1}{2}g^{2}W^{+}_{\mu}W^{-}_{\mu}W^{+}_{\nu}W^{-}_{\nu}+\frac{1}{2}g^{2}
W^{+}_{\mu}W^{-}_{\nu}W^{+}_{\mu}W^{-}_{\nu}
+g^2c^{2}_{w}(Z^{0}_{\mu}W^{+}_{\mu}Z^{0}_{\nu}W^{-}_{\nu}-Z^{0}_{\mu}Z^{0}_{\mu}W^{+}_{\nu}
W^{-}_{\nu})
+g^2s^{2}_{w}(A_{\mu}W^{+}_{\mu}A_{\nu}W^{-}_{\nu}-A_{\mu}A_{\mu}W^{+}_{\nu}
W^{-}_{\nu})
+g^{2}s_{w}c_{w}(A_{\mu}Z^{0}_{\nu}(W^{+}_{\mu}W^{-}_{\nu}-W^{+}_{\nu}W^{-}_{\mu})
-2A_{\mu}Z^{0}_{\mu}W^{+}_{\nu}W^{-}_{\nu}) .
\end{math}
\end{center}
We show that they can be written as a sum of terms of the following
form.
\begin{enumerate}
  \item Mass terms for the $W^\pm$ and the
$Z^0$
  \item Yang-Mills interaction $-\frac
14\,F^a_{\mu\nu}\,F_a^{\mu\nu}$ for the gauge fields
$B_\mu,W_\mu^a,g_\mu^a$
  \item Feynman gauge fixing terms $\cL_{feyn}$ for all
gauge fields \index{gauge!Feynman}
\end{enumerate}

\medskip
\begin{lem}\label{namelesslemma}
One has
\begin{equation}\label{massleftover}
\cL_{g}=-M^{2}W^{+}_{\mu}W^{-}_{\mu} -\frac{1}{2c^{2}_{w}}
M^{2}Z^{0}_{\mu}Z^{0}_{\mu}-\frac
14\,F^a_{\mu\nu}\,F_a^{\mu\nu}-\frac
12\sum_a(\sum_\mu\partial_{\mu}G^a_{\mu})^2
\end{equation}
\end{lem}

\proof It is enough to show that the expression
\begin{center}
\begin{math}
-\partial_{\nu}W^{+}_{\mu}\partial_{\nu}W^{-}_{\mu}
-\frac{1}{2}\partial_{\nu}(c_w\,Z^{0}_{\mu}+s_w\,A_\mu)\partial_{\nu}(c_w\,Z^{0}_{\mu}+s_w\,A_\mu)
-igc_{w}(\partial_{\nu}Z^{0}_{\mu}(W^{+}_{\mu}W^{-}_{\nu}-W^{+}_{\nu}W^{-}_{\mu})
-Z^{0}_{\nu}(W^{+}_{\mu}\partial_{\nu}W^{-}_{\mu}-W^{-}_{\mu}\partial_{\nu}W^{+}_{\mu})
+Z^{0}_{\mu}(W^{+}_{\nu}\partial_{\nu}W^{-}_{\mu}-W^{-}_{\nu}\partial_{\nu}W^{+}_{\mu}))
-igs_{w}(\partial_{\nu}A_{\mu}(W^{+}_{\mu}W^{-}_{\nu}-W^{+}_{\nu}W^{-}_{\mu})
-A_{\nu}(W^{+}_{\mu}\partial_{\nu}W^{-}_{\mu}-W^{-}_{\mu}\partial_{\nu}W^{+}_{\mu})
+A_{\mu}(W^{+}_{\nu}\partial_{\nu}W^{-}_{\mu}-W^{-}_{\nu}\partial_{\nu}W^{+}_{\mu}))
-\frac{1}{2}g^{2}W^{+}_{\mu}W^{-}_{\mu}W^{+}_{\nu}W^{-}_{\nu}+\frac{1}{2}g^{2}
W^{+}_{\mu}W^{-}_{\nu}W^{+}_{\mu}W^{-}_{\nu}
+g^2c^{2}_{w}(Z^{0}_{\mu}W^{+}_{\mu}Z^{0}_{\nu}W^{-}_{\nu}-Z^{0}_{\mu}Z^{0}_{\mu}W^{+}_{\nu}
W^{-}_{\nu})
+g^2s^{2}_{w}(A_{\mu}W^{+}_{\mu}A_{\nu}W^{-}_{\nu}-A_{\mu}A_{\mu}W^{+}_{\nu}
W^{-}_{\nu})
+g^{2}s_{w}c_{w}(A_{\mu}Z^{0}_{\nu}(W^{+}_{\mu}W^{-}_{\nu}-W^{+}_{\nu}W^{-}_{\mu})
-2A_{\mu}Z^{0}_{\mu}W^{+}_{\nu}W^{-}_{\nu})
\end{math}
\end{center}
coincides with the Yang--Mills action of the $\SU(2)$-gauge field.

In fact, the kinetic terms will then combine with those of the
$B$-field, namely
$$
-\frac{1}{2}\partial_{\nu}(-s_w\,Z^{0}_{\mu}+c_w\,A_\mu)\partial_{\nu}(-s_w\,Z^{0}_{\mu}+c_w\,A_\mu).
$$

One can rewrite the above in terms of $W_{\mu 3}=\,s_w\,A_\mu
+\,c_w\,Z_\mu^0$. This gives
\begin{center}
\begin{math}
-\partial_{\nu}W^{+}_{\mu}\partial_{\nu}W^{-}_{\mu}
-\frac{1}{2}\partial_{\nu}W_{\mu 3}\partial_{\nu}W_{\mu 3}
-ig(\partial_{\nu}W_{\mu
3}(W^{+}_{\mu}W^{-}_{\nu}-W^{+}_{\nu}W^{-}_{\mu}) -W_{\nu
3}(W^{+}_{\mu}\partial_{\nu}W^{-}_{\mu}-W^{-}_{\mu}\partial_{\nu}W^{+}_{\mu})
+W_{\mu
3}(W^{+}_{\nu}\partial_{\nu}W^{-}_{\mu}-W^{-}_{\nu}\partial_{\nu}W^{+}_{\mu}))
-\frac{1}{2}g^{2}W^{+}_{\mu}W^{-}_{\mu}W^{+}_{\nu}W^{-}_{\nu}+\frac{1}{2}g^{2}
W^{+}_{\mu}W^{-}_{\nu}W^{+}_{\mu}W^{-}_{\nu} +g^2(W_{\mu
3}W^{+}_{\mu}W_{\nu 3}W^{-}_{\nu}-W_{\mu 3}W_{\mu 3}W^{+}_{\nu}
W^{-}_{\nu}) .
\end{math}
\end{center}
Using $W_\mu^+=\frac{W_{\mu 1}-iW_{\mu 2}}{\sqrt 2}$ and
$W_\mu^-=\frac{W_{\mu 1}+iW_{\mu 2}}{\sqrt 2}$, one checks that it
coincides with the Yang-Mills action functional $-\frac
14\,F^a_{\mu\nu}\,F_a^{\mu\nu}$ of the $\SU(2)$-gauge field $W_{\mu
j}$.

More precisely, let
$$ \nabla_\mu=\partial_\mu-i\frac g2
W^\mu_a\sigma_a .
$$
One then has
$$
[\nabla_\mu,\nabla_\nu]=-i\frac g2(\partial_\mu W^\nu_a-\partial_\nu
W^\mu_a)\sigma_a+(-i\frac g2)^2(W^\mu_b W^\nu_c \sigma_b\,
\sigma_c-W^\nu_c W^\mu_b \sigma_c\, \sigma_b)
$$
and, with $\sigma_b\, \sigma_c- \sigma_c\, \sigma_b=2i\,
\epsilon_{abc}\,\sigma_a$, this gives
\begin{equation}\label{ymcurvterms}
F_{\mu\nu}^a= \partial_\mu W^\nu_a-\partial_\nu W^\mu_a + g
\,\epsilon_{abc}\,W^\mu_b W^\nu_c .
\end{equation}
One then checks directly that the above expression coincides with
\begin{equation}\label{ymfullterms}
-\frac 14\,F^a_{\mu\nu}\,F_a^{\mu\nu}-\frac
12\sum_a(\sum_\mu\partial_{\mu}W^a_{\mu})^2 .
\end{equation}

Notice that the addition of the Feynman gauge fixing term $ -\frac
12(\sum_\mu\partial_{\mu}G^{\mu})^2$ to the kinetic term $-\frac
14\,|dG|^2$ of the Yang-Mills action for each of the gauge fields
$G^{\mu}$ gives kinetic terms of the form $ -\frac
12\,\partial_{\nu}G^{\mu}\,\partial_{\nu}G^{\mu}$ and very simple
propagators.

This addition of the gauge fixing term is not obtained from the
spectral action, but has to be added afterwards together with the
ghost fields.
\endproof

\subsection{The minimal coupling of the Higgs field}\hfill\medskip
\label{sectminimalcoupling}

We add the mass terms \eqref{massleftover} to the minimal coupling
terms of the Higgs fields with the gauge fields which is of the form
\begin{equation}\label{mcterms}
\cL_{Hg}=
\end{equation}
\begin{center}
\begin{math}
-\frac{1}{2}\partial_{\mu}H\partial_{\mu}H
-\partial_{\mu}\phi^{+}\partial_{\mu}\phi^{-}
-\frac{1}{2}\partial_{\mu}\phi^{0}\partial_{\mu}\phi^{0}-gMW^{+}_{\mu}W^{-}_{\mu}H-\frac{1}{2}g\frac{M}{c^{2}_{w}}Z^{0}_{\mu}Z^{0}_{\mu}H
-\frac{1}{2}ig\left(W^{+}_{\mu}(\phi^{0}\partial_{\mu}\phi^{-}
-\phi^{-}\partial_{\mu}\phi^{0})
-W^{-}_{\mu}(\phi^{0}\partial_{\mu}\phi^{+}
-\phi^{+}\partial_{\mu}\phi^{0})\right)
+\frac{1}{2}g\left(W^{+}_{\mu}(H\partial_{\mu}\phi^{-}
-\phi^{-}\partial_{\mu}H)
+W^{-}_{\mu}(H\partial_{\mu}\phi^{+}-\phi^{+}\partial_{\mu}H)\right)
+\frac{1}{2}g\frac{1}{c_{w}}(Z^{0}_{\mu}(H\partial_{\mu}\phi^{0}-\phi^{0}\partial_{\mu}H)
-ig\frac{s^{2}_{w}}{c_{w}}MZ^{0}_{\mu}(W^{+}_{\mu}\phi^{-}-W^{-}_{\mu}\phi^{+})
   +igs_{w}MA_{\mu}(W^{+}_{\mu}\phi^{-}-W^{-}_{\mu}\phi^{+})
-ig\frac{1-2c^{2}_{w}}{2c_{w}}Z^{0}_{\mu}(\phi^{+}\partial_{\mu}\phi^{-}
-\phi^{-}\partial_{\mu}\phi^{+})
+igs_{w}A_{\mu}(\phi^{+}\partial_{\mu}\phi^{-}-\phi^{-}\partial_{\mu}\phi^{+})
-\frac{1}{4}g^{2}W^{+}_{\mu}W^{-}_{\mu}
\left(H^{2}+(\phi^{0})^{2}+2\phi^{+}\phi^{-}\right)
-\frac{1}{8}g^{2}\frac{1}{c^{2}_{w}}Z^{0}_{\mu}Z^{0}_{\mu}
\left(H^{2}+(\phi^{0})^{2}+2(2s^{2}_{w}-1)^{2}\phi^{+}\phi^{-}\right)
-\frac{1}{2}g^{2}\frac{s^{2}_{w}}{c_{w}}Z^{0}_{\mu}\phi^{0}(W^{+}_{\mu}\phi^{-}+W^{-}_{\mu}\phi^{+})
-\frac{1}{2}ig^{2}\frac{s^{2}_{w}}{c_{w}}Z^{0}_{\mu}H(W^{+}_{\mu}\phi^{-}-W^{-}_{\mu}\phi^{+})
+\frac{1}{2}g^{2}s_{w}A_{\mu}\phi^{0}(W^{+}_{\mu}\phi^{-}+W^{-}_{\mu}\phi^{+})
+\frac{1}{2}ig^{2}s_{w}A_{\mu}H(W^{+}_{\mu}\phi^{-}-W^{-}_{\mu}\phi^{+})
-g^{2}\frac{s_{w}}{c_{w}}(2c^{2}_{w}-1)Z^{0}_{\mu}A_{\mu}\phi^{+}\phi^{-}
-g^{2}s^{2}_{w}A_{\mu}A_{\mu}\phi^{+}\phi^{-}+M\,(\frac{1}{c_{w}}Z^{0}_{\mu}\partial_{\mu}\phi^{0}+W^{+}_{\mu}
\partial_{\mu}\phi^{-}+W^{-}_{\mu}
\partial_{\mu}\phi^{+}) .
\end{math}
\end{center}
This is, by construction, a sum of terms labeled by $\mu$. Each of
them contains three kinds of terms, according to the number of
derivatives. We now compare this expression with the minimal
coupling terms which we get from the spectral action.

\begin{lem}\label{minimalcouplingcomparison} With the notation
\eqref{gaugeform1} of Lemma \ref{lemmalgfandlhf}, the minimal
coupling terms \eqref{mcterms} are given by
\begin{equation}\label{mcterms1}
\cL_{Hg}=\,-\frac 12\,|D_\mu\,\varphi|^2
\end{equation}
with $D_\mu\varphi$ given by \eqref{minimalcoupling}, with $g_2=g$,
$g_1=g'$.
\end{lem}

\proof We have from \eqref{minimalcoupling}
$$
D_{\mu} \, \varphi  =  \,
\partial_{\mu} \, \varphi + \frac{i }{ 2} \, g \, W_{\mu}^{ \alpha} \,\varphi\, \sigma^{ \alpha}
 - \frac{i }{ 2} \, g' \, B_{\mu} \, \varphi \,,\quad g'=\tan (\theta_w)\,g
 $$
where, by Lemma \ref{lemmalgfandlhf}, we have
$$\varphi=(\varphi_1,\varphi_2)= (\frac{2M}{g}+
H-i\phi^{0},-\,i\,\sqrt{2}\,\phi^{+})\,,\quad
B_\mu=\,c_w\,A_\mu-s_w\,Z^0_\mu\,\,,\quad W_{\mu 3}=\,s_w\,A_\mu
+\,c_w\,Z_\mu^0$$ and the matrix $W_{\mu}^{ \alpha} \, \sigma^{
\alpha}$ is given by
$$
W_{\mu}^{ \alpha} \, \sigma^{ \alpha}=\, \left[
  \begin{array}{cc}
    s_w\,A_\mu
+\,c_w\,Z_\mu^0 & W_{\mu}^{1} -i\,W_{\mu}^{2}\\
    W_{\mu}^{1} +i\,W_{\mu}^{2} & -s_w\,A_\mu
-\,c_w\,Z_\mu^0 \\
  \end{array}
\right] = \, \left[
  \begin{array}{cc}
    s_w\,A_\mu
+\,c_w\,Z_\mu^0 & \sqrt 2\,W_{\mu}^{+} \\
    \sqrt 2\,W_{\mu}^{-} & -s_w\,A_\mu
-\,c_w\,Z_\mu^0 \\
  \end{array}
\right] .
$$

The kinetic terms are simply
$$
-\frac{1}{2}\partial_{\mu}H\partial_{\mu}H
-\partial_{\mu}\phi^{+}\partial_{\mu}\phi^{-}
-\frac{1}{2}\partial_{\mu}\phi^{0}\partial_{\mu}\phi^{0}
$$
and one checks that they are obtained.

\smallskip

Let us consider the terms with no derivatives. The combination
$W_{\mu}^{ \alpha} \,\varphi\, \sigma^{ \alpha}$ is given by
$$
 ((\frac{2M}{g}+ H-i\phi^{0})(s_w\,A_\mu
+\,c_w\,Z_\mu^0)-2i\,\phi^{+}\,W_{\mu}^{-}, (\frac{2M}{g}+
H-i\phi^{0})\sqrt 2\,W_{\mu}^{+}+i\,\sqrt{2}\,\phi^{+}(s_w\,A_\mu
+\,c_w\,Z_\mu^0)) .
$$
The term $B_{\mu} \, \varphi$ is given by
$$
B_{\mu} \, \varphi=\,((\frac{2M}{g}+
H-i\phi^{0})(c_w\,A_\mu-s_w\,Z^0_\mu),-\,i\,\sqrt{2}\,
 \phi^{+}(c_w\,A_\mu-s_w\,Z^0_\mu)) .
$$

The dangerous term in $M\,A_\mu$ (which would give a mass to the
photon) has to disappear. This follows from $g'=\tan (\theta_w)\,g$.
This means that we consider the expression $W_{\mu}^{ \alpha}
\,\varphi\, \sigma^{ \alpha}-\tan (\theta_w)\,B_{\mu} \, \varphi$.
It gives
\begin{equation}\label{connectionsm}
W_{\mu}^{ \alpha} \,\varphi\, \sigma^{ \alpha}-\tan
(\theta_w)\,B_{\mu} \, \varphi=(X_1,X_2)=
\end{equation}
$$
((\frac{2M}{g}+
H-i\phi^{0})\frac{1}{c_w}\,Z_\mu^0-2i\,\phi^{+}\,W_{\mu}^{-},
(\frac{2M}{g}+ H-i\phi^{0})\sqrt
2\,W_{\mu}^{+}+i\,\sqrt{2}\,\phi^{+}(2\,s_w\,A_\mu
+\,(c_w-\frac{s_w^2}{c_w})\,Z_\mu^0)) .
$$
One has to multiply by $\frac g2 \sqrt {-1}$ and then take $-\frac
12$ of the norm square. The direct computation gives
\begin{center}
\begin{math}
-M^{2}W^{+}_{\mu}W^{-}_{\mu} -\frac{1}{2c^{2}_{w}}
M^{2}Z^{0}_{\mu}Z^{0}_{\mu}
-gMW^{+}_{\mu}W^{-}_{\mu}H-\frac{1}{2}g\frac{M}{c^{2}_{w}}Z^{0}_{\mu}Z^{0}_{\mu}H
-ig\frac{s^{2}_{w}}{c_{w}}MZ^{0}_{\mu}(W^{+}_{\mu}\phi^{-}-W^{-}_{\mu}\phi^{+})
   +igs_{w}MA_{\mu}(W^{+}_{\mu}\phi^{-}-W^{-}_{\mu}\phi^{+})
-\frac{1}{4}g^{2}W^{+}_{\mu}W^{-}_{\mu}
\left(H^{2}+(\phi^{0})^{2}+2\phi^{+}\phi^{-}\right)
-\frac{1}{8}g^{2}\frac{1}{c^{2}_{w}}Z^{0}_{\mu}Z^{0}_{\mu}
\left(H^{2}+(\phi^{0})^{2}+2(2s^{2}_{w}-1)^{2}\phi^{+}\phi^{-}\right)
-\frac{1}{2}g^{2}\frac{s^{2}_{w}}{c_{w}}Z^{0}_{\mu}\phi^{0}(W^{+}_{\mu}\phi^{-}+W^{-}_{\mu}\phi^{+})
-\frac{1}{2}ig^{2}\frac{s^{2}_{w}}{c_{w}}Z^{0}_{\mu}H(W^{+}_{\mu}\phi^{-}-W^{-}_{\mu}\phi^{+})
+\frac{1}{2}g^{2}s_{w}A_{\mu}\phi^{0}(W^{+}_{\mu}\phi^{-}+W^{-}_{\mu}\phi^{+})
+\frac{1}{2}ig^{2}s_{w}A_{\mu}H(W^{+}_{\mu}\phi^{-}-W^{-}_{\mu}\phi^{+})
-g^{2}\frac{s_{w}}{c_{w}}(1-2s^{2}_{w})Z^{0}_{\mu}A_{\mu}\phi^{+}\phi^{-}
-g^{2}s^{2}_{w}A_{\mu}A_{\mu}\phi^{+}\phi^{-} .
\end{math}
\end{center}
Taking into account the terms \eqref{massleftover}, the terms with
no derivatives in \eqref{mcterms} are
\begin{center}
\begin{math}
-M^{2}W^{+}_{\mu}W^{-}_{\mu} -\frac{1}{2c^{2}_{w}}
M^{2}Z^{0}_{\mu}Z^{0}_{\mu}
-gMW^{+}_{\mu}W^{-}_{\mu}H-\frac{1}{2}g\frac{M}{c^{2}_{w}}Z^{0}_{\mu}Z^{0}_{\mu}H
-ig\frac{s^{2}_{w}}{c_{w}}MZ^{0}_{\mu}(W^{+}_{\mu}\phi^{-}-W^{-}_{\mu}\phi^{+})
   +igs_{w}MA_{\mu}(W^{+}_{\mu}\phi^{-}-W^{-}_{\mu}\phi^{+})
-\frac{1}{4}g^{2}W^{+}_{\mu}W^{-}_{\mu}
\left(H^{2}+(\phi^{0})^{2}+2\phi^{+}\phi^{-}\right)
-\frac{1}{8}g^{2}\frac{1}{c^{2}_{w}}Z^{0}_{\mu}Z^{0}_{\mu}
\left(H^{2}+(\phi^{0})^{2}+2(2s^{2}_{w}-1)^{2}\phi^{+}\phi^{-}\right)
-\frac{1}{2}g^{2}\frac{s^{2}_{w}}{c_{w}}Z^{0}_{\mu}\phi^{0}(W^{+}_{\mu}\phi^{-}+W^{-}_{\mu}\phi^{+})
-\frac{1}{2}ig^{2}\frac{s^{2}_{w}}{c_{w}}Z^{0}_{\mu}H(W^{+}_{\mu}\phi^{-}-W^{-}_{\mu}\phi^{+})
+\frac{1}{2}g^{2}s_{w}A_{\mu}\phi^{0}(W^{+}_{\mu}\phi^{-}+W^{-}_{\mu}\phi^{+})
+\frac{1}{2}ig^{2}s_{w}A_{\mu}H(W^{+}_{\mu}\phi^{-}-W^{-}_{\mu}\phi^{+})
-g^{2}\frac{s_{w}}{c_{w}}(2c^{2}_{w}-1)Z^{0}_{\mu}A_{\mu}\phi^{+}\phi^{-}
-g^{2}s^{2}_{w}A_{\mu}A_{\mu}\phi^{+}\phi^{-} .
\end{math}
\end{center}
Thus, there is only one difference with respect to the above, namely
the  replacement $(2c^{2}_{w}-1)\mapsto (1-2s^{2}_{w})$ in the
$13$'th term. This has no effect since $s^{2}_{w}+c^{2}_{w}=1$.

\smallskip

We now need to take care of the terms with one derivative. With the
notation as above, we compute the cross terms of
$$
-\frac 12
\,|(\partial_{\mu}\varphi_1,\partial_{\mu}\varphi_2)+\frac{ig}{2}\,
 (X_1,X_2)|^2 ,
$$
\ie the terms
$$
-\frac 12
\,(\partial_{\mu}\bar\varphi_1\,\frac{ig}{2}X_1-\partial_{\mu}\varphi_1\,\frac{ig}{2}\bar
X_1+\partial_{\mu}\bar\varphi_2\,\frac{ig}{2}X_2-\partial_{\mu}\varphi_2\,\frac{ig}{2}\bar
X_2) .
$$
The computation gives
\begin{center}
\begin{math}
-\frac{1}{2}ig\left(W^{+}_{\mu}(\phi^{0}\partial_{\mu}\phi^{-}
-\phi^{-}\partial_{\mu}\phi^{0})
-W^{-}_{\mu}(\phi^{0}\partial_{\mu}\phi^{+}
-\phi^{+}\partial_{\mu}\phi^{0})\right)
+\frac{1}{2}g\left(W^{+}_{\mu}(H\partial_{\mu}\phi^{-}
-\phi^{-}\partial_{\mu}H)
+W^{-}_{\mu}(H\partial_{\mu}\phi^{+}-\phi^{+}\partial_{\mu}H)\right)
+\frac{1}{2}g\frac{1}{c_{w}}(Z^{0}_{\mu}(H\partial_{\mu}\phi^{0}-\phi^{0}\partial_{\mu}H)
-ig\frac{1-2c^{2}_{w}}{2c_{w}}Z^{0}_{\mu}(\phi^{+}\partial_{\mu}\phi^{-}
-\phi^{-}\partial_{\mu}\phi^{+})
+igs_{w}A_{\mu}(\phi^{+}\partial_{\mu}\phi^{-}-\phi^{-}\partial_{\mu}\phi^{+})
+M\,(\frac{1}{c_{w}}Z^{0}_{\mu}\partial_{\mu}\phi^{0}+W^{+}_{\mu}
\partial_{\mu}\phi^{-}+W^{-}_{\mu}
\partial_{\mu}\phi^{+}) ,
\end{math}
\end{center}
which agrees with the sum of terms with one derivative in
\eqref{mcterms}.
\endproof

\subsection{The Higgs field self interaction}\hfill\medskip
\label{secthiggsself}

The Higgs self coupling terms of the standard model are of the form
\begin{equation}\label{higgsquart}
\cL_{H}=
\end{equation}
\begin{center}
\begin{math}
-\frac{1}{2}m^{2}_{h}H^{2}
-\beta_{h}\left(\frac{2M^{2}}{g^{2}}+\frac{2M}{g}H+\frac{1}{2}(H^{2}+\phi^{0}\phi^{0}+2\phi^{+}\phi^{-
})\right)
  +\frac{2M^{4}}{g^{2}}\alpha_{h}
  -g\alpha_h
M\left(H^3+H\phi^{0}\phi^{0}+2H\phi^{+}\phi^{-}\right)
-\frac{1}{8}g^{2}\alpha_{h}
\left(H^4+(\phi^{0})^{4}+4(\phi^{+}\phi^{-})^{2}
+4(\phi^{0})^{2}\phi^{+}\phi^{-}
+4H^{2}\phi^{+}\phi^{-}+2(\phi^{0})^{2}H^{2}\right) .
\end{math}
\end{center}

\medskip
\begin{lem}\label{Higs4lem}
Let $\varphi$ be given by \eqref{higgsrecast3} and assume that
\begin{equation}\label{higgsquart3}
\alpha_{h}=\frac{m^{2}_{h}}{4\,M^2} .
\end{equation}
Then one has
\begin{equation}\label{higgsquart3.5}
\cL_H=-\frac{1}{8}g^{2}\alpha_{h}\,|\varphi|^4+\,(\alpha_{h}\,M^2\,
 -\frac{\beta_h}{2})\,|\varphi|^2 .
\end{equation}
\end{lem}

\proof The expression \eqref{higgsquart} can be simplified in terms
of the field $\psi$. The quartic term is simply given by
$$
-\frac{1}{8}g^{2}\alpha_{h}
\left(H^4+(\phi^{0})^{4}+4(\phi^{+}\phi^{-})^{2}
+4(\phi^{0})^{2}\phi^{+}\phi^{-}
+4H^{2}\phi^{+}\phi^{-}+2(\phi^{0})^{2}H^{2}\right)=-\frac{1}{8}g^{2}
 \alpha_{h}|\psi|^4 ,
$$
since
$$
|\psi|^2=|\psi_1|^2+|\psi_2|^2=H^2+(\phi^{0})^{2}
 +2\, \phi^{+}\phi^{-} .
$$
The cubic term is
$$
-g\alpha_h M\left(H^3+H\phi^{0}\phi^{0}+2H\phi^{+}\phi^{-}\right)=
-g\alpha_h M\,H\,|\psi|^2 ,
$$
which arises in the expansion of
\begin{equation}\label{higgsquart1}
-\frac{1}{8}g^{2}\alpha_{h}\,|\varphi|^4 ,
\end{equation}
with $\varphi$ given by \eqref{higgsrecast3}, so that
$$
|\varphi|^2=\,|\psi|^2+\frac{4M}{g}\,H+\frac{4M^2}{g^2}
$$
and
$$
|\varphi|^4=\,|\psi|^4+ \frac{8M}{g}\,H\,|\psi|^2+\frac{16
M^2}{g^2}\,H^2+\frac{8M^2}{g^2}\,|\psi|^2+\frac{16 M^4}{g^4}
+\frac{32 M^3}{g^3}\,H \, .
$$
Thus, the natural invariant  expression with no tadpole (\ie with
the expansion in $H$ at an extremum) is
\begin{equation}\label{higgsquart1.5}
-\frac{1}{8}g^{2}\alpha_{h}\,|\varphi|^4
 +\,\alpha_{h}\,M^2\,|\varphi|^2 .
\end{equation}
It expands as
\begin{equation}\label{higgsquart2}
-\frac{1}{8}g^{2}\alpha_{h}|\psi|^4 -g\alpha_h
M\,H\,|\psi|^2-2\alpha_{h}\,M^2\,H^2+\frac{2M^{4}}{g^{2}}
 \alpha_{h} ,
\end{equation}
which takes care of the constant term
$+\frac{2M^{4}}{g^{2}}\alpha_{h}$ in \eqref{higgsquart}. Thus, we
get
\begin{equation}\label{LHsofar}
\cL_{H}=(-\frac{1}{8}g^{2}\alpha_{h}\,|\varphi|^4+\,\alpha_{h}\,M^2\,|\varphi|^2)
+(2\alpha_{h}\,M^2-\frac{1}{2}m^{2}_{h})H^{2}
-\frac{\beta_h}{2}\,|\varphi|^2 ,
\end{equation}
since the quadratic ``tadpole" term in \eqref{higgsquart} is
\begin{equation}\label{betahtadpole}
-\beta_{h}\left(\frac{2M^{2}}{g^{2}}+\frac{2M}{g}H+\frac{1}{2}(H^{2}+\phi^{0}\phi^{0}+2\phi^{+}\phi^{-
})\right)=-\frac{\beta_{h}}{2}\,|\varphi|^2 .
\end{equation}
The assumption \eqref{higgsquart3} of the lemma implies that the
coefficient of the term in $H^2$ in \eqref{LHsofar} vanishes.
\endproof

\begin{rem}\label{tadrem}{\rm
The tadpole term \eqref{betahtadpole} is understandable, since in
renormalizing the terms one has to maintain the vanishing of the
term in $H$. The assumption \eqref{higgsquart3} is a standard
relation giving the Higgs mass (\cf \cite{VDiag}).}
\end{rem}

\medskip
\subsection{The coupling with gravity}\hfill\medskip
\label{sectcouplinggrav}

By construction the spectral action delivers the standard model
minimally coupled with Einstein gravity. Thus   the Lagrangian of
the standard model of \S \ref{sectstandardmodel} is now written
using the Riemannian metric $g_{\mu\nu}$ and the corresponding Dirac
operator $\dirac_M$ in curved space-time. We shall check below that
the Einstein term (the scalar curvature) admits the correct physical
sign and size for the functional integral in Euclidean signature.
The addition of the minimally coupled standard model gives the
Einstein equation when one writes the equations of motion by
differentiating with respect to $g_{\mu\nu}$ (\cf for instance
\cite{Weinberg} Chapter 12 \S 2).

\smallskip
The spectral action contains one more term that couple gravity with
the standard model, namely the term in $ \,R\,\higgs^2$. This term
is unavoidable as soon as one considers gravity simultaneously with
scalar fields as explained in \cite{feynmgrav}. The only other new
term is the Weyl curvature term
\begin{equation}\label{weylsquareterm}
-\frac{3\,f_0 }{ 10\,\pi^2}\,\int \  \, C_{\mu \nu \rho \sigma} \,
C^{\mu \nu \rho \sigma}\, \sqrt g\,d^4 x
\end{equation}

\bigskip
\bigskip

This completes the proof of Theorem \ref{maintheorem}.
\endproof

\bigskip
\section{Phenomenology and predictions}

\subsection{Coupling constants at unification}\hfill\medskip
\label{sectunif}

The relations
$$ g_2^2 = g_3^2 = \frac 53 g_1^2 $$
we derived in \eqref{coeffymterm} among the gauge coupling constants
coincide with those obtained in grand unification theories (\cf
\cite{CMGMP} and \cite{MoPa} \S 9). This indicates that the action
functional \eqref{rescaledbosaction} should be taken as the {\it
bare action} at the unification cutoff scale $\Lambda$ and we first
briefly recall how this scale is computed.

\smallskip
The electromagnetic coupling constant is given by \eqref{weinberg1}
and is $g\,\sin(\theta_w)$. The fine structure constant
$\alpha_{em}$ is thus given by
\begin{equation}\label{finestructure}
\alpha_{em}=\,\sin(\theta_w)^2\,\alpha_2\,,\quad \alpha_i =
\frac{g_i^2 }{ 4\pi}
\end{equation}
Its infrared value is $\sim 1/137.036$ but it is running as a
function of the energy and increases to  the  value $\alpha_{em}
(M_Z) = 1/128.09$ already, at the energy $M_Z\sim 91.188$ GeV.

\smallskip

Assuming the ``big desert'' hypothesis, the running of the three
couplings $\alpha_i$ is known. With 1-loop corrections only, it is
given by (\cite{knecht}, \cite{ACKMPRW})
\begin{equation}\label{runningalphas}
\beta_{g_i}=(4\pi)^{-2}\,b_i\,g_i^3, \ \ \ \text{ with } \ \
b=(\frac{41}{6},-\frac{19}{6},-7),
\end{equation}
so that (\cite{ross})
\begin{eqnarray}
   \alpha_1^{-1} (\Lambda) &=& \,  \alpha_1^{-1} ( M_Z) -
\frac{41 }{ 12\pi} \, \log \, \frac{ \Lambda }{ M_Z} \label{rengroup1}\\
  \alpha_2^{-1} (\Lambda) &=&\,
 \alpha_2^{-1} ( M_Z ) +\frac{19 }{ 12\pi} \, \log \, \frac{ \Lambda }{ M_Z}\nonumber \\
  \alpha_3^{-1} (\Lambda) &=&  \,  \alpha_3^{-1} (  M_Z) + \frac{42
}{ 12\pi} \, \log \, \frac{ \Lambda }{ M_Z} \nonumber
\end{eqnarray}
where  $M_Z$ is the mass of the $Z^0$ vector boson. For 2-loop
corrections see \cite{ACKMPRW}.

\smallskip

\begin{figure}
\begin{center}
\includegraphics[scale=0.75]{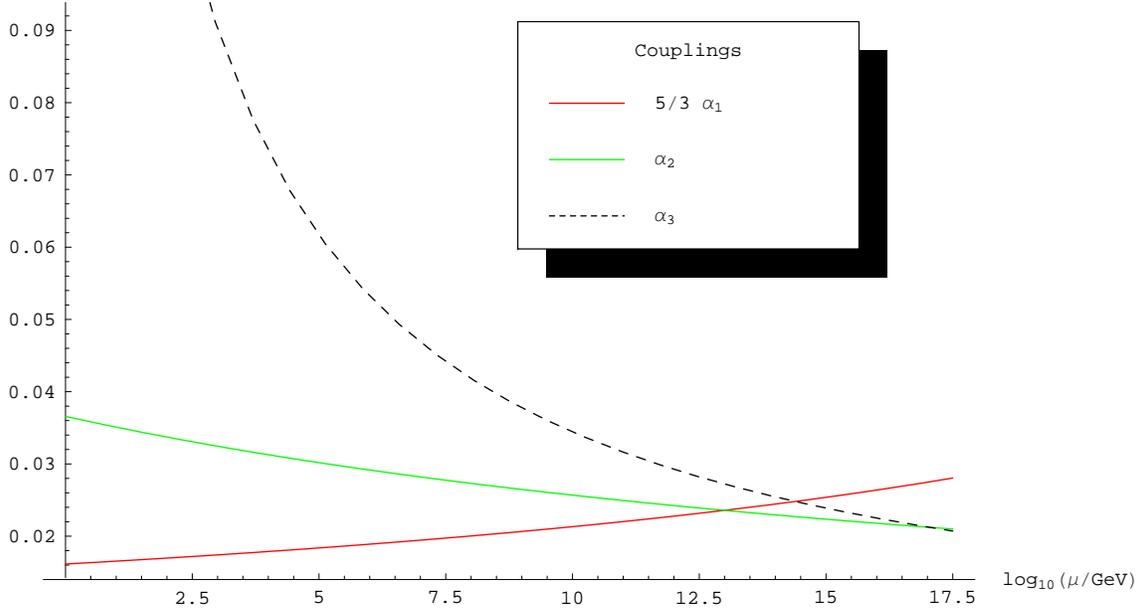}
\end{center}
\caption{The running of the three couplings. \label{runcoupling}}
\end{figure}

It is known that the predicted unification of the coupling constants
does not hold exactly, which points to the existence of new physics,
in contrast with the ``big desert'' hypothesis. In fact, if one
considers the actual experimental values
\begin{equation}\label{measuredalpha}
g_1(M_Z)=0.3575, \ \ \ g_2(M_Z)=0.6514, \ \ \ g_3(M_Z)=1.221 ,
\end{equation}
one obtains the values
\begin{equation}\label{measuredalphai}
\alpha_1(M_Z)=0.0101,\ \ \  \alpha_2(M_Z)=0.0337,\ \ \
\alpha_3(M_Z)=0.1186 .
\end{equation}
Thus, one sees that the graphs of the running of the three constants
$\alpha_i$ do not meet exactly, hence do not specify a unique
unification energy (\cf Figure \ref{runcoupling} where the
horizontal axis labels the logarithm in base $10$ of the scale
measured in GeV).

\subsection{The Higgs scattering parameter and the Higgs mass}
\hfill \medskip\label{secthiigsmass}

When written in terms of $\higgs$, and using \eqref{coeffymterm},
the quartic term
$$ \frac{f_0 }{ 2\,\pi^2}
\int\,b\, |\varphi|^4 \, \sqrt g \,d^4 x= \frac{\pi^2 }{2\,
f_0}\frac{b}{ a^2}\,\int\, |\higgs|^4 \, \sqrt g \,d^4 x
$$
gives a further relation in our theory between the $\tilde\lambda
|\higgs|^4$ coupling and the gauge couplings to be imposed at the
scale $\Lambda$. This is of the form
\begin{equation}\label{renhiggsmass1}
 \tilde\lambda(\Lambda) = \, g_{3}^2 \, \frac{b}{ a^2}.
\end{equation}

\smallskip

We introduce the following notation. For $v=\frac{2M}{g}$ we define
the elements $(y^\sigma_\cdot)$ with $\sigma=1,2,3$ the generation
index and $\cdot=u,d,\nu,e$ by the relation
\begin{equation}\label{ymlambda}
\frac{v}{\sqrt 2} (y^\sigma_\cdot)=(m^\sigma_\cdot),
\end{equation}
where the $(m^\sigma_\cdot)$ are defined as in
\eqref{massrelations}. In particular, $y^\sigma_u$ for $\sigma=3$
gives the top quark Yukawa coupling. We also set
\begin{equation}\label{tlogLambdamu}
t=\log(\frac{\Lambda}{M_Z}) \ \ \ \text{ and } \ \ \ \mu=M_Z e^t.
\end{equation}
We consider the Yukawa couplings $(y^\sigma_\cdot)$ as depending on
the energy scale through their renormalization group equation (\cf
\cite{ACKMPRW}, \cite{CEIN}, \cite{Pilaf}). We consider in
particular the top quark case $y^\sigma_u(t)$ for $\sigma=3$. The
running of the top quark Yukawa coupling $y_t=y^\sigma_u(t)$, with
$\sigma=3$, is governed by the equation (\cf \cite{Sher} equation
(2.143) and equation (A9) of \cite{ACKMPRW})
\begin{equation}\label{ytrun}
 \frac{dy_t }{ dt}  = \frac{1}{ 16\pi^2} \left[ \frac 92 y_t^3 - \left(  a \, g_1^2 + b \,
g_2^2 + c\,g_3^2 \right) y_t \right]\,,\quad (a,b,c)= (\frac{17 }{
12},\frac{9}{ 4},8)
\end{equation}

\smallskip

The relation \eqref{renhiggsmass1} could be simplified if we assume
that the top quark Yukawa coupling is much larger than all the other
Yukawa couplings. In this case equation \eqref{renhiggsmass1}
simplifies. In fact, one gets $a\sim  3\, m_{top}^2$ and $b\sim 3\,
m_{top}^4$, where $m_{top}=m^\sigma_u$, with $\sigma=3$ in the
notation of \eqref{massrelations}, so that
\begin{equation}\label{renhiggsmass2}
 \tilde\lambda ( \Lambda) \sim \frac 43\,\pi\,  \alpha_3 ( \Lambda) \, .
\end{equation}
This agrees with \cite{cc1} equation (3.31). In fact, the
normalization of the Higgs field there is as in the l.h.s.~of
\eqref{mohapvelt} which gives  $\lambda(\mu)=4\tilde\lambda(\mu)$,
with $\mu$ as in \eqref{tlogLambdamu}. In terms of the Higgs
scattering parameter $\alpha_h$ of the standard model,
\eqref{renhiggsmass2} reads
\begin{equation}\label{renhiggsmassbis}
 \alpha_h(\Lambda) \sim \frac 83\,
\end{equation}
which agrees with \cite{knecht} equation (1). Therefore, the value
of $\lambda=4\,\tilde \lambda$ at the unification scale of
$\Lambda=10^{17}$ GeV is $ \lambda_0 \sim 0.356$ showing that one
does not go outside the perturbation domain.

Equation \eqref{renhiggsmass2} can be used, together with the RG
equations for $ \lambda$ and $y^\sigma_u(t)$, with $\sigma=3$, to
determine the Higgs mass at the low-energy scale $M_Z$.

For simplicity of notation, in the following we write
\begin{equation}\label{yt}
y_t = y^\sigma_u(t), \ \ \ \text{ with } \ \ \sigma =3.
\end{equation}

We have (\cf \cite{Sher} equations (2.141), (2.142), (4.2) and the
formula (A15) of \cite{ACKMPRW}) the equation
\begin{equation}\label{evolutionkt}
  \frac{d\lambda }{ dt}  = \lambda  \gamma   +
\frac{1}{ 8\pi^2} (12\lambda^2 +B) \end{equation}
 where
\begin{eqnarray}\label{gammaB}
  \gamma  &=& \, \frac{1}{16\pi^2} (12\,y_t^2 - 9\, g_2^2 -
3\,g_1^2) \\
  B  &=& \,  \frac{3}{ 16} (3\,g_2^4
+ 2 g_1^2 \, g_2^2 +g_1^4) -3\,y_t^4  \, . \nonumber
\end{eqnarray}

\begin{figure}
\begin{center}
\includegraphics[scale=0.7]{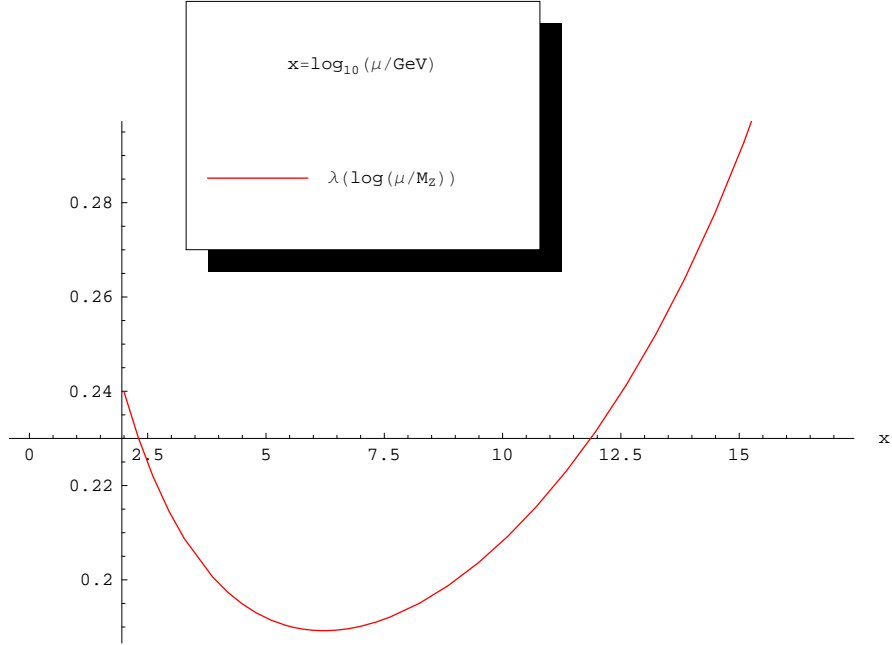}
\end{center}
\caption{The running of the quartic Higgs coupling.
\label{runninghiggs}}
\end{figure}

The Higgs mass is then given by
\begin{equation}\label{higgsmassren}
m_H^2=\,8\lambda\,\frac{M^2}{g^2}\,,\quad
m_H=\sqrt{2\lambda}\,\frac{2M}{g}
\end{equation}

The numerical solution to these equations with the boundary value
$\lambda_0= 0.356$ at $\Lambda=10^{17}$ GeV gives $\lambda(M_Z)\sim
0.241$ and a Higgs mass of the order of  $170$ GeV.  We refer to
\cite{Iochum} and to \cite{knecht} for the analysis of variants of
the model.

\begin{rem}\label{nutaurem}{\rm The estimate of equation \eqref{renhiggsmass2} is obtained
under the assumption that the Yukawa coupling for the top quark is
the dominant term and the others are negligible. However, due to the
see-saw mechanism discussed in \S \ref{sectseesaw} below, one should
expect that the Yukawa coupling for the tau neutrino is also large
and of the same order as the one for the top quark. Thus, the factor
of $4/3$ in \eqref{renhiggsmass2} should be corrected to $1$ as in
\eqref{yunu3} below. One can check by direct calculation that this
does not affect substantially the estimate we obtain for the Higgs
mass which is then around $168$ GeV.}\end{rem}

\bigskip
\subsection{Neutrino mixing and the see-saw mechanism}\hfill
\medskip\label{sectseesaw}

Let us briefly explain how the see-saw mechanism appears in our
context. Let $D=D(Y)$ be as in \eqref{dofm}. The restriction of
$D(Y)$ to the subspace of $\cH_F$ with basis the
$(\nu_R,\nu_L,\bar\nu_R,\bar\nu_L)$ is given by a matrix of the form
\begin{equation}\label{seesawmatrix}
\left[
  \begin{array}{cccc}
    0 & M_\nu^* & M_R^* & 0 \\
    M_\nu & 0 & 0 & 0 \\
    M_R & 0 & 0 & \bar{M}_\nu^* \\
    0 & 0 & \bar{M}_\nu & 0 \\
  \end{array}
\right]
\end{equation}
where $M_\nu=\frac{2M}{g} K_\nu$ with $K_\nu$ as in \eqref{Knu}.

The largest eigenvalue of $M_R$ is set to the order of the
unification scale by the equations of motion of the spectral action
as in the following result.

\begin{lem}\label{MRxkR}
Assume that the matrix $M_R$ is a multiple of a fixed matrix $k_R$,
\ie is of the form $M_R=x\,k_R$. In flat space, and assuming that
the Higgs vacuum expectation value is negligible with respect to
unification scale, the equations of motion of the spectral action
fix $x$ to be either $x=0$ (unstable) or satisfying
\begin{equation}\label{x2eq}
x^2=\frac{2\,f_2\,\Lambda^2\,\Tr(k_R^*k_R)}{f_0\,\Tr((k_R^*k_R)^2)}.
\end{equation}
\end{lem}

\proof The value of $x$ is fixed by the equations of motion of the
spectral action
\begin{equation}\label{eqmotion}
\partial_u \Tr(f(D_A/\Lambda))=0 ,
\end{equation}
with $u=x^2$.

One can see from \eqref{momentslabels} that $x$ only appears in the
coefficients $c$, $d$, and $e$. In the variation \eqref{eqmotion},
the terms in the spectral action \eqref{bossm} of Theorem
\ref{spectralactmf} containing the coefficient $c$ and $e$ produce
linear terms in $x^2$, proportional to the scalar curvature $R$ and
the square $|\varphi|^2$ of the Higgs vacuum expectation value, and
an additional linear term coming from the cosmological term. The
cosmological term also contains the coefficient $d$, which depends
quadratically on $x^2$. In flat space, and under the assumption that
$|\varphi|^2$ is sufficiently small, \eqref{eqmotion} then
corresponds to minimizing the cosmological term.

This gives
\begin{equation}\label{eqmotion2}
\partial_x(-f_2\,\Lambda^2\,c+\frac{
f_0}{4}\,d)=0,\ \ \ \  c=x^2\,\Tr(k_R^*k_R), \ \ \ \
d=x^4\,\Tr((k_R^*k_R)^2) .
\end{equation}
Thus, we get $M_R=x k_R$ with $x$ satisfying \eqref{x2eq}. In other
words we see that
\begin{equation}\label{mrstarmr}
M_R^*M_R=\frac{2\,f_2\,\Lambda^2}{f_0}\,\frac{k_R^*k_R\,\Tr(k_R^*k_R)}{\Tr((k_R^*k_R)^2)}
\end{equation}
\endproof

The Dirac mass $M_\nu$ is of the order of the Fermi energy $v$ and
hence much smaller. The eigenvalues of the matrix
\eqref{seesawmatrix} are then given, simplifying to one generation,
by
\begin{equation}\label{eigenMv}
\frac 12\,(\pm m_R\, \pm \sqrt{m_R^2+4\,v^2}),
\end{equation}
where $m_R$ denotes the eigenvalues of $M_R$, which is of the order
of $\Lambda$ by the result of Lemma \ref{MRxkR}, see
\eqref{mrstarmr}.

This gives two eigenvalues very close to  $\pm m_R$ and two others
very close to $\pm \frac{v^2}{m_R}$ as can be checked directly from
the determinant of the matrix \eqref{seesawmatrix}, which is equal
to $|M_\nu|^4\sim v^4$ (for one generation).

\begin{rem}\label{Sakh}{\rm
This is compatible with the scenario proposed by Fukigita and
Yanagida (\cf \cite{MoPa}) following the ideas of Sakharov and
t'Hooft, to explain the asymmetry between matter and antimatter in
the universe. }
\end{rem}

Typical estimates for the large masses of the right handed neutrinos
\ie the eigenvalues of $M_R$ are given (\cf \cite{MoPa}) by
\begin{equation}\label{mRmasses}
(m_R)_1\geq 10^7 GeV\,,\quad  (m_R)_2\geq 10^{12} GeV\,,\quad
(m_R)_3\geq 10^{16} GeV\,.
\end{equation}

\medskip
\subsection{The fermion--boson mass relation}\hfill
\medskip\label{sectmassrel}

There are two different normalizations for the Higgs field in the
literature.
\begin{enumerate}
  \item In Veltman \cite{VDiag} the kinetic term has a factor of $\frac 12$
  \item In Mohapatra--Pal it has a factor of $1$ (\cf \cite{MoPa} equation (1.43))
\end{enumerate}

One passes from one to the other by
\begin{equation}\label{mohapvelt}
\varphi_{mp}=\frac{1}{\sqrt 2}\,\varphi_{velt}
\end{equation}
In \cite{cc1} we used the second convention. Let us then stick to
that for the definition of the Yukawa couplings
$(y^\sigma_\cdot)(t)$ which is then given by \eqref{ymlambda} above.

The mass of the top quark is governed by the top quark Yukawa
coupling $y_t=y^\sigma_u(t)$ with $\sigma=3$ by the equation
\begin{equation}\label{topmass}
m_{top}(t)=\frac{1}{\sqrt 2}\frac{2M}{g}\,y_t =\frac{1}{\sqrt
2}\,v\,y_t ,
\end{equation}
where $v=\frac{2M}{g}$ is the vacuum expectation value of the Higgs
field.  The running of the top quark Yukawa coupling
$y_t=y^\sigma_u(t)$, with $\sigma=3$, is governed by  equation
\eqref{ytrun}.

\medskip

In terms of the Yukawa couplings $(y^\sigma_\cdot)$ of
\eqref{ymlambda}, the mass constraint \eqref{massrelation2} reads as
\begin{equation}\label{massrelation3}
\frac{v^2}{2}\sum_\sigma\,(y^\sigma_\nu)^2
+(y^\sigma_e)^2+3\,(y^\sigma_u)^2+3\, (y^\sigma_d)^2=\,2\,g^2\,v^2 ,
\end{equation}
with $v=\frac{2M}{g}$ the vacuum expectation value of the Higgs, as
above.

\medskip

In the traditional notation for the Standard Model the combination
$$
Y_2=\sum_\sigma\,(y^\sigma_\nu)^2+(y^\sigma_e)^2+3\,(y^\sigma_u)^2+3\,
(y^\sigma_d)^2 $$ is denoted by $Y_2=Y_2(S)$ (\cf \cite{ACKMPRW}).
Thus,  the mass constraint \eqref{massrelation2}  is of the form
\begin{equation}\label{massrelation3bis}
Y_2(S)=4\,g^2 .
\end{equation}

\begin{figure}
\begin{center}
\includegraphics[scale=0.6]{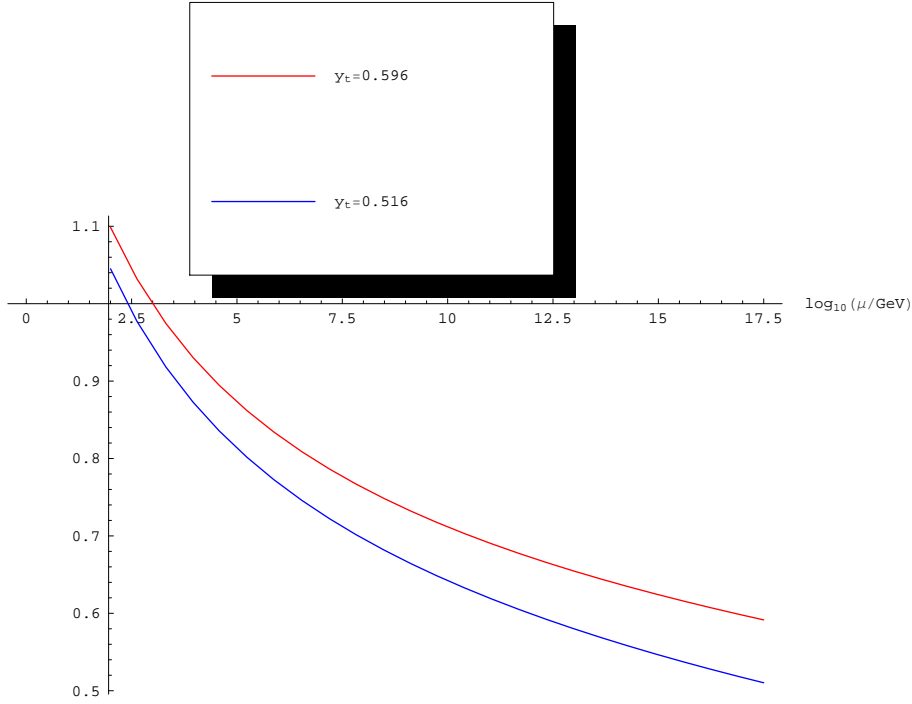}
\end{center}
\caption{The running of the top quark Yukawa coupling.
\label{running}}
\end{figure}

Assuming that it holds at a unification scale of $10^{17}$ GeV and
neglecting all other Yukawa couplings with respect to the top quark
$y^\sigma_u$, with $\sigma=3$, we get the following approximate form
of \eqref{massrelation2},
\begin{equation}\label{approx}
y^\sigma_u=\frac{2}{\sqrt 3}\,g , \ \ \ \text{ with } \ \sigma =3.
\end{equation}
The value of $g$ at a unification scale of $10^{17}$ GeV is $\sim
0.517$. Thus, neglecting the $\tau$ neutrino Yukawa coupling, we get
the simplified relation
\begin{equation}\label{approx1}
y_t=\,\frac{2}{\sqrt 3}\,g\sim 0.597\,,\quad t\sim 34.6 \, .
\end{equation}
Thus, in first approximation, numerical integration of the
differential equation \eqref{ytrun} with the boundary condition
\eqref{approx1} gives the value $y_0= \sim 1.102$ and a top quark
mass of the order of $\frac {1}{\sqrt 2} y_0\,v \sim \,173.683\,y_0$
GeV.

The see-saw mechanism, however, suggests that the Yukawa coupling
for the $\tau$ neutrino is of the same order as the top quark Yukawa
coupling. Indeed, even if the tau neutrino mass has an upper bound
of the order of (\cf \cite{MoPa})
$$ m_{\nu_\tau}\leq 18.2 \ \text{MeV}, $$
the see-saw mechanism allows for a large Yukawa coupling term by the
relation \eqref{eigenMv} and \eqref{mRmasses}. It is then natural to
take the Yukawa coupling $y^\sigma_{\nu}$, with $\sigma=3$ for the
tau neutrino to be the same, at unification, as that of the top
quark. This introduces in \eqref{approx1} a correction factor of
$\sqrt \frac 34$. In fact, for $x_t=y^\sigma_{\nu}(t)$ and
$y_t=y^\sigma_u(t)$, with $\sigma=3$, we now have
\begin{equation}\label{yunu3}
Y_2(S)\sim x_t^2 + 3 y_t^2 \sim \frac 43 \cdot 3 y_t^2=4
\,y_t^2\Rightarrow  y_t\sim g
\end{equation}
This has the effect of lowering the value of $y_0$ to $y_0\sim 1.04
$, which yields an acceptable value for the top quark mass, given
that we neglected all other Yukawa couplings except for the top and
the tau neutrino.

\medskip

\subsection{The gravitational terms}\hfill\medskip
\label{sectgravity}

We now discuss the behavior of the gravitational terms in the
spectral action, namely
\begin{equation}\label{Sgravterms}
\int  \biggl( \frac{1}{ 2\kappa_0^2} \, R + \alpha_0 \, C_{\mu \nu
\rho \sigma} \, C^{\mu \nu \rho \sigma} + \gamma_0 + \tau_0 \, R^*
R^*   - \xi_0\, R \,\vert \higgs \vert^2  \biggl)\sqrt g\, d^4 x.
\end{equation}

\begin{figure}
\begin{center}
\includegraphics[scale=0.7]{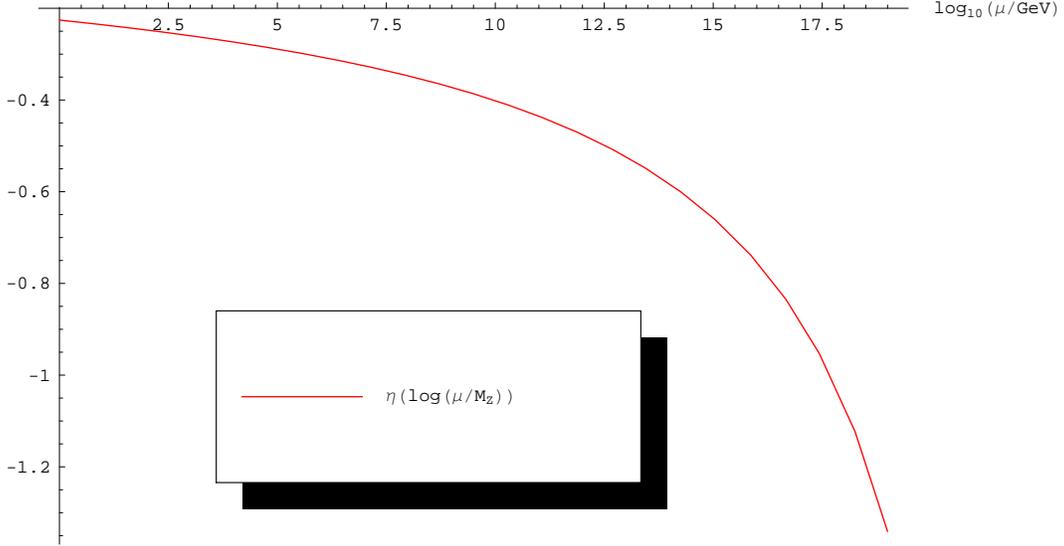}
\end{center}
\caption{The running of the Weyl curvature term in
\eqref{weylsquareterm1}. \label{runcouplinggrav1}}
\end{figure}

\begin{figure}
\begin{center}
\includegraphics[scale=0.75]{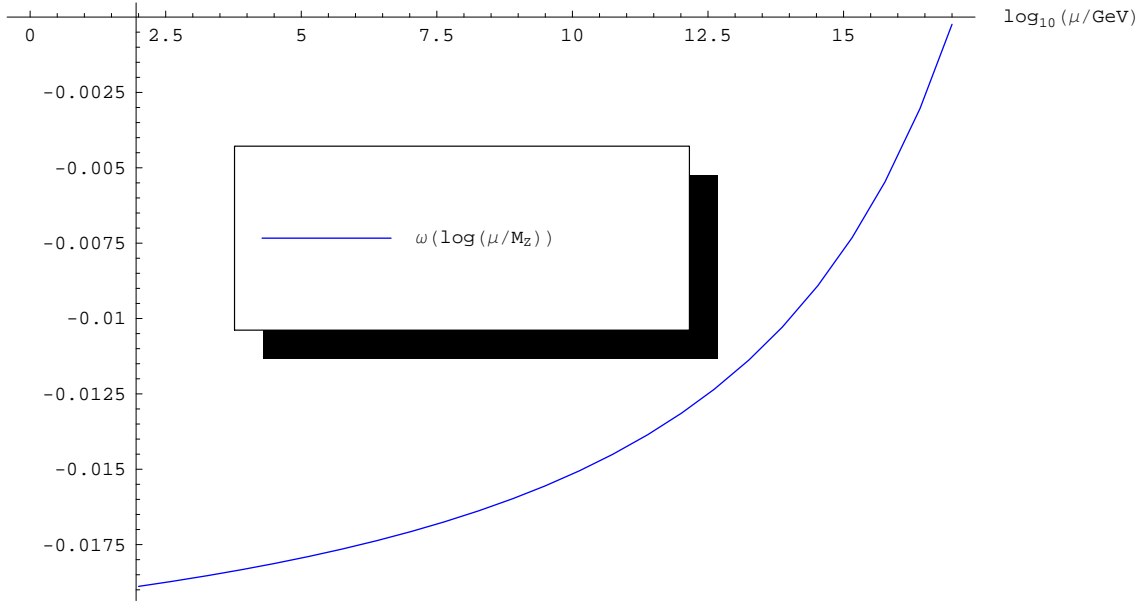}
\end{center}
\caption{The running of the ratio of the coefficients of the $R^2$
term and the Weyl term in \eqref{weylsquareterm1}.
\label{runcouplinggrav2}}
\end{figure}

\begin{figure}
\begin{center}
\includegraphics[scale=0.75]{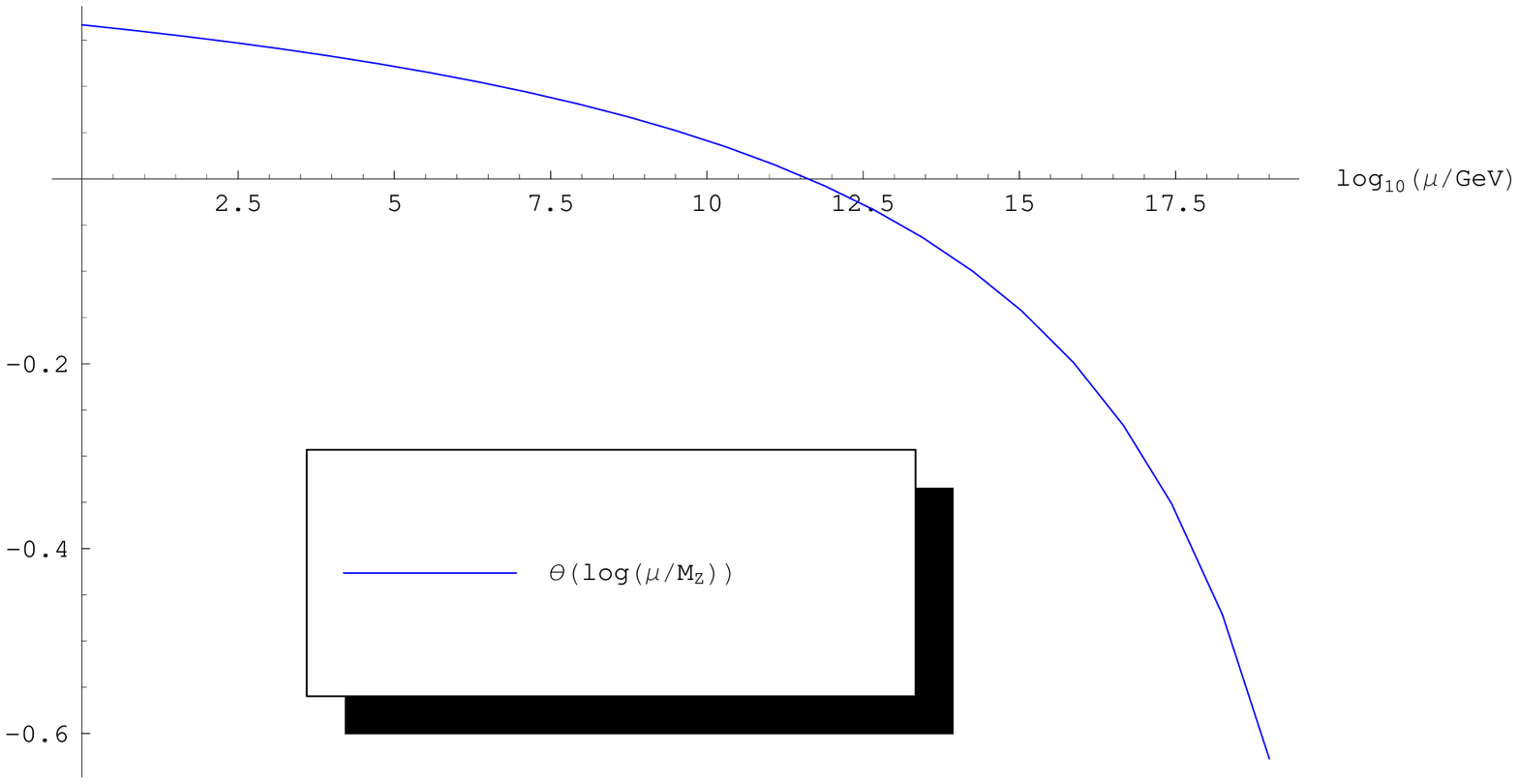}
\end{center}
\caption{The running of the ratio of the coefficients of the
topological term and the Weyl term in \eqref{weylsquareterm1}.
\label{runcouplinggrav3}}
\end{figure}

The traditional form of the Euclidean higher derivative terms that
are quadratic in curvature  is (see \eg \cite{Dono}, \cite{CoPer})
\begin{equation}\label{weylsquareterm1}
\int \  \, \biggl( \frac{1}{2\eta}\,C_{\mu \nu \rho \sigma} \,
C^{\mu \nu \rho
\sigma}-\frac{\omega}{3\eta}\,R^2+\frac{\theta}{\eta}\,E\,\biggr)\,
\sqrt g\,d^4 x ,
\end{equation}
with $E=R^* R^*$ the topological term which is the integrand in the
Euler characteristic
\begin{equation}\label{eulerchar}
\chi(M)=\ \frac{1}{32 \pi^2}\int \ E\,\sqrt g\,d^4 x =\frac{1}{32
\pi^2} \int \ \, R^* R^*  \sqrt g\,d^4 x
\end{equation}
 The running of the coefficients of the Euclidean
higher derivative terms in \eqref{weylsquareterm1}, determined by
the renormalization group equation, is gauge independent and is
given by (see \eg \cite{Avramidi} equations 4.49 and 4.71 and
\cite{Dono}, \cite{CoPer})
\begin{align*}
\beta_{\eta} &  =-\frac{1}{\left(  4\pi\right)  ^{2}}\frac{133}{10}\eta^{2}\\
\beta_{\omega} &  =-\frac{1}{\left(  4\pi\right)
^{2}}\frac{25+1098\,\omega
+200\,\omega^{2}}{60}\eta\\
\beta_{\theta} &  =\frac{1}{\left(  4\pi\right)  ^{2}}\frac{7\left(
56-171\,\theta\right)  }{90}\eta
\end{align*}
while the graphs are shown in Figures \ref{runcouplinggrav1},
\ref{runcouplinggrav2}, \ref{runcouplinggrav3}. Notice that the
infrared behavior of these terms approaches the fixed point
$\eta=0$, $\omega=-0.0228$,  $\theta=0.327$. The coefficient $\eta$
goes to zero in the infrared limit, sufficiently slowly, so that, up
to scales of the order of the size of the universe, its inverse
remains $O(1)$. On the other hand, $\eta(t)$, $\omega(t)$ and
$\theta(t)$ have a common singularity at an energy scale of the
order of $10^{23}$ GeV, which is above the Planck scale. Moreover,
within the energy scales that are of interest to our model $\eta(t)$
is neither too small nor too large (it does not vary by more than a
single order of magnitude between the Planck scale and infrared
energies).

\medskip

The only known experimental constraints on the values of the
coefficients of the quadratic curvature terms $R_{\mu\nu}R^{\mu\nu}$
and $R^2$ at low energy are very weak and predict that their value
should not exceed $10^{74}$ (\cf \eg \cite{Dono}). In our case, this
is guaranteed by the running described above. Note that we have
neglected the coupling $R\,\higgs^2$ with the Higgs field which
ought to be taken into account in a finer analysis.

\medskip

The coefficient of the Einstein  term is of the form
\begin{equation}\label{coeffeinsteinterm1}
  \frac{1}{ \kappa_0^2} = \  \frac{96\,f_2\,\Lambda^2
-f_0\,c}{ 12\pi^2} .
\end{equation}

With the above notation, by the result of Lemma \ref{MRxkR}, we get
$$
c=\,x^2\,\Tr(k_R^*k_R)=
\frac{2\,f_2\,\Lambda^2\,(\Tr(k_R^*k_R))^2}{f_0\,\Tr((k_R^*k_R)^2)}.
$$
Thus, the range of variation of $\frac{ (\Tr(k_R^*k_R))^2}{
\Tr((k_R^*k_R)^2)}$ for $N$ generations is the interval $[1,N]$. In
particular, with $N=3$ we get
\begin{equation}\label{coeffeinsteinterm}
\frac{90\,f_2\,\Lambda^2}{ 12\pi^2}\leq\frac{1}{ \kappa_0^2}\leq
\frac{94\,f_2\,\Lambda^2}{ 12\pi^2}
\end{equation}
This estimate is not modified substantially if one takes into
account the contribution from the $R\,\higgs^2$ term using  the
vacuum expectation value of the Higgs field. Thus we see that
independently of the choice of $k_R$ the coefficient $\kappa_0^{-2}$
of the Einstein term $\frac 12\,\int\,R\,\sqrt g\,d^4x$ is positive
and of the order of $f_2\,\Lambda^2$. Thus the result is similar to
what happened for the Einstein-Yang-Mills system \cite{cc2} and the
sign is the correct one. As far as the size is concerned let us now
compare the value we get for $\kappa_0$ with the value given by
Newton's constant. In our case we get
$$
\kappa_0^{-1}\sim \Lambda\,\sqrt f_2
$$
Thus if we take for $\Lambda$ the energy scale of the meeting point
of the electroweak and strong couplings, namely $\Lambda\sim
1.1\times 10^{17}$ GeV, we get
$$
\kappa_0^{-1}\sim 1.1\times 10^{17}\,\sqrt f_2\,{\rm GeV}
$$
On the other hand using the usual form of the gravitational action
\begin{equation}\label{EHaction1}
S(g)=\frac{1}{16\pi G}\, \int_M R \, dv \,,
\end{equation}
 and the experimental value of Newton's
constant at ordinary scales one gets the coupling constant
$$
\kappa_0(M_Z)=\,\sqrt{8\pi G}\,,\quad \kappa_0^{-1}\sim
1.221\,10^{19}/\sqrt{8\pi}\sim 2.43\times 10^{18} \,{\rm GeV}.
$$

One should expect that the Newton constant runs at higher energies
(\cf \eg \cite{Dono}, \cite{Perc}, \cite{PaTo}) and increases at
high energy when one approaches the Planck scale. Thus the ratio
\begin{equation}\label{EHactionrun}
\rho=\kappa_0(\Lambda)/\kappa_0(M_Z)
\end{equation}
for $\Lambda\sim 1.1\times 10^{17}$ GeV, which measures the running
at unification scale, should be larger than $1$.

\smallskip

By the normalization of the kinetic terms of the gauge fields, one
has \eqref{coeffymterm}
 $$
f_0=\,\frac{\pi^2}{2\,g^2}=\,\frac{\pi}{8\,\alpha_2(\Lambda)}\sim
18.45 \, .
$$
Thus
$$
1.1\times 10^{17}\,\sqrt f_0 \sim  4.726\times 10^{17}
$$
It follows  that if
\begin{equation}\label{EHactionrun1}
f_2/f_0= \tau^2/\rho^2\,, \  \  \tau   \sim 5.1
\end{equation}
 one obtains the correct physical value for
the Newton constant. In fact starting with a test function  $g$ such
that $g_2=g_0$, equality \eqref{EHactionrun1} holds provided one
performs the transformation
$$
g\mapsto f\,,\quad f(x)=g(\frac{\rho \, x}{\tau})\,.
$$

\smallskip

\bigskip

\bigskip

\subsection{The cosmological term}\label{cosmosect}\hfill\medskip

The cosmological term depends, in our model, on the remaining
parameter $f_4$.

\begin{lem}\label{cosmocon}
Under the hypothesis of Lemma \ref{MRxkR}, the cosmological term
gives
\begin{equation}\label{cosmoterm}
\frac{1}{\pi^2} \left( 48 f_4 - \frac{\Tr (k_R^* k_R)^2}{\Tr((k_R^*
k_R)^2)} \frac{f_2^2}{f_0}\right) \Lambda^4.
\end{equation}
\end{lem}

\proof In \eqref{cosmologicalterm} we have the cosmological term
$$ \frac{1}{\pi^2}
\left( 48 f_4 \Lambda^4 - f_2 c \Lambda^2 + \frac{f_0}{4} d \right),
$$ where the coefficients $c$ and $d$ are given by
$$ c= \Tr(\mass_R^*\mass_R) \ \ \ \text{ and } \ \ \ d= \Tr((\mass_R^*\mass_R)^2)). $$
We use the result of Lemma \ref{MRxkR} and \eqref{mrstarmr}. We
obtain
$$ c= \frac{2\,f_2\,\Lambda^2}{f_0} \frac{\Tr(k_R^*k_R)^2} {\Tr((k_R^*k_R)^2)} \ \ \
\text{ and } d= \frac{4 f_2^2 \Lambda^4}{f_0^2}
\frac{\Tr(k_R^*k_R)^2} {\Tr((k_R^*k_R)^2)}. $$
\endproof

\smallskip

The positivity of the $f_j$, and the freedom in choosing the $f_4$
makes it possible to adjust the value of the cosmological term.
Notice that, if one assumes that the function $f$ is decreasing (and
positive as usual), then the Schwartz inequality gives the
constraint
$$ f_2^2 \leq f_0 f_4. $$
The Schwartz inequality also gives the estimate
$$ \frac{\Tr(k_R^*k_R)^2} {\Tr((k_R^*k_R)^2)} \leq 3 $$
in \eqref{cosmoterm}. Thus, for a decreasing positive function, this
cosmological term is positive. Of course to obtain the physical
cosmological constant one needs to add to this term the contribution
from the vacuum expectation value of the various fields which give
an additional contribution of the order of
$(96-28)\frac{1}{32\pi^2}\Lambda^4$ and generate a fine tuning
problem to ensure that the value of the cosmological constant at
ordinary scale is small. It is natural in this context to replace
the cut-off $\Lambda$ by a dynamical dilaton field as in \cite{cc3},
\cf \S \ref{dilrem} below.

\bigskip

\subsection{The tadpole term and the naturalness problem}\label{tadpolesect}
\hfill\medskip

The naturalness problem for the standard model arises from the
quadratically divergent corrections to the tadpole term
\begin{equation}\label{tadpoleterm1}
 \delta\beta_h\sim \Lambda^2\,\sum \,c_n\,\log(\Lambda/M_Z)^n
\end{equation}
that are required in order to maintain the Higgs vacuum expectation
value at the electroweak scale (\cf \cite{Reina} \S II.C.4). In our
set-up the only natural scale is the unification scale. Thus, an
explanation for the weak scale still remains to be found. We shall
not attempt to address this problem here but make a few remarks.

\subsubsection{Naturalness and fine tuning}

 When the cutoff
regularization method is used a number of diagrams involving the
Higgs fields are actually quadratically divergent and thus generate
huge contributions to the tadpole bare term. To be more specific,
one has the following quadratically divergent diagrams:
\begin{itemize}
  \item Minimal coupling with $W$ and $B$ fields
  \item Quartic self-coupling of Higgs fields
  \item Yukawa couplings with fermions
  \end{itemize}
If we want to fix the Higgs vacuum at $\frac{2M}{g}$ in the standard
model we need to absorb the huge quadratic term in $\Lambda$  in the
tadpole term of the action. The tadpole constant $\beta_h$ then
acquires a quadratically divergent contribution
\begin{equation}\label{tadpoleterm1}
\frac 12 \delta\beta_h\sim \frac{\Lambda^2}{ 32\pi^2}\,q(t) \,, \
t=\log(\Lambda/M_Z)\,,
\end{equation}
where (\cf \cite{EJ}, \cite{Kolda}, \cite{Reina})
\begin{equation}\label{barehiggsmasscutoff1}
q(t)= \, \frac{9}{ 4} \, g_2^2 + \frac{3}{ 4} \, g_1^2 + 6\lambda -
6\,y_t^2 \,,
\end{equation}
where, as above, $y_t=y^\sigma_u(t)$, with $\sigma=3$ is the top
quark Yukawa coupling. This form of \eqref{barehiggsmasscutoff1}
holds under the assumption that the contribution coming from the top
quark is the dominant term in the Yukawa coupling (see, however, the
previous discussion on the term $y^\sigma_\nu(t)$ with $\sigma=3$ in
\S \ref{sectmassrel}).

\begin{figure}
\begin{center}
\includegraphics[scale=0.8]{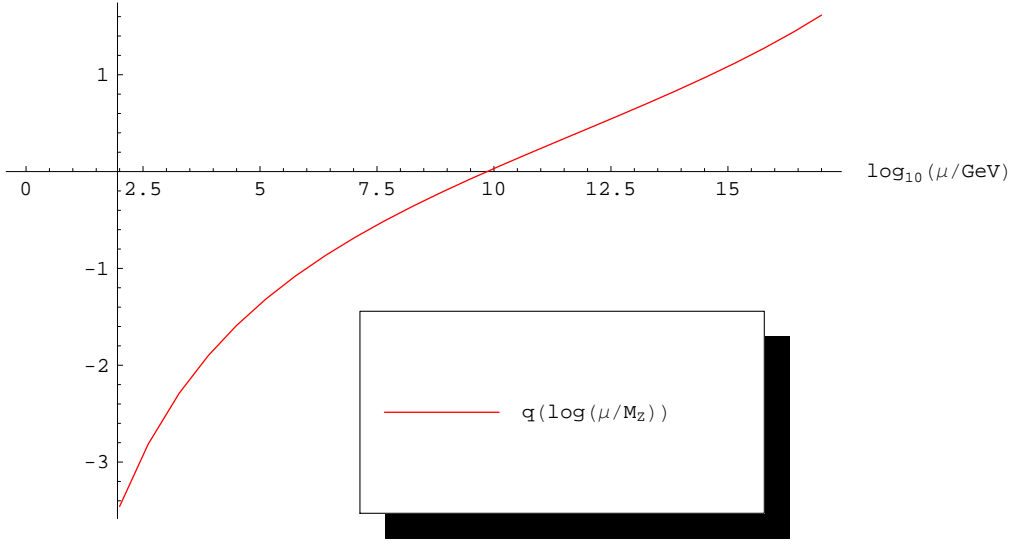}
\end{center}
\caption{The running of the tadpole term. \label{runningmass}}
\end{figure}

One can check that the contribution $y_t$ is sufficiently large in
the standard model so that, for small $t$, $q(t)$ is negative.
However, as shown in Figure \ref{runningmass} the expression $q(t)$
changes sign at energies of the order of $10^{10}$ GeV, and is then
positive, with a value at unification $\sim 1.61$.

\smallskip
While the plot \ref{runningmass} uses the known experimental values,
one can show directly that our boundary conditions at unification
scale $t_{unif}$ also imply that $q(t_{unif})>0$.  In fact it is
better to replace $3y_t^2$ by $Y_2$ and we can then use our mass
relation at unification in the form \eqref{massrelation3bis}
$$
Y_2=4\,g^2
$$
Also at unification we have a precise form of $\lambda$ namely
\eqref{renhiggsmass1}, together with $\lambda=4\tilde\lambda$ and
get
$$
\lambda=\,4\,g^2\,\frac{b}{a^2}
$$
We can thus rewrite \eqref{barehiggsmasscutoff1} as (with $g=g_2$)
\begin{equation}\label{barehiggsmasscutoffbis}
q(t_{unif})= \, \frac{9}{ 4} \, g^2 + \frac{3}{ 4} \, g_1^2 +
24\,g^2\,\frac{b}{a^2} - 8\,g^2 \,,
\end{equation}
We can now use the inequality
$$
\frac{b}{a^2}\geq \frac 14
$$
which holds even with a large tau neutrino Yukawa coupling, to get
\begin{equation}\label{barehiggsmasscutoffbis1}
q(t_{unif})\geq \, \frac{9}{ 4} \, g^2 + \frac{3}{ 4} \, g_1^2 +
24\,g^2\,\frac{1}{4} - 8\,g^2 =\frac{1}{ 4} \, g^2 + \frac{3}{ 4} \,
g_1^2> 0\,.
\end{equation}

\smallskip

\subsubsection{Sign of the quadratic term}

In the spectral action we also have a similar term which is
quadratic in $\Lambda$ namely the term $-\mu_0^2\,\higgs^2$ of
\eqref{rescaledbosaction} where $\mu_0^2 = 2\frac{f_2\,\Lambda^2}{
f_0}-\frac{e}{a}$.  We show   that, under the simplifying hypothesis
of Lemma \ref{MRxkR}, the coefficient of $\Lambda^2$ in $\mu_0^2$ in
the spectral action is generally positive but can be small and have
an arbitrary sign provided there are at least two generations and
one chooses suitable Yukawa and Majorana mass matrices. The reason
why we can use Lemma \ref{MRxkR} is that we are interested in small
values of $\mu_0^2$, a more refined analysis would be required to
take care of the general case. By Lemma \ref{MRxkR} we have $M_R=x
k_R$ with $x$ as in \eqref{x2eq}.

\begin{lem}
Under the hypothesis of Lemma \ref{MRxkR}, the coefficient of the
Higgs quadratic term $-\mu_0^2\,\higgs^2$ in the spectral action is
given by
  \begin{equation}\label{muzerosimple}
\mu_0^2=\,2\,\Lambda^2\,\frac{f_2}{f_0}\,(1-X)=\,(1-X)\,\frac{4\,g^2\,\Lambda^2}{\pi^2}\,f_2
\end{equation}
 where
\begin{equation}\label{muzerosimple1}
X=\frac{\Tr(k_R^*k_R\,k_\nu^*k_\nu)\,\Tr(k_R^*k_R)}{\Tr(k_\nu^*k_\nu
+k_e^*k_e+3(k_u^*k_u+k_d^*k_d))\Tr((k_R^*k_R)^2)}
\end{equation}
\end{lem}

\proof One has $\mu_0^2 = \ 2\,\frac{f_2\,\Lambda^2}{
f_0}-\frac{e}{a}$ with $e$ and $a$ as in \eqref{momentslabels}.

Using \eqref{mrstarmr} and \eqref{rescaledyukawamasses} we then get
the first equality in \eqref{muzerosimple}. The second follows from
\eqref{coeffymterm}.
\endproof

\medskip
In order to compare $X$ with $1$ we need to determine the range of
variation of the largest eigenvalue of
$\frac{k_R^*k_R\,\Tr(k_R^*k_R)}{\Tr((k_R^*k_R)^2)}$ as a function of
the number of generations.

\begin{lem} \label{rangeofvar} The range of variation of the largest eigenvalue,
$$
\rho(k_R)=\,||\frac{k_R^*k_R\,\Tr(k_R^*k_R)}{\Tr((k_R^*k_R)^2)}||
$$
for $k_R\in M_N(\C)$, is the interval
$$
[1,\frac 12(1+\sqrt N)]
$$
\end{lem}

\medskip

\proof Notice first that one has
$$
\Tr((k_R^*k_R)^2)\leq \Tr(k_R^*k_R)\,||k_R^*k_R ||
$$
so that the inequality $\geq 1$ follows. Moreover this lower bound
is reached exactly when $k_R^*k_R$ is a multiple of an idempotent
which means that $k_R$ is a multiple of a partial isometry. To
understand the upper bound,
 we can assume that
$k_R^*k_R$ is diagonal with eigenvalues $\lambda_j^2$. We just need
to understand the range of variation of $
F_N(\lambda)=\frac{\lambda_1^2\,\sum\,\lambda_j^2}{\sum\,\lambda_j^4}
$. Using Lagrange multipliers one gets that, at an extremum, all the
$\lambda_j^2$ for $j\neq 1$ are equal. Thus, one just needs to get
the range of variation of the simpler function
$f_N(u)=\frac{u^2(u^2+N-1)}{u^4+N-1}$. Computing the value of $f_N$
at the maximum $u^2=1+\sqrt{N}$ yields the required answer.
\endproof

\medskip

We thus see that the maximal value for $X$ obtained by replacing
$k_R^*k_R$ by its maximal eigenvalue, yields the inequality
\begin{equation}\label{basicinequ}
 X\leq \,\frac{(1+\sqrt
N)}{2}\,\frac{\Tr(k_\nu^*k_\nu)}{\Tr(k_\nu^*k_\nu
+k_e^*k_e+3(k_u^*k_u+k_d^*k_d))}
\end{equation}

As we show now the range of variation of the simplified quadratic
term (\ie  the right hand side of equation \eqref{muzerosimple})
depends on the number $N$ of generations.
\medskip

\begin{prop} \label{propnumberofgen} Let $N$ be the number of generations.
\begin{enumerate}
  \item If $N=1$, or if $k_R$ is a scalar multiple of a partial isometry,
  the quadratic term \eqref{muzerosimple} is positive and its size of the order
  of $\frac{f_2\,\Lambda^2}{f_0}$.
  \item If $N\geq 2$, the quadratic term \eqref{muzerosimple} can vanish
  and have arbitrary sign, provided one chooses $k_R$, $k_\nu$ appropriately.
\end{enumerate}
\end{prop}

\proof 1) By lemma \ref{rangeofvar}  we have $\rho(k_R)=1$ and thus
by \eqref{muzerosimple1},
$$
X\leq \frac{\Tr(k_\nu^*k_\nu)}{\Tr(k_\nu^*k_\nu
+k_e^*k_e+3(k_u^*k_u+k_d^*k_d)) }<1
$$
 Thus
 $$
 \mu_0^2=\,2\,\Lambda^2\,\frac{f_2}{f_0}\,(1-X)
 $$
  is
positive and of the same order as $\frac{f_2\,\Lambda^2}{f_0}$.

  \smallskip

  2) We take $N=3$ and explain how to choose $k_R$, $k_\nu$ etc...
  so that the coefficient of the quadratic term vanishes. We choose
  $k_R$ such that the eigenvalues of $k_R^*k_R$ are of the form
  $(1+\sqrt 3,1,1)$. Then as in lemma \ref{rangeofvar} the
  eigenvalues of $\frac{k_R^*k_R\,\Tr(k_R^*k_R)}{\Tr((k_R^*k_R)^2)}$
  are $\frac 12(1+\sqrt 3,1,1)$. We can now choose
  $k_\nu$ in such a way that it is is diagonal in the same basis as
  $k_R^*k_R$ with a single order one eigenvalue on the first basis vector
  while the two other eigenvalues are small. It
  follows that
  $$
  X\sim \frac 12(1+\sqrt 3)\frac{\Tr(k_\nu^*k_\nu)}{\Tr(k_\nu^*k_\nu
+k_e^*k_e+3(k_u^*k_u+k_d^*k_d)) }\sim 1
  $$
provided that
$$
\frac 12(\sqrt 3 -1)\,\Tr(k_\nu^*k_\nu)\sim \Tr(
k_e^*k_e+3(k_u^*k_u+k_d^*k_d))
$$
\endproof

\medskip
Neglecting the Yukawa couplings except for the tau neutrino and the
top quark, one gets $k_{\nu_\tau} \sim 2.86 \,k_{top}$. While the
seesaw mechanism allows for a large Yukawa matrix for the neutrinos,
the above relation yields a Yukawa coupling for the tau neutrino
which is quite a bit larger than the expected one as in GUT theories
where it is similar to the top Yukawa coupling. In summary we have
shown that  $\mu_0^2>0$ except under the above special choice of
Yukawa coupling matrices. We have been working under the simplifying
hypothesis of Lemma \ref{MRxkR} and to eliminate that, a finer
analysis involving the symmetry breaking of the potential in the
variables $x$ and $\varphi$ (after promoting $x$ to a scalar field)
would be necessary.

\smallskip
\subsubsection{The dilaton field}\label{dilrem}
In fact there is another scalar field which plays a natural role in
the above set-up and which has been neglected for simplicity in the
above discussion. Indeed as in \cite{cc3} it is natural when
considering
 the spectral action (in particular on non-compact spaces) to replace
 the cut-off $\Lambda$ by a dynamical dilaton field. We refer to \cite{cc3}
 for the computation of the spectral action with dilaton and its
 comparison with the Randall-Sundrum model. Its extension to the
 present set-up is straightforward using the technique of
 \cite{cc3}. One obtains a model which is closely related to the
 model of scale invariant extended inflation of \cite{kolb}.

 \smallskip
\subsubsection{Geometric interpretation}\label{cosmorem}

Our geometric interpretation of the standard model gives a picture
of space-time as the product of an ordinary spin manifold (in
Euclidean signature) by a finite noncommutative geometry F. The
geometry of $F$ is specified by its Dirac operator $D_F$ whose size
is governed by the vacuum expectation value of the Higgs field. In
other words it is the (inverse of the) size of the space $F$ that
specifies the electroweak scale. It is thus tempting to look for an
explanation for the smallness of the ratio $M_Z/M_P$ along the same
lines as inflation as an explanation for the large size of the
observable universe in Planck units.

\bigskip

\section{Appendix: Gilkey's Theorem}\hfill
\medskip\label{gilkeysthm}

The square of the Dirac operator appearing in the spectral triple of
a noncommutative space is written in a form suitable to apply the
standard local formulas for the heat expansion (see \cite{Gilkey} \S
4.8). We   now briefly recall the statement of Gilkey's Theorem
(\cite{Gilkey} Theorem 4.8.16). One starts with a compact Riemannian
manifold $M$ of dimension $m$, with metric $g$ and one lets $F$ be a
vector bundle on $M$ and $P$ a differential operator acting on
sections of $F$ and with leading symbol given by the metric tensor.
Thus locally one has,
\begin{equation}
P=-\left(
g^{\mu\nu}I\,\partial_{\mu}\partial_{\nu}+A^{\mu}\partial_{\mu
}+B\right)  ,\label{opdata}
\end{equation}
where $g^{\mu\nu}$ plays the role of the inverse metric, $I$ is the
unit matrix, $A^{\mu}$ and $B$ are endomorphisms of the bundle $F$.
The Seeley-De witt coefficients are the terms $a_n (x,P)$ in the
heat expansion, which is of the form
\begin{equation}\label{heat1}
\Tr \,e^{-tP} \sim \sum_{n\geq 0} t^{\frac{n-m}{ 2}} \int_M \,a_n
(x,P) \, dv (x)
\end{equation}
where $m$ is the dimension of the manifold  and $dv (x) = \sqrt{
\det\,g_{\mu \nu}} \, d^m \, x$ where $g_{\mu \nu}$ is the metric on
$M$.

\medskip
By Lemma 4.8.1 of \cite{Gilkey} the operator $P$ is uniquely written
in the form
\begin{equation}\label{canP}
P=\,\nabla^*\,\nabla-\,\cE
\end{equation}
where $\nabla$ is a connection on $F$, $\nabla^*\,\nabla$ the
connection Laplacian and where $\cE$ is an endomorphism of $F$. The
explicit formulas for the connection $\nabla$ and the endomorphism
$\cE$ are
\begin{equation}\label{nablamu}
\nabla_\mu=\,\partial_\mu\,+\,\omega'_\mu
\end{equation}
\begin{equation}\label{omegamu}
\omega'_{\mu} = \frac 12 \, g_{\mu \nu} ({ A}^{\nu} +
\Gamma^{\nu}\cdot {\rm id} )
\end{equation}
\begin{equation}\label{endE}
 \cE = { B} - g^{\mu \nu} (\partial_{\mu} \, \omega'_{\nu}
+ \omega'_{\mu} \, \omega'_{\nu} - \Gamma_{\mu \nu}^{\rho} \,
\omega'_{\rho})
\end{equation}
Where one lets $\Gamma_{\mu\nu}^{\rho}\left( g\right)  $ be the
Christoffel symbols of the Levi-Civita connection of the metric $g$
and
$$
\Gamma^{\rho}\left(  g\right)
=g^{\mu\nu}\Gamma_{\mu\nu}^{\rho}\left( g\right)
$$
One lets $\Omega$ be the curvature of the connection $\nabla$ so
that (\cf \cite{Gilkey} Lemma 4.8.1),
\begin{equation}\label{curvnabla}
\Omega_{\mu \nu} = \partial_{\mu} \, \omega'_{\nu} - \partial_{\nu}
\, \omega'_{\mu} + [\omega'_{\mu} ,\, \omega'_{\nu}]
\end{equation}

The Seeley-de Witt coefficients $a_n (P)$ vanish for odd values of
$n$. The first three $a_n$'s for $n$ even have the following
explicit form in terms of the Riemann curvature tensor $R$, the
curvature $\Omega$ of the connection $\nabla$ and the endomorphism
$\cE$,

\medskip
\begin{thm}\label{gilkeythm}  \cite{Gilkey} One has :
\begin{eqnarray}
  a_0 (x,P) &=& (4\pi)^{-m/2} \Tr ({\rm id}) \\
  a_2 (x,P) &=& (4\pi)^{-m/2} \Tr \left(-\frac{R}{ 6} \, {\rm id} +\cE \right) \\
  a_4 (x,P)  &=& (4\pi)^{-m/2}\frac{1 }{ 360} \Tr (-12 R;_{\mu}
{}^{\mu} + 5R^2 - 2R_{\mu \nu} \, R^{\mu \nu}  \\
  &+& 2R_{\mu \nu
\rho \sigma} \, R^{\mu \nu \rho \sigma} - 60 \,R\,\cE + 180 \,\cE^2
+ 60
\,\cE;_{\mu} {}^{\mu} \nonumber \\
   &+& 30 \,\Omega_{\mu \nu} \, \Omega^{\mu \nu})\nonumber
\end{eqnarray}
\end{thm}

\begin{rem} \label{eminusr} Notice that $\cE$ only appears through the
terms
\begin{equation}\label{eminusr1}
\Tr \left(-\frac{R}{ 6} \, {\rm id} +\cE \right)\,,\quad \Tr
\left((-\frac{R}{ 6} \, {\rm id} +\cE)^2 \right)
\end{equation}
and the boundary term $\Tr(\cE;_{\mu} {}^{\mu})$.
\end{rem}

Here $R;_{\mu}{}^{\mu}=\nabla_\mu \nabla^\mu R$ and similarly
$\cE;_{\mu} {}^{\mu}=\nabla_\mu \nabla^\mu \cE$.

\medskip
\subsection{The generalized Lichnerowicz formula}\label{lichneform}\hfill\medskip

Let $M$ be a compact Riemannian spin manifold of dimension $m$, $S$
the spinor bundle with the canonical riemannian connection
$\nabla_S$. Let $V$ be a hermitian vector bundle over $M$ with a
compatible connection $\nabla_V$. One lets $\dirac_V$ be the Dirac
operator on $S\otimes V$ endowed with the tensor product connection
(\cite{Lawson} Proposition 5.10)
\begin{equation}\label{compoundconnection}
\nabla(\xi\otimes v)=\,(\nabla_S\,\xi)\otimes v+\,\xi \otimes
(\nabla_V\,v)
\end{equation}
Let then $R_V$ be the bundle endomorpism of the bundle $S\otimes V$
defined by
\begin{equation}\label{curvendo}
R_V(\xi\otimes v)=\,\frac
12\,\sum_{j,\,k=1}^m\,(\gamma_j\gamma_k\,\xi)\otimes (R(V)_{jk}\,v)
\end{equation}
where $R(V)$ is the curvature tensor of the bundle $V$.

\smallskip
One then has (\cite{Lawson} Theorem 8.17)

\begin{thm} let $s=-R$ be the scalar curvature of $M$, then the Dirac operator $\dirac_V$
satisfies
\begin{equation}\label{licnegen}
\dirac_V^2=\nabla^*\,\nabla+ \frac 14\,s+\,R_V
\end{equation}
where $\nabla^*\,\nabla$ is the connection Laplacian of $S\otimes
V$.
\end{thm}

\smallskip
Notice that all three terms of the right hand side of
\eqref{licnegen} are self-adjoint operators by construction. In
particular $R_V$ is self-adjoint. One can write $R_V$ in the
following form where the terms in the sum are pairwise orthogonal
for the natural inner product on the Clifford algebra (induced by
the Hilbert-Schmidt inner product $\langle A,B\rangle=\Tr(A^*B)$ in
the spin representation)
\begin{equation}\label{curvendo1}
R_V =\,\sum_{j<k}\,\gamma_j\gamma_k \otimes R(V)_{j\,k}
\end{equation}

\bigskip
\subsection{The asymptotic expansion and the residues}
\hfill\medskip

The spectral action can be expanded in decreasing powers of the
scale $\Lambda$ in the form
\begin{equation}
\mathrm{Trace}\,(f(D/\Lambda))\sim\,\sum_{k\in\,\Pi^{+}}\,f_{k}\,\Lambda
^{k}\,{\int\!\!\!\!\!\!-}\,|D|^{-k}\,+\,f(0)\,\zeta_{D}(0)+\,o(1),
\label{expansion}
\end{equation}
where the function $f$ only appears through the scalars
\begin{equation}
f_{k}=\,\int_{0}^{\infty}f(v)\,v^{k-1}\,dv. \label{coeff}
\end{equation}
The term independent of the parameter $\Lambda$ is the value at
$s=0$ (regularity at $s=0$ is assumed) of the zeta function
\begin{equation}
\zeta_{D}(s)=\,\mathrm{Tr}\,(|D|^{-s})\,. \label{zeta}
\end{equation}

The terms involving negative powers of $\Lambda$ involve the full
Taylor expansion of $f$ at $0$.

\medskip

Let us briefly review the classical relation between residues and
the heat kernel expansion in order to check the numerical
coefficients.

For the positive operator $\Delta=D^2$ one has,
\begin{equation}
|D|^{-s}=\Delta^{-s/2} = \frac{1}{\Gamma \left( \frac{s}{2} \right)}
\int_0^{\infty} e^{-t \Delta} \, t^{s/2-1} \, dt \label{laplace}
\end{equation}
and the relation between the asymptotic expansion,
\begin{equation}
{\rm Trace} \, (e^{-t\Delta}) \sim \sum \, a_{\alpha} \, t^{\alpha}
\qquad (t \rightarrow 0) \label{heat}
\end{equation}
and the $\zeta$ function,
\begin{equation}
\zeta_D (s) = {\rm Trace} \, (\Delta^{-s/2}) \label{zetaD}
\end{equation}
is given by the following result.

\begin{lem}
\begin{itemize}
\item A non-zero term $a_{\alpha}$ with $\alpha < 0$ gives a {\it pole} of
$\zeta_D$ at $-2\alpha$ with
\begin{equation}
 {\rm Res}_{s=-2\alpha} \, \zeta_D (s) = \frac{2 \, a_{\alpha}}{\Gamma
(-\alpha)} \label{respole}
\end{equation}
  \item The absence of $\log t$ terms gives regularity  at $0$ for $\zeta_D$ with
\begin{equation}
\zeta_D (0) = a_0 \, . \label{zeta0}
\end{equation}
\end{itemize}
\end{lem}

\proof We just check the coefficients,
 replacing ${\rm Trace} \, (e^{-t\Delta})$ by $a_{\alpha} \, t^{\alpha}$
and using
$$
\int_0^1\,t^{\alpha+s/2-1}\, dt=\,(\alpha+s/2)^{-1}
$$
one gets the first statement. The second follows from the
equivalence
$$
\frac{1}{\Gamma \left( \frac{s}{2} \right)}\sim \frac s 2\,,\quad
s\to 0
$$
so that only the pole part at $s=0$ of
$$
\int_0^{\infty} \Tr( e^{-t \Delta}) \, t^{s/2-1} \, dt
$$
contributes to the value $\zeta_D (0)$. But this pole part is given
by $$ a_0 \int_0^1\,t^{s/2-1}\, dt=\,a_0\frac 2 s$$ so that one gets
\eqref{zeta0}.
\endproof

\begin{rem}\label{f0f2f4}{\rm
The relations \eqref{respole} and \eqref{zeta0} in particular show
that our coefficients $f_0$, $f_2$ and $f_4$ are related to the
coefficients of the asymptotic expansion of the spectral action as
written in \cite{cc2} in the following way. Our $f_0$ is the $f_4$
of \cite{cc2}. Our $f_2$ is $1/2$ of the $f_2$ of \cite{cc2}. Our
$f_4$ is $1/2$ of the $f_0$ of \cite{cc2}. In fact our
$f(u)=\chi(u^2)$, for $\chi$ as in (2.14) of \cite{cc2}.}
\end{rem}


\begin{thebibliography}{99}

\bibitem{ACKMPRW} H.Arason, D.J.~Castano, B.~Kesthlyi, E.J.Piard,
P.Ramond, B.D.Wright, {\em Renormalization-group study of the
standard model and its extensions: the standard model}, Phys. Rev.
D, 46 (1992) N.9, 3945--3965.

\bibitem{Atiyah} M.F. Atiyah. $K$-Theory. {\it Benjamin, New York,} 1967

\bibitem{Avramidi} I. G. Avramidi  {\em  Covariant methods for the calculation of the effective action
in quantum field theory and investigation of higher-derivative
quantum gravity}. PhD Thesis, Moscow University 1986, hep-th/9510140

\bibitem{Barrett} J. W.~Barrett  {\em  A Lorentzian version of the
non-commutative geometry of the standard model of particle physics},
hep-th/0608221

\bibitem{Iochum}  L.~Carminati, B.~Iochum, T.~Schucker, {\em Noncommutative Yang-Mills and Noncommutative Relativity: A Bridge Over Trouble Water}, Eur.Phys.J. C8 (1999) 697--709.

\bibitem{CEIN} J.A.~Casas, J.R.~Espinosa, A.~Ibarra, I.~Navarro, {\em General RG Equations for Physical Neutrino Parameters and their Phenomenological Implications}, Nucl.Phys. B573 (2000) 652--684.

\bibitem{cc1}  A.~Chamseddine, A.~Connes, {\em  Universal Formula for Noncommutative Geometry Actions: Unification of
Gravity and the Standard Model}, Phys. Rev. Lett. 77, 486804871
(1996).

\bibitem{cc2} A.~Chamseddine, A.~Connes, {\em  The Spectral Action Principle}, Comm. Math. Phys. 186, 731-750 (1997).

\bibitem{cc3} A.~Chamseddine, A.~Connes, {\em  Scale Invariance in the Spectral Action}, J.Math.Phys.47:063504
(2006).
\bibitem{CFF} A.~Chamseddine, G. ~Felder and J.~Fr\"ohlich, {\em   Unified Gauge theories in Noncommutative Geometry},
Phys.Lett.B296:109-116,(1992).


\bibitem{CMGMP} D.~Chang, R.N.~Mohapatra, J.M.~Gipson, R.E.~Marshak,
M.K.~Parida, {\em Experimental tests of new $SO(10)$ grand
unification}, Phys. Rev. D 31 (1985) N.7, 1718-1732.

\bibitem{CoPer} A.~Codello, R.~Percacci, {\em Fixed points of higher derivative gravity}, hep-th/0607128.

\bibitem{Coleman} S.~Coleman, {\em Aspects of symmetry}, Selected Erice
Lectures, Cambridge University Press, 1985.

\bibitem{Co-book}  A.~Connes, {\it Noncommutative geometry},
 Academic Press (1994).

\bibitem{Coreal} A.~Connes, {\em  Non commutative geometry and
reality},   Journal of Math. Physics  36 no. 11 (1995).

 \bibitem{CoSM} A.~Connes, {\em Gravity coupled with matter and the
foundation of noncommutative geometry}, Comm. Math. Phys. (1995)

\bibitem{CoSMneu} A.~Connes, {\em Noncommutative Geometry and the standard model with neutrino
mixing}, hep-th/0608226.

\bibitem{cc4} A.~Connes, A.~Chamseddine,  {\em Inner fluctuations of the spectral
action}, hep-th/0605011.

\bibitem{CoMM} A.~Connes, M.~Marcolli {\em Noncommutative geometry
from quantum fields to motives}, Book in preparation.

\bibitem{dab}  L.~D\c{a}browski, A.~Sitarz, {\em
Dirac operator on the standard Podle\'s quantum sphere},
Noncommutative Geometry and Quantum Groups, Banach Centre
Publications 61, Hajac, P.~M. and Pusz, W. (eds.), Warszawa: IMPAN,
2003, pp. 49--58.

\bibitem{Dono} J.F. Donoghue, {\em General relativity as an effective field theory: the
leading quantum corrections}, Phys. Rev. D, Vol.50 (1994) N.6,
3874-3888.

\bibitem{EJ} M.B.~Einhorn, D.R.T.~Jones, {\em
Effective potential and quadratic divergences}, Phys. Rev. D, Vol.46
(1992) N.11, 5206-5208.

\bibitem{feynmgrav} R.~Feynman, {\em Feynman lectures on gravitation},
Perseus books (1995).

\bibitem{FGV} H.~Figueroa, J.M.~Gracia-Bond\'ia, J.~Varilly, {\em
Elements of Noncommutative Geometry}, Birkh\"auser, 2000.

\bibitem{FroGaw} J.~Fr\"ohlich, K.~Gawedski, {\em Conformal
field theory and geometry of strings}, in ``Mathematical quantum
theory. I. Field theory and many-body theory" pp. 57--97, CRM Proc.
Lecture Notes, 7, AMS, 1994.

\bibitem{Gilkey} P.~Gilkey, {\em Invariance Theory, the  Equation and
the Atiyah-Singer Index Theorem}, Wilmington, Publish or Perish,
1984.

\bibitem{gbis} J.~Gracia-Bonda, B.~Iochum, T.~Schucker, {\em The standard model in noncommutative
geometry and fermion doubling}. Phys. Lett. B 416 no. 1-2 (1998),
123--128.

\bibitem{kolb} R.~Holman, E.~Kolb, S.~Vadas, Y.~Wang, {\em Scale invariant extended inflation}.
Phys. Review. D 43 no. 12 (1991), 3833--3845.

\bibitem{InSu} T.Inagaki, H.Suzuki, {\em Majorana and Majorana--Weyl
fermions in lattice gauge theory}, JHEP 07 (2004) 038, 30 pages.

\bibitem {kastler} D.~Kastler, {\em  Noncommutative geometry and fundamental
physical interactions: The Lagrangian level}, Journal. Math. Phys.
41 (2000), 3867-3891.

\bibitem{Kolda} C.~Kolda, H.~Murayama, {\em
The Higgs mass and new physics scales in the minimal standard
model}, JHEP 007 (2000) 035, 21 pages.

\bibitem{knecht} M.~Knecht, T.~Schucker {\em Spectral action and big
desert} Phys. Lett. B 640, 272 (2006).

\bibitem{reuter} O.~Lauscher, M.~Reuter,
{\em  Asymptotic Safety in Quantum Einstein Gravity: nonperturbative
renormalizability and fractal spacetime structure}, hep-th/0511260

\bibitem{Lawson} H. B.~Lawson, M-L.~Michelsohn {\em Spin geometry},
Princeton Mathematical Series, 38. Princeton University Press,
Princeton, NJ, 1989.

\bibitem{schulaz} S. Lazzarini, T.~Schucker,
{\em  A Farewell To Unimodularity},  Phys.Lett.B510, 277 (2001).

\bibitem{lizzi} F.~Lizzi, G.~Mangano, G.~Miele, G.~Sparano, {\em
Fermion Hilbert space and Fermion Doubling in the Noncommutative
Geometry Approach to Gauge Theories}, Phys.Rev.D55, 6357 (1997).

\bibitem{MoPa} R.N.~Mohapatra, P.B.~Pal, {\em Massive neutrinos in
physics and astrophysics}, World Scientific, 2004.

\bibitem{NiWa} P.~van Nieuwenhuizen, A.~Waldron, {\em On Euclidean
spinors and Wick rotations}, Phys. Lett. B 389 (1996) 29--36.

\bibitem{PaTo} L.~Parker, D.J.~Toms, {\em Renormalization-group
analysis of grand unified theories in curved spacetime}, Phys. Rev.
D, Vol.29 (1984) N.8, 1584--1608.

\bibitem{Perc} R.~Percacci, {\em Renormalization group,
systems of units and the hierarchy problem}, hep-th/0409199.

\bibitem{Pilaf} A.~Pilaftsis, {\em Gauge and scheme dependence
of mixing renormalization}, Phys. Rev. D Vol.65 (2002) N.11 16pp.

\bibitem{Ramond} P.~Ramond, {\em Field theory: a modern primer}, Addison-Wesley, 1990.

\bibitem{Reina} L.~Reina, {\em TASI 2004 Lecture notes on Higgs
boson physics}, hep-ph/0512377.

\bibitem{ross} G.~Ross, {\em Grand unified theories}, Frontiers in Physics Series,
Vol.60, Benjamin, New York, (1985).

\bibitem{Sher} M.~Sher, {\em Electroweak Higgs potential and vacuum
stability}, Phys. Rep. Vol.179 (1989) N.5-6,  273--418.

\bibitem{VDiag} M.~Veltman, {\em Diagrammatica: the path to Feynman
diagrams}, Cambridge Univ. Press, 1994.

\bibitem{Weinberg} S.~Weinberg, {\em Gravitation and Cosmology}, John Wiley and sons (1972).



\end{thebibliography}
\end{document}